\documentclass[iop]{emulateapj}

\usepackage{color}
\usepackage{amssymb, amsmath}

\usepackage[utf8]{inputenc}
\usepackage{lipsum}
\usepackage{pifont}
\usepackage{amsmath, amssymb, bm}
\usepackage{balance}
\usepackage{multirow}
\usepackage{mathtools}
\usepackage{enumitem}
\usepackage{booktabs}
\usepackage{appendix}

\shorttitle{Statistical properties of paired fixed fields}
\shortauthors{Francisco Villaescusa-Navarro et al.}

\usepackage[usenames,dvipsnames]{xcolor}
\usepackage{hyperref}
\hypersetup{
    colorlinks = true,
    citecolor = {MidnightBlue},
    linkcolor = {BrickRed},
    urlcolor = {blue}		
}

\newcommand{\be}{\begin{equation}}
\newcommand{\ee}{\end{equation}}
\newcommand{\ba}{\begin{eqnarray}}
\newcommand{\ea}{\end{eqnarray}}

\begin{document}

\title{Statistical properties of paired fixed fields}

\author{Francisco Villaescusa-Navarro$^{1,\dagger}$, Sigurd Naess$^1$, Shy Genel$^{1,2}$, Andrew Pontzen$^{3}$, Benjamin Wandelt$^{1,4,5}$, Lauren Anderson$^{1}$,  Andreu Font-Ribera$^{2}$, Nicholas Battaglia$^{1,6,7}$, 
David N. Spergel$^{7,1}$\\}
\affil{$^{1}${Center for Computational Astrophysics, Flatiron Institute, 162 5th Avenue, 10010, New York, NY, USA}}
\affil{$^{2}${Columbia Astrophysics Laboratory, Columbia University, 550 West 120th Street, New York, NY 10027, USA}}
\affil{$^3${Department of Physics and Astronomy, University College London, 132 Hampstead Road, London, NW1 2PS, United Kingdom}}
\affil{$^4$Institut d'Astrophysique de Paris 98bis, Boulevard Arago, 75014 Paris, France}
\affil{$^5$Sorbonne Universites, Institut Lagrange de Paris 98 bis Boulevard Arago, 75014 Paris, France}
\affil{$^6$Department of Astronomy, Cornell University, Ithaca, NY, 14853, USA}
\affil{$^7$Department of Astrophysical Sciences, Princeton University, Peyton Hall, Princeton NJ 08544-0010, USA}
\altaffiltext{$\dagger$}{fvillaescusa@flatironinstitute.org}

\begin{abstract}
The initial conditions of cosmological simulations are commonly drawn from a Gaussian ensemble. The limited number of modes inside a simulation volume gives rise to statistical fluctuations known as \textit{sample variance}, limiting the accuracy of simulation predictions. Fixed fields offer an alternative initialization strategy; they have the same power spectrum as standard Gaussian fields but without intrinsic amplitude scatter at linear order. Paired fixed fields consists of two fixed fields with opposite phases that cancel phase correlations which otherwise induce second-order scatter in the non-linear power spectrum. We study the statistical properties of those fields for 19 different quantities at different redshifts through a large set of 600 N-body and 506 state-of-the-art magneto-hydrodynamic simulations covering a wide range of scales, mass and spatial resolutions. We find that paired fixed simulations do not introduce a bias on any of the examined quantities. We quantify the statistical improvement brought by these simulations, over standard ones, on different power spectra such as matter, halos, CDM, gas, stars, black-holes and magnetic fields, finding that they can reduce their variance by factors as large as $10^6$. We quantify the improvement achieved by fixing and by pairing, showing that sample variance in some quantities can be highly suppressed by pairing after fixing. Paired fixed simulations do not change the scatter in quantities such as the probability distribution function of matter density, or the halo, void or stellar mass functions. We argue that procedures aiming at reducing the sample variance of those quantities are unlikely to work. Our results show that paired fixed simulations do not affect either mean relations or scatter of galaxy properties, and suggest that the information embedded in 1-pt statistics is highly complementary to that in clustering. 
\end{abstract} 

\keywords{large-scale structure of universe -- methods: numerical -- methods: statistical }

\section{Introduction}
\label{sec:introduction}

The standard model of cosmology is a well established theoretical framework that explains with great success a large and diverse range of cosmological observables. The parameters of the model represent fundamental physics quantities such as the nature of dark energy, the density of dark matter or the sum of the neutrino masses. The goal of current and upcoming cosmological surveys is to determine the value of those parameters with the highest accuracy possible, in order to improve our knowledge of fundamental physics. 

The amount of information that can be extracted from cosmological surveys depends on the accuracy of the theoretical model. For instance, theoretical predictions are very accurate and fast-to-compute in the linear regime, but the amount of information that can be extracted with them is limited, since that regime can only accurately describe the largest scales. Perturbation theory \citep{Bernardeau_2002} is an ideal tool to make accurate theoretical predictions in the mildly non-linear regime. However, theoretical predictions in the fully non-linear regime require numerical simulations. 

Ideally, the best way to extract cosmological information would be by evaluating the likelihood in every point of the parameter space by using the theoretical prediction from cosmological hydrodynamic simulations. This procedure has been impractical so far \citep[see however][for similar efforts with the Ly$\alpha$-forest]{Palanque_2015} due to several factors: 1) the volume of the parameter space can be very large, requiring many simulations for sampling it; 2) a very large number of simulations are needed to compute the covariance matrix in each point of the parameter space; 3) simulations covering representative survey volumes with the required mass resolution are computationally expensive; 4) each simulation has an intrinsic variance, commonly called \textit{sample variance}, arising from the limited number of modes it contains, such that many simulations are needed to compute the mean. 

The first point can be addressed by running simulations on a subset of strategic locations in the parameter space \citep[see e.g.][]{Coyote}. For the second and third points, a large amount of work has been carried out to speed up the running time of N-body simulations and to evaluate the covariance matrix, at the expense of accuracy \citep{PTHALOS, Kitaura_2013, PATCHY, Tassev_2013,Tassev_2015,L-PICOLA, EZmocks, FastPM, Monaco_2002, Taffoni_2002, Monaco_2002b, Monaco_2013, Chuang_2015,Rizzo_2017}. Those methods, however, do not include the non-linear effects of baryons. 

The scope of this paper is to investigate how to mitigate the fourth point, i.e. the intrinsic sample variance attached to each simulation. We focus our attention on paired fixed fields, introduced in \cite{Pontzen_2016, RP_16}. Those fields can be obtained from Gaussian density fields by performing certain operations on the amplitudes and/or the phases of their modes. \cite{RP_16} showed that numerical simulations run with those fields as initial conditions lead to quantities, such as the matter power spectrum, with a much lower variance than those obtained from traditional Gaussian fields. 

The purpose of this work is to further investigate the properties of paired fixed fields and 1) identify the quantities for which paired fixed fields help in reducing the intrinsic statistical scatter, 2) quantify the statistical improvement, 3) study whether a bias is introduced in any quantities.

We carry out our study using a large set of 600 N-body simulations with different box sizes and mass and spatial resolutions. We use them to study the impact of paired fixed simulations on the matter, halo and halo-matter power spectra, the halo bias, the probability distribution function of matter density, the halo mass function and the void mass function.

We then study the statistical properties of paired fixed simulations using a set of $\sim500$ state-of-the-art magneto-hydrodynamic simulations. We investigate the properties of the above quantities, along with the power spectra of the other components: gas, cold dark matter (CDM), stars, black-holes and magnetic fields. We also study the impact of paired fixed fields on the star-formation rate history, on the stellar mass function and on several internal galaxy properties such as radii or maximum circular velocity.

This paper is organized as follows. In section \ref{sec:definitions} we define Gaussian, paired Gaussian, fixed and paired fixed fields. The set of numerical simulations run for this project is described in section \ref{sec:methods}, where we also explain the tools we use to carry out the statistical analysis. We present the results from our N-body and hydrodynamic simulations in sections \ref{sec:Nbody} and \ref{sec:hydro_intermediate} and \ref{sec:hydro_small} for large, intermediate and small scales, respectively. In section \ref{sec:1pt_stats} we investigate whether we can generate fields with reduced sample variance in both their the 1-pt and 2-pt statistics. Finally, we draw the main conclusions of this paper in section \ref{sec:conclusions}.


\section{Definitions}
\label{sec:definitions}

We now define what Gaussian, paired Gaussian, fixed and paired fixed fields are. For a given density field $\rho(\vec{x})$, the density contrast is defined as
\be
\delta(\vec{x})=\frac{\rho(\vec{x})-\bar{\rho}}{\bar{\rho}}~,
\ee
where $\bar{\rho}=\langle \rho(\vec{x})\rangle$. We express its value in Fourier-space as
\be
\delta(\vec{k})=\frac{1}{(2\pi)^3}\int {\rm d}^3\vec{x}e^{-i\vec{k}\cdot\vec{x}}\delta(\vec{x})=Ae^{i\theta}
\ee
where $A$ is the mode's amplitude and $\theta$ is its phase. We notice that the value of both $A$ and $\theta$ depend on the particular wavelength, $\vec{k}$, considered. Since the density field is real the modes satisfy $\delta(-\vec{k})=\delta^*(\vec{k})$.  The power spectrum of the field is defined as
\be
\langle \delta(\vec{k}_1)\delta^*(\vec{k}_2)\rangle = \delta^D(\vec{k}_1-\vec{k}_2)P(k_1)
\ee
and for a simulation of box size $L$ and volume $V=L^3$ the above equation reads
\be
P(k_1)=\frac{(2\pi)^3}{V}\delta_{\vec{k}_1,\vec{k}_2}\langle \delta(\vec{k}_1)\delta^*(\vec{k}_2)\rangle~.
\ee
In a Gaussian density field $\theta$ is a random variable distributed uniformly between 0 and $2\pi$ whereas $A$ follows a Rayleigh distribution 
\be
p(A){\rm d}A=\frac{A}{\sigma^2}e^{-A^2/2\sigma^2}{\rm d}A~,
\label{Eq:Rayleigh}
\ee
with $\sigma^2=VP(k)/(16\pi^3)$. The mean value of the mode amplitude is
\be
\langle A \rangle = \int_0^\infty \frac{A^2}{\sigma^2}e^{-A^2/2\sigma^2}{\rm d}A=\sqrt{\frac{VP(k)}{32\pi^2}}~.
\label{Eq:meanA}
\ee
A density field built as above will satisfy 
\be
\langle \delta(\vec{k})\delta^*(\vec{k})\rangle=\langle A^2\rangle=\int_0^\infty \frac{A^3}{\sigma^2}e^{-A^2/2\sigma^2}{\rm d}A=\frac{VP(k)}{(2\pi)^3}.
\ee
A Gaussian field is completely described by its 2-pt correlation function, or power spectrum. 

It is interesting to consider a different distribution for the amplitudes of the modes that fulfills two conditions: 1) the amplitude of the power spectrum is the same as in Gaussian fields, i.e.~$\langle \delta(\vec{k})\delta^*(\vec{k})\rangle=VP(k)/(2\pi)^3$ and 2) it has no intrinsic scatter. The following distribution satisfies these two conditions:
\be
p(A){\rm d}A = \delta^D\left(A-\sqrt{\frac{VP(k)}{(2\pi)^3}}\right){\rm d}A~.
\label{Eq:fixed}
\ee
We note that in such fields, the value we assign to each mode with wavenumber $\vec{k}$ is not the mean of the Rayleigh distribution (see Eq. \ref{Eq:meanA}). We emphasize that fields constructed with amplitudes drawn from the above distribution are not Gaussian.

\begin{figure*}
\includegraphics[width=1\textwidth]{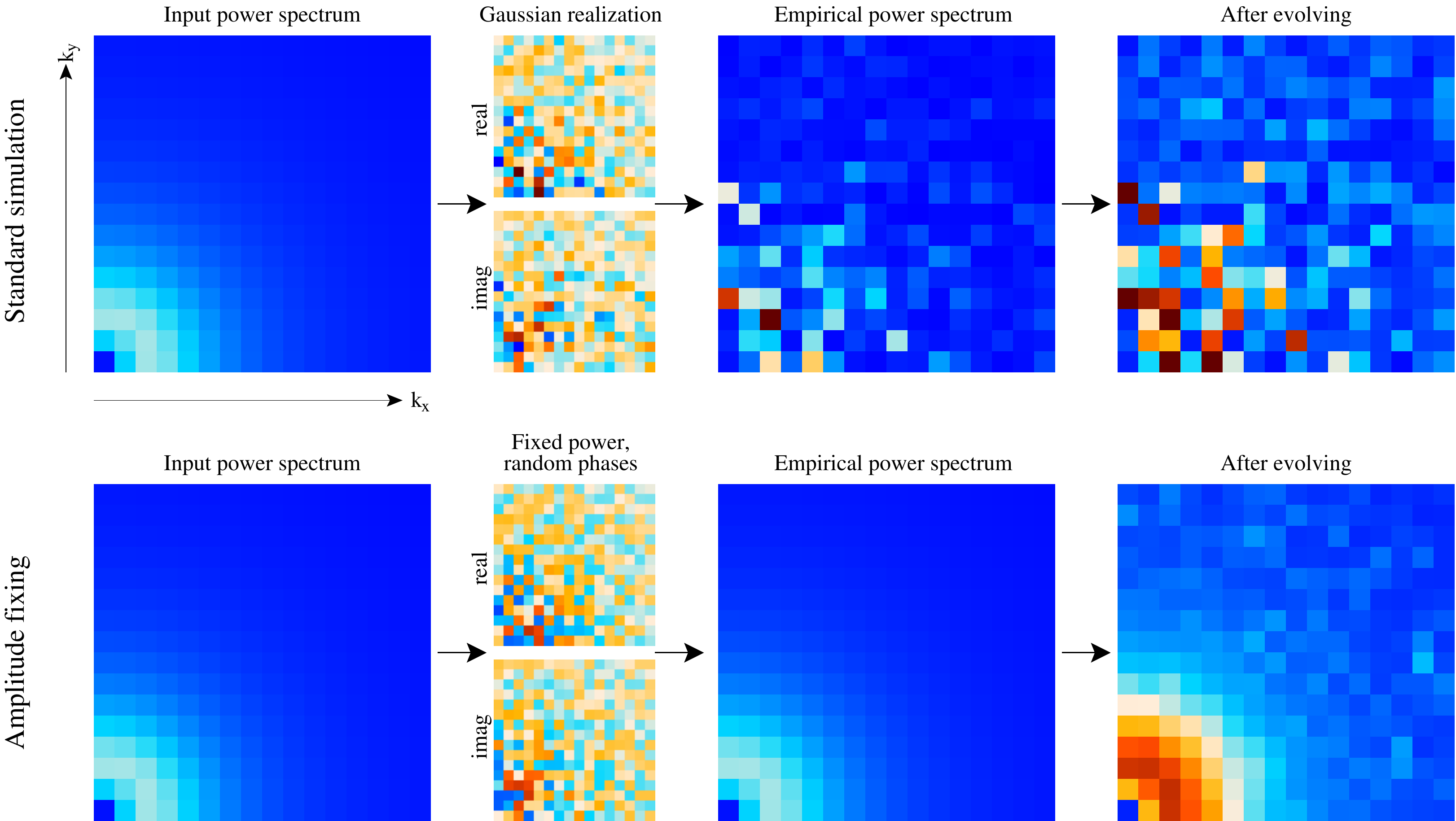}
\caption{Toy example illustrating sample variance in standard (top row) and
fixed (bottom row) simulations. In both cases, a random density
field (center-left, here shown as the real and and imaginary
Fourier components) is drawn from an input power spectrum (leftmost),
but because the fixed simulations fix the complex amplitude, they
avoid introducing sample variance into the empirical power spectrum
measured from the random realization (center-right). 
The rightmost panels show a sketch of the effect
of further evolution: linear scales simply grow proportionally,
preserving whatever sample variance was present, while on smaller
scales, non-linearities and mode mixing reintroduce some of the
suppressed cosmic variance in the fixed simulation (see
figure~\ref{fig:Mode_mixing}).}
\label{fig:Fixed_fields}
\end{figure*}

We define Gaussian, paired Gaussian, fixed and paired fixed fields as follows:
\begin{itemize}
\item {\bf Gaussian field}: A field with $\delta(\vec{k})=Ae^{i\theta}$, where $A$ follows the Rayleigh distribution of Eq. \ref{Eq:Rayleigh}.
\item {\bf Paired Gaussian field}: A pair of Gaussian fields, $\delta_1(\vec{k})=Ae^{i\theta}$ and $\delta_2(\vec{k})=Ae^{i(\theta+\pi)}=-\delta_1(\vec{k})$, where the values of $A$ and $\theta$ are the same for the two fields and $A$ follows the Rayleigh distribution of Eq. \ref{Eq:Rayleigh}.
\item {\bf Fixed field}: A field with $\delta(\vec{k})=Ae^{i\theta}$, where $A$ follows the distribution of Eq. \ref{Eq:fixed}.
\item {\bf Paired fixed field}: A pair of fields, $\delta_1(\vec{k})=Ae^{i\theta}$ and $\delta_2(\vec{k})=Ae^{i(\theta+\pi)}=-\delta_1(\vec{k})$, where the values of $A$ and $\theta$ are the same for the two fields and $A$ follows the distribution of Eq. \ref{Eq:fixed}.
\end{itemize}
In all the above fields $\theta$ is a random variable distributed uniformly between 0 and $2\pi$. Any of the above fields satisfies the Hermitian condition: $\delta(-\vec{k})=\delta^*(\vec{k})$. In Fig. \ref{fig:Fixed_fields} we show the 2D power spectrum from a Gaussian and fixed field and its comparison with the input power spectrum. We also show schematically the effects of non-linear evolution.

Paired Gaussian fields were introduced in \cite{Pontzen_2016}. Fixed fields have been relatively well known \citep[see e.g.][]{Viel_2010}. Paired fixed fields were first studied in \cite{RP_16}. 

In this paper we have run simulations where the initial conditions have been generated using the above fields. We refer to these simulations as standard, paired, fixed and paired fixed simulations. 

We note that although Gaussian and fixed fields share, by construction, the same power spectrum, they differ in higher order correlations like the trispectrum. For instance, the variance of the power spectrum 
\begin{eqnarray}
\sigma(P(k))&=&\langle \left(P(k)-\bar{P}(k)\right)^2 \rangle\nonumber\\
&=&\frac{(2\pi)^6}{V^2}\langle \delta(\vec{k}) \delta^*(\vec{k})\delta(\vec{k}) \delta^*(\vec{k}) \rangle - \bar{P}^2(k)
\end{eqnarray}
is equal to $P(k)$ for Gaussian fields but is identically zero for fixed fields. For this reason, we expect that the scatter in the matter power spectrum of fixed (and paired fixed) fields will be lower than in Gaussian fields. One of the purposes of this paper is to study the reduction on the scatter of a considered quantity achieved by fixed and paired fixed simulations.

On the other hand, the value of some quantities, e.g.~the probability distribution function (pdf) of the density field, will depend on the value of the n-point correlation function. Since the value of these functions may be different in Gaussian and fixed fields, we expect that fixed, and paired fixed, simulations may introduce a bias on the value of those quantities. \cite{RP_16} argued using perturbation theory that observable quantities should be unbiased but the accuracy of that statement in the non-linear evolution remains to be tested. Thus, the other key point of this work is to quantify the magnitude of that bias.

\section{Methods}
\label{sec:methods}

In this section we describe the numerical simulations run for this work. We also explain the statistical analysis we carry out to quantify 1) whether paired fixed simulations introduce a bias on the considered quantity and 2) the statistical improvement achieved over standard simulations.

\subsection{Numerical simulations}
\label{sec:simulations}

\begin{table*}
\begin{center}
\resizebox{1\textwidth}{!}{\begin{tabular}{| c | c | c | c | c | c | c | c | c | c | c |}
\hline
Name & Type & Code & \# Standard & \# paired fixed & $N^{1/3}_{\rm CDM}$ & $N^{1/3}_{\rm gas}$ & $m_{\rm CDM}$ & $m_{\rm gas}$ & $\epsilon$&Box size\\
& & & realizations & realizations & & & ($h^{-1}M_\odot$) & ($h^{-1}M_\odot$) & ($h^{-1}{\rm kpc}$) & ($h^{-1}{\rm Mpc}$)\\
\hline
N1000 & N-body & {\sc Gadget3} & 100 & 100 & 512 & - & $6.6\times10^{11}$ &- & 50 &1000\\
\hline
N20 & N-body & {\sc Gadget3} & 100 & 100 & 256 & - & $4.2\times10^{7}$ &- & 2 & 20\\
\hline
H200 & Hydrodynamic & {\sc Arepo} & 26 & 15 & 640 & 640 & $2.3\times10^9$ & $4.2\times10^8$ & 8 & 200 \\
\hline
H20 & Hydrodynamic & {\sc Arepo} & 250 & 100 & 256 & 256 & $3.6\times10^7$& $6.5\times10^6$& 2&20\\
\hline
\end{tabular}}
\end{center}
\caption{Specifications of the simulations run for this paper. The value of the cosmological parameters is the same for all simulations: $\Omega_{\rm m}=0.3175$, $\Omega_{\rm b}=0.049$, $\Omega_\nu=0$, $n_s=0.96$, $h=0.67$, $\sigma_8=0.834$. We have four different sets of simulations, with different box sizes and numbers of particles. The first letter of the name set represents whether it is an N-body (N) or hydrodynamic (H) simulation, while the number thereafter is the box size in $h^{-1}$ Mpc. Note that one paired fixed realization corresponds to two simulations with random phases flipped by $\pi$. The physics included in our magneto-hydrodynamic simulations is: radiative cooling, star formation, metal enrichment, galactic winds, black hole accreting and feedback. The H200 simulations are similar to the IllustrisTNG300-3 simulation.}
\label{table:sims}
\end{table*}

A large number of realizations is needed to study the statistical properties of paired fixed simulations. Thus, in this work we have run an unusually large number, $\sim1000$, of standard and paired fixed simulations. 

The purpose of this paper is to investigate the properties of those fields across a large range of scales, from linear to fully non-linear scales. Doing so with a single set of simulations would require the simulations to have a large box size and a large number of particles. Having a sensible number of those simulations will be computationally expensive. Thus, we decided to run three different sets of simulations that encompass three different ranges of scales: 1) N-body simulations with box sizes of 1000 $h^{-1}$Mpc at low mass resolution, 2) hydrodynamic simulations with boxes of 200 $h^{-1}$Mpc at intermediate mass resolution and 3) both N-body and hydrodynamic simulations with boxes of 20 $h^{-1}$Mpc at high mass resolution. 

All our simulations share the same value of the cosmological parameters, $\Omega_{\rm m}=0.3175$, $\Omega_{\rm b}=0.049$, $\Omega_\nu=0$, $n_s=0.96$, $h=0.67$, $\sigma_8=0.834$, that are in agreement with the results by Planck \citep{Planck_2015}. We have generated the initial conditions by displacing and assigning peculiar velocities to particles initially laid down on a regular grid by using the Zel'dovich approximation at $z=99$. The initial power spectrum and growth rates are computed by rescaling the $z=0$ matter power spectrum and transfer functions according to the method described by \cite{Zennaro_2017}, i.e.~we account for both the scale-dependence of the growth factor and growth rate in simulations with 2 fluids.

The N-body simulations were run using the {\sc Gadget-III} code, last described in \cite{Gadget}. They consist of two different sets. One set follows the evolution of $512^3$ CDM particles in a periodic box of 1000 comoving $h^{-1}{\rm Mpc}$ while in the other $256^3$ CDM particles are evolved in a box size of 20 comoving $h^{-1}{\rm Mpc}$. The gravitational softening is set to 50 and 2 comoving $h^{-1}$kpc, respectively. We call these sets N1000 and N20, and we use them to study the statistical properties of paired fixed fields on large  and small scales (and for very massive and low-mass objects), respectively. Each set contains 300 simulations: 100 standard simulations and 100 pairs of fixed simulations. We will show results obtained from the N1000 set, while the N20 is mainly used to cross-check the results of the H20 simulation set that we describe below.

We also have two different sets of magneto-hydrodynamic simulations, run with the {\sc arepo} code \citep{Arepo}. In one, we follow the evolution of $640^3$ CDM plus $640^3$ gas particles in a periodic box of 200 comoving $h^{-1}$ Mpc, while in the other we have a box of 20 comoving $h^{-1}$Mpc with $256^3$ CDM plus $256^3$ gas particles. Both use the IllustrisTNG models of galaxy formation, which include gas radiative cooling, star-formation, metal enrichment, galactic winds, and black hole accretion and feedback \citep{WeinbergerR_16a,PillepichA_16a}. The numerical methods and subgrid physics models build upon the Illustris simulation model \citep{Vogelsberger_2013, Vogelsberger_2014a, Vogelsberger_2014b, Torrey_2014, Genel_2014}. The softening lengths are 8 and 2 comoving $h^{-1}$kpc, respectively. We name these sets H200 and H20, correspondingly. H200 has 56 simulations, 26 standard and 15 pairs, while H20 is made of 450 simulations: 250 standard and 100 pairs. We use the H200 simulations, which are very close to the TNG300-3 simulation \citep{MarinacciF_17a,NaimanJ_17a,NelsonD_17a,PillepichA_17a,SpringelV_17a}, to study the improvement on  different power spectra (matter, CDM, gas) introduced by paired fixed simulations on intermediate scales. The H20 set is used to study the properties of paired fixed simulations on very small scales and to investigate the impact of those fields on galaxy properties. A summary of our simulation suite is shown in Table \ref{table:sims}.

Snapshots are saved at different redshifts, from $z=15$ down to $z=0$. In this work we focus on redshifts 0, 1 and 5. Dark matter halos are identified using the Friends-of-Friends algorithm \citep{FoF} with a value of the linking length parameter $b=0.2$. In the hydrodynamic simulations we identify galaxies through the {\sc subfind} algorithm \citep{subfind}. We use the algorithm described in \cite{Arka_2016} to identify voids in the matter distribution of our snapshots.

\subsection{Formalism}
\label{subsec:analysis_tools}

Here we describe the formalism we use to carry out the statistical analysis for each quantity considered in this paper. The most important goals of this work are to 1) study whether paired fixed simulations introduce a bias with respect to standard simulations and 2) quantify the statistical improvement achieved by fixed and paired fixed simulations in comparison with standard simulations.

\begin{figure*}
\begin{center}
\includegraphics[width=0.42\textwidth]{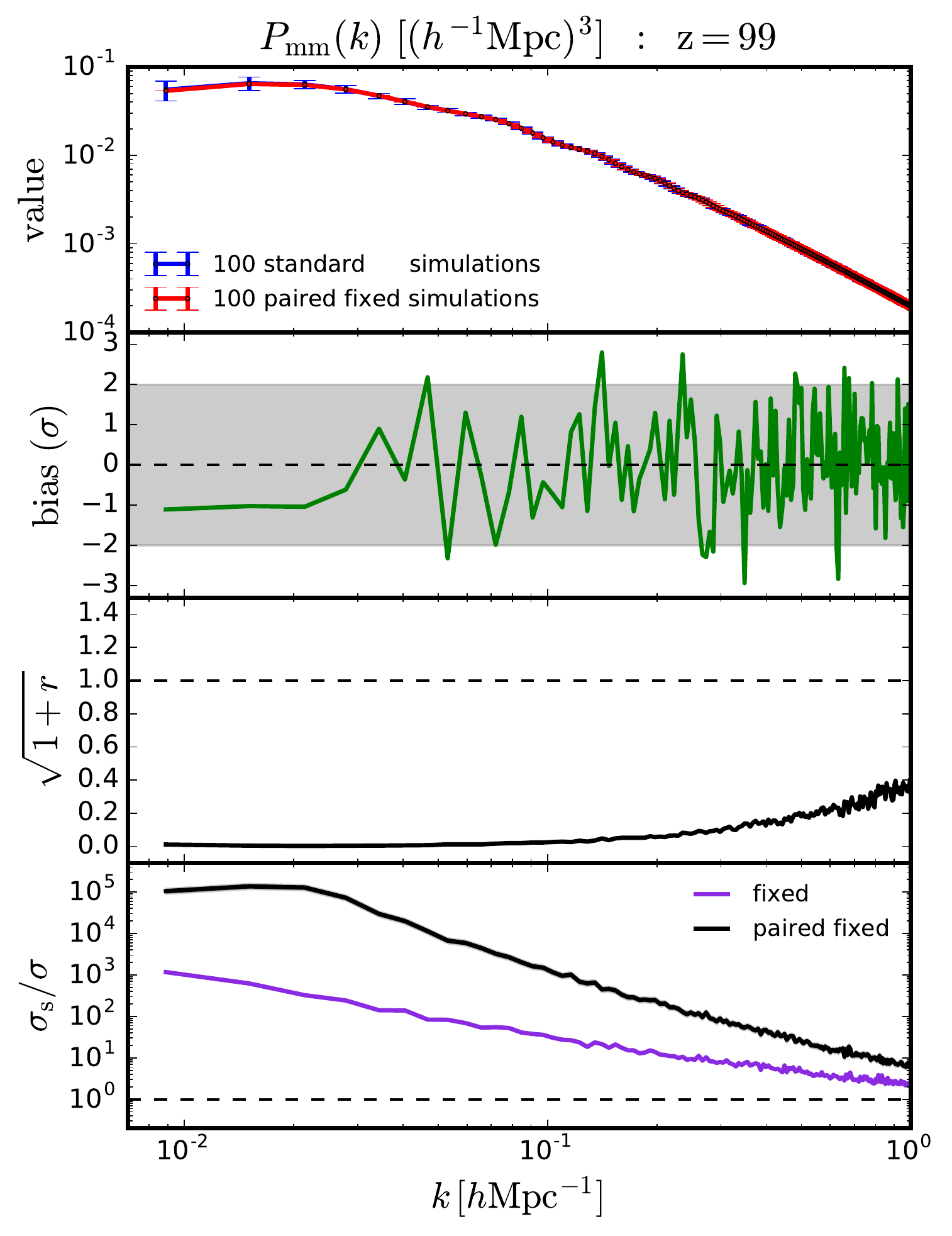}
\includegraphics[width=0.42\textwidth]{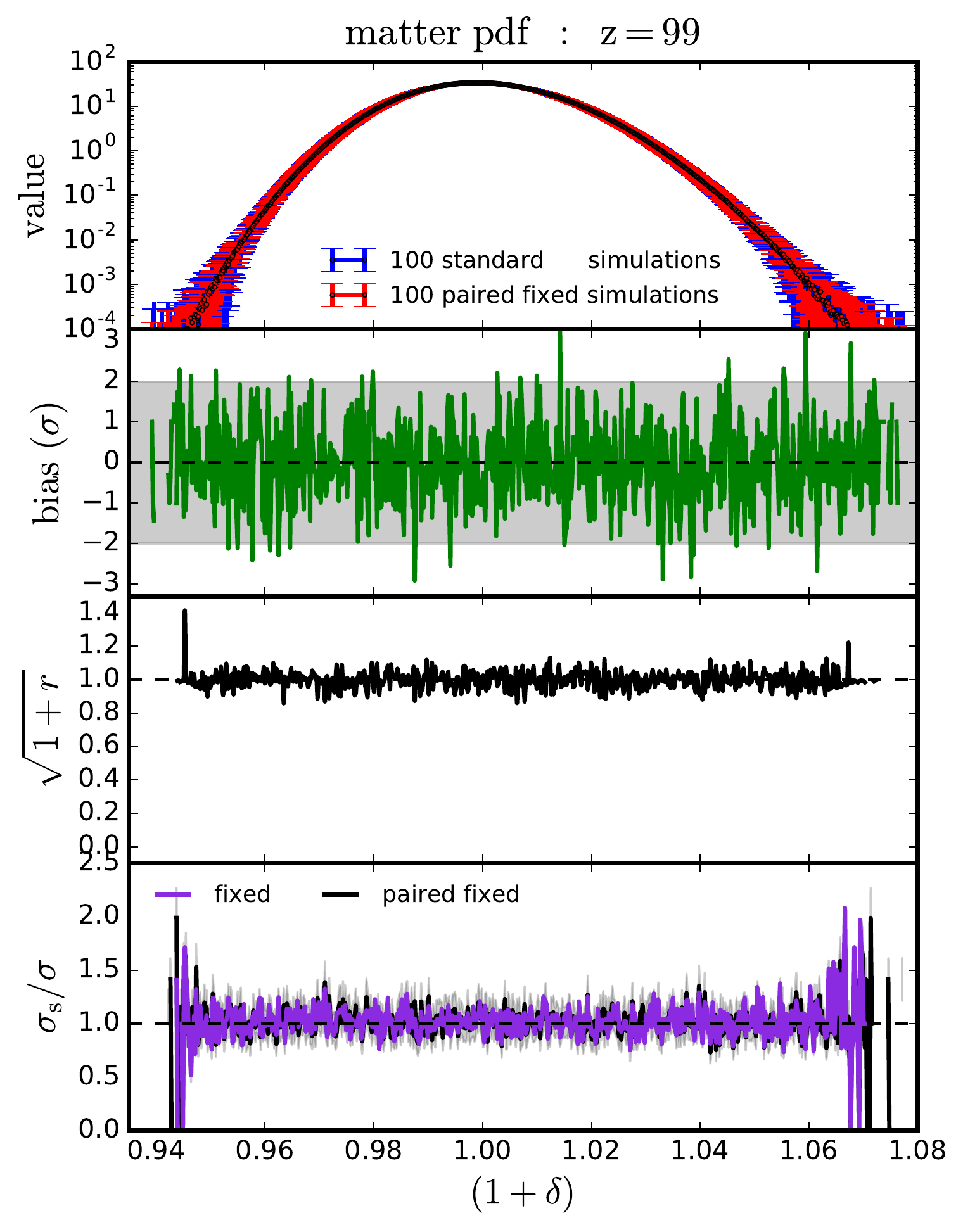}
\caption{Most of the plots shown in this paper have the same structure, that we describe in detail here. {\bf First panel:} Mean and standard deviation of the considered quantity from the standard (blue) and paired fixed simulations (red). {\bf Second panel}: An important aspect of our study is to investigate whether paired fixed simulations introduce a bias on the considered quantity. In this panel we show the difference between the means from the two simulation sets, divided by the expected error on the difference (see text for details). This panel quantifies thus the statistical agreement between both data sets. Any point within the grey region indicates that the agreement between means is within $2\sigma$. The black dashed line indicates a $0\sigma$ bias. {\bf Third panel:} The other important quantify in paired fixed simulations is the statistical improvement they achieve with respect to standard simulations. For the considered quantity, the variance from standard and paired fixed simulations is given by $\sigma_{\rm s}^2$ and $\sigma_{\rm pf}^2$, respectively. We can express $\sigma_{\rm pf}$ as $\sigma_{\rm pf}=\sigma_{\rm f}\sqrt{(1+r)}$, where $\sigma_{\rm f}$ is the standard deviation of each individual pair and $r$ is the cross-correlation coefficient between the pairs. Expressing the variance in that way is very helpful as the improvement achieved by fixing the amplitude and flipping the phase is embedded in the value of $\sigma_{\rm f}$ and $r$, respectively. In this panel we show the value of $\sqrt{1+r}$. 0 values mean that the errors are completely anti-correlated and a large statistical improvement can be achieved. When the quantities from the two pairs are independent (dashed black line) $r=0$ and $\sqrt{1+r}=1$ and no statistical improvement is brought by pairing. If both simulations are completely correlated the value of $\sqrt{1+r}$ is $\sqrt{2}$ and the normalized variance worsen. {\bf Fourth panel:} This panel shows the ratios between the standard deviation from the standard and paired fixed simulations $\sigma_{\rm s}/\sigma_{\rm pf}$ (black line) and from the standard and fixed simulations $\sigma_{\rm s}/\sigma_{\rm f}$ (purple line). That ratio indicates the statistical improvement achieved by the fixing and pair fixing over traditional simulations. The grey region around the black line in that panel represents the $1\sigma$ uncertainty on the ratio. The dashed line indicate 1 and can be interpreted as no statistical improvement with respect to standard simulations. We can see how fixed and paired fixed simulations largely reduce the scatter on the matter power spectrum, while they leave the variance on the matter density pdf unaffected. }
\label{fig:Pk_ICs}
\end{center}
\end{figure*}

Throughout the paper we show plots that share the same structure and contain information on the above two statistical properties. An example of such a plot appears in the left panel of Fig.~\ref{fig:Pk_ICs}. 

We compute each quantity for each standard and paired fixed realization in the considered simulation set. For example, the left panel of Fig.~\ref{fig:Pk_ICs} considers the matter power spectrum at $z=99$. We denote by $X_{\rm s,i}$ and $X_{\rm pf,i}$ the value of that quantity from the realization i of the standard and paired fixed simulations, respectively. We compute the value of $X_{\rm pf,i}$ as
\be
X_{\rm pf,i}=\frac{1}{2}[X_{\rm pf,i,1}+X_{\rm pf,i,2}]
\ee
where $X_{\rm pf,i,1}$ and $X_{\rm pf,i,2}$ are the considered quantity in each simulation of a paired fixed realization\footnote{In the case of the non-linear power spectrum this demonstrably cancels phase correlation errors at leading order. For other quantities, it may be possible to construct improved estimators using cross-correlations between simulations, but this is beyond the scope of the present work.}. From $X_{\rm s,i}$ and $X_{\rm pf,i}$ we estimate the mean and variance of each simulation type as

\begin{eqnarray}
        \bar{X}_{\alpha}&\equiv&\langle X_{\alpha}\rangle=\frac{1}{N}\sum_{i=1}^NX_{\rm \alpha,i}\\[2ex]
        \sigma_{\rm \alpha}^2&\equiv&\left\langle(X_\alpha-\bar{X}_\alpha)^2 \right\rangle=\frac{\sum_{i=1}^N(X_{\rm \alpha,i}-\bar{X}_{\rm \alpha})^2}{N-1}   
\end{eqnarray}
where $\alpha=\{{\rm s, pf}\}$. The upper panel always shows the mean and standard deviation from the standard and paired fixed simulations in blue and red, respectively. 

In the second panel of each figure we quantify the bias introduced by the paired fixed simulations with respect to standard simulations. We calculate it by computing
\be
\frac{\bar{X}_{\rm s}-\bar{X}_{\rm pf}}{\sigma_{\rm s-pf}}
\ee
where $\sigma_{\rm s-pf}$ is the expected error on the difference between the means from the standard and paired fixed simulations. In this paper we have assumed that all the considered quantities are normally distributed. In that case, the expected error on the difference of the means is\footnote{Since the standard and paired fixed simulations have different random seeds, the covariance between them vanishes.}
\be
\sigma^2_{\rm s - pf}=\frac{\sigma^2_{\rm s}}{N_{\rm s}}+\frac{\sigma^2_{\rm pf}}{N_{\rm pf}}~,
\label{Eq:mean_diff_variance}
\ee
where $N_{\rm s}$ and $N_{\rm pf}$ are the number of standard and paired fixed realizations. We note that this is a reasonable assumption for power spectra, where the amplitude in a given $k$-bin receives contributions from many different independent modes. For halo/void mass functions and pdfs, a more appropriate distribution will be a Poissonian. However, in this work we only show results for bins that contain many halos/voids/cells. In that case, the Poisson distribution is well approximated by a Gaussian.

The green line in the second panel measures thus the bias introduced by the paired fixed procedure, with respect to the standard simulations, in $\sigma$ units. The grey band indicates where the bias is less than $2\sigma$.

In the third and fourth panels we quantify the statistical improvement achieved by the paired fixed simulations with respect to the standard simulations. The \textit{normalized variance}\footnote{While fixed simulations only contain one simulation, paired fixed contain two. For quantities in which pairing and fixing do not help, we will still see an improvement when using paired fixed simulations simply because we are estimating the quantity through two simulations instead of one. We correct for that by computing the normalized variance, so that we can compare directly $\sigma_{\rm pf}$, $\sigma_{\rm f}$ and $\sigma_{\rm p}$ (see appendix \ref{sec:analytics} for further details).} of the paired fixed simulations can be expressed as
\be
\sigma_{\rm pf}^2=\sigma_{\rm f}^2(1+r)~,
\label{eq:sigma_pf}
\ee
where $\sigma_{\rm f}$ is the variance of individual fixed simulations
\be
\sigma_{\rm f}^2=\langle (X_{\rm pf,1}-\bar{X}_{\rm pf})^2\rangle=\langle (X_{\rm pf,2}-\bar{X}_{\rm pf})^2\rangle
\ee
and $r$ is the cross-correlation coefficient between $X_{\rm pf,1}$ and $X_{\rm pf,2}$
\be
r=\frac{1}{\sigma_{\rm f}^2N} \sum_{i=1}^N (X_{\rm pf,i,1}-\bar{X}_{\rm pf,1})(X_{\rm pf,i,2}-\bar{X}_{\rm pf,2}).
\ee
We note that $\sigma_{\rm f}^2$ can be interpreted as the variance obtained by fixing the amplitude without doing pairing (i.e.~the variance of fixed simulations), while the value of $r$ measures the correlation between the two sets of pairs and $1+r$ can be interpreted as the statistical improvement on the variance achieved by pairing.

The third panel of Fig.~\ref{fig:Pk_ICs} shows the value of $\sqrt{1+r}$. If the two pair quantities are independent, $r=0$, and $\sqrt{1+r}=1$. In this case pairing does not bring any improvement and the variance of paired fixed simulations will be just the variance of fixed simulations\footnote{Notice that we expect an improvement of $1/\sqrt{2}$ in the variance if we compute a quantity with two independent measurements instead of one (see appendix \ref{sec:analytics}). However, in this work we are interested in the net gain, so we reabsorb that improvement in our definition.}. If the two pairs are completely correlated, $r=1$, pairing does actually worsens the results. This happens because the second simulation adds no information and is therefore wasted. Finally, if both pair quantities are completely anti-correlated, $r=-1$, the variance of the paired fixed simulations reduces to 0. This can be understood taking into account that if both pair quantities are completely anti-correlated, as $X_{\rm pf,1}$ increases its value, $X_{\rm pf,2}$ shrinks such that $\frac{1}{2}(X_{\rm pf,1}+X_{\rm pf,1})$ remains constant. Thus, the lower the value of $r$, the larger the improvement brought  by pairing. We emphasize that this is the improvement achieved by pairing once fixed. In other words, the value of $r$ from just paired simulations that are not fixed can be different from that of paired fixed (see appendix \ref{sec:analytics} for further details). We provide explanations for the actual values in that figure in section \ref{sec:Nbody}.

Finally, in the fourth panel of Fig.~\ref{fig:Pk_ICs} we show the ratios between the standard deviations of the standard and paired fixed simulations (solid black line), $\sigma_{\rm s}/\sigma_{\rm pf}$, and between the standard and fixed simulations (solid purple line), $\sigma_{\rm s}/\sigma_{\rm f}$. The purple line quantifies the statistical improvement achieved by fixing the amplitude while the black line represents the gain obtained by fixing and pairing. We notice that the black line can be obtained from the purple line and the line in the third panel through Eq. \ref{eq:sigma_pf}. The dashed horizontal line in the fourth panel shows a value of 1, indicating the level where fixed and paired fixed simulations do not bring any statistical improvement over standard simulations. The black line is also surrounded by a grey shaded region (hard to see in Fig. \ref{fig:Pk_ICs} due to the large dynamic range), indicating the associated error on the standard deviation ratio, which we estimate as (see appendix \ref{sec:variance_ratios})
\be
\sigma\left(\frac{\sigma_{\rm s}}{\sigma_{\rm pf}}\right)=\frac{1}{2}\left(\frac{\sigma_{\rm s}}{\sigma_{\rm pf}}\right)\sqrt{\frac{2}{N_{\rm s}} + \frac{2}{N_{\rm pf}}}
\label{Eq:error_std}
\ee
We only show it for the black line for clarity.

\section{Large scales: N-body}
\label{sec:Nbody}

In this section we study the statistical properties of paired fixed fields on large scales using the N1000 N-body simulations. The halo catalogues are comprised of all halos with masses above $\simeq1.5\times10^{13}~h^{-1}M_\odot$.

\subsection{Initial conditions}
\label{subsec:ICs}

We start by quantifying the improvement achieved by paired fixed simulations at the level of initial conditions, as we naively expect that non-linear evolution can, in general, only degrade it. We focus our analysis on the matter power spectrum and on the matter density pdf.

\subsubsection{Clustering}

For each realization of the standard and paired fixed simulations in the N1000 set we have computed the matter power spectrum at $z=99$. 

We show the results in the left panel of Fig.~\ref{fig:Pk_ICs}. We find an excellent agreement between the results of both simulation sets, with paired fixed no introducing a bias on the results. Note that a few points show a bias larger than $2\sigma$; this is expected under the assumption that the data is independent and normally distributed, which implies that $\simeq5\%$ of the points should exhibit a bias larger than $2\sigma$.

From the third panel we can see that the power spectra from the two simulations of each paired fixed realization are highly anti-correlated on almost all scales. We note that the deviation of $\sqrt{1+r}$ from 0 is due primarily to aliasing. We have explicitly tested this by computing the power spectra using a grid with fewer cells. This anti-correlation is the origin of the large improvement that we obtain by pairing once we fix the amplitude, as we will see below.

From the fourth panel we can see how fixed simulations highly reduce the sample variance present in the standard simulations: from a factor of $\simeq10^3$ at $k\simeq10^{-2}~h{\rm Mpc}^{-1}$ to a few at $k=1~h{\rm Mpc}^{-1}$. We find that the improvement worsens at smaller scales. 
This is an effect of the way the power spectrum is measured in an individual box; as we move to smaller scales, there are rapidly increasing number of modes per $k$-bin. Thus, the measured power spectrum asymptotes to the ensemble average at high k and no initial improvement is achieved by fixing the power in this limit.

Paired fixed simulations further reduce the sample variance amplitude with respect to fixed simulations, with ratios as large as $10^5$ on the largest scales we probe. The improvement brought by pairing has its origin in the fact that the first order non-linear perturbations are cancelled \citep{Pontzen_2016}. Even at $z=99$, the Zel'dovich approximation has introduced such non-linearities. 

\subsubsection{Probability distribution function}

We now investigate another key quantity to understand our results at lower redshifts: the probability distribution function (pdf) of the matter density field in real-space.

For each initial condition realization of the standard and paired fixed simulations we have computed the matter density field by assigning particle positions to a grid with $128^3$ cells using the cloud-in-cell (CIC) mass-assignment scheme. We have then computed the pdf as the fraction of cells with matter overdensity, $1+\delta=\rho/\bar{\rho}$, in the interval $[\rho,\rho+d\rho]/\bar{\rho}$. We show the results of our statistical analysis in the right panel of Fig.~\ref{fig:Pk_ICs}.

As already pointed out in \cite{RP_16}, the pdf of paired fixed simulations shows a good agreement with that from standard simulations, as can be seen from the first panel. From the second panel we can see that paired fixed simulations do not introduce a bias on the matter density pdf with respect to the results from standard simulations. 

In the third panel we show the cross-correlation coefficient between the pairs of the paired (orange) and paired fixed (black) simulations. We find that in both cases the value $r$ is compatible with $0$ ($\sqrt{1+r}=1$), meaning that the results of both pairs are independent from each other. Thus, pairing does not help in reducing the variance on the matter density pdf from the standard simulations.

We show the statistical improvement achieved by fixed and paired fixed simulations, with respect to standard simulations, in the fourth panel. We find that all simulation types exhibit the same scatter as standard simulations. We do not find improvements on the variance amplitude for fixed or paired fixed simulations, meaning that fixing the amplitude does not reduce the pdf fluctuations either. In some ways this is a blessing: the local statistical properties of a fixed field are identical to the properties of its Gaussian counterpart and therefore one can expect local physics such as galaxy formation to proceed correctly in a fixed Universe.

Thus, we conclude that while paired fixed simulations can reduce the scatter on the power spectrum of the initial conditions by large factors, the pdf does not benefit from this and its scatter remains unchanged. We will see below that other quantities tightly related to the pdf, such as the halo or void mass functions, the stellar mass function or intrinsic galaxies properties will not exhibit significant statistical improvement when estimated using paired fixed simulations. 

We find only modest improvements on the variance of paired fixed simulations on the halo mass function, matter density pdf and star-formation rate history, when analyzing the H20 simulations, as we will see in section \ref{sec:hydro_small}.

\subsection{Clustering}

\begin{figure*}
\begin{centering}
\includegraphics[width=0.33\textwidth]{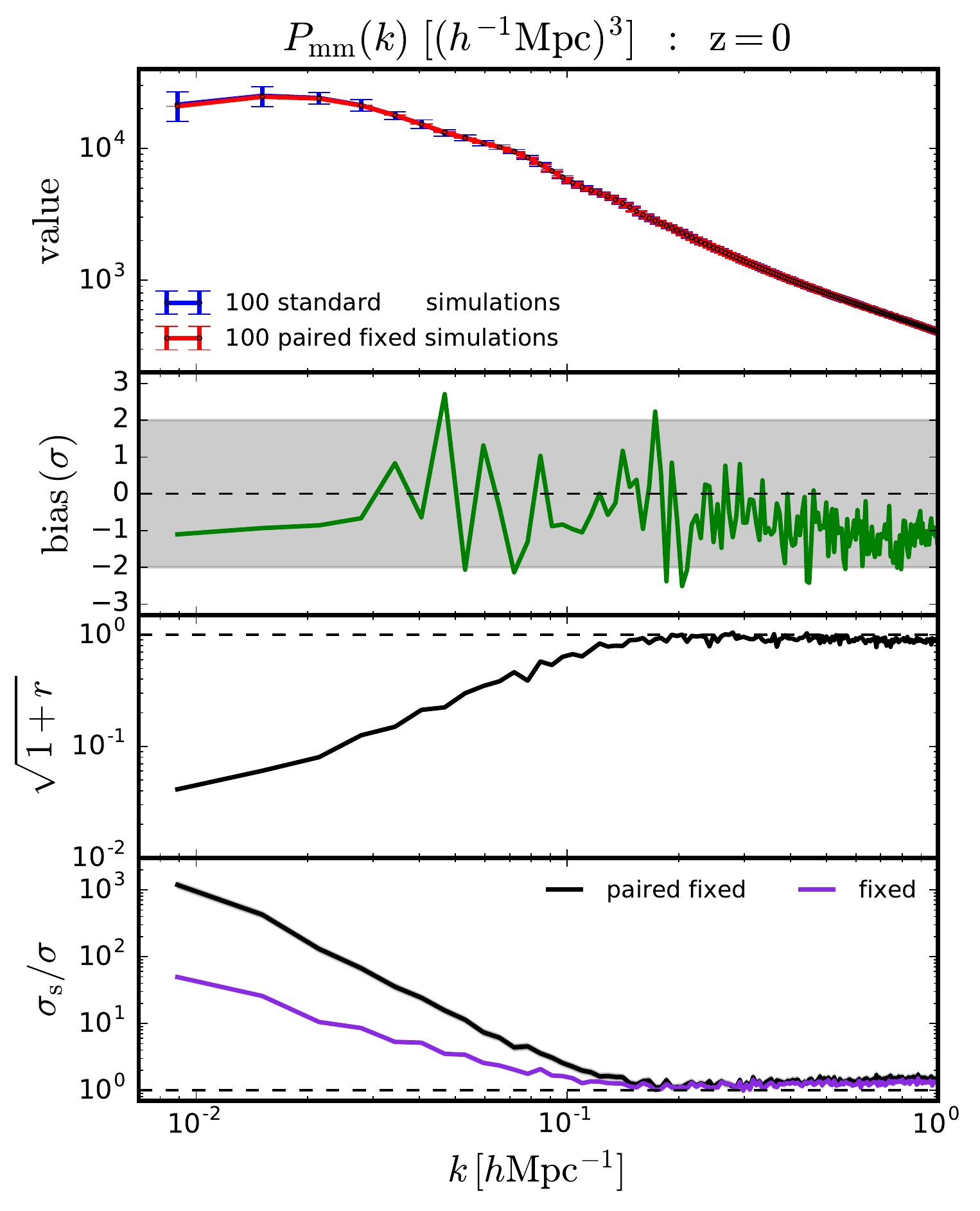}
\includegraphics[width=0.33\textwidth]{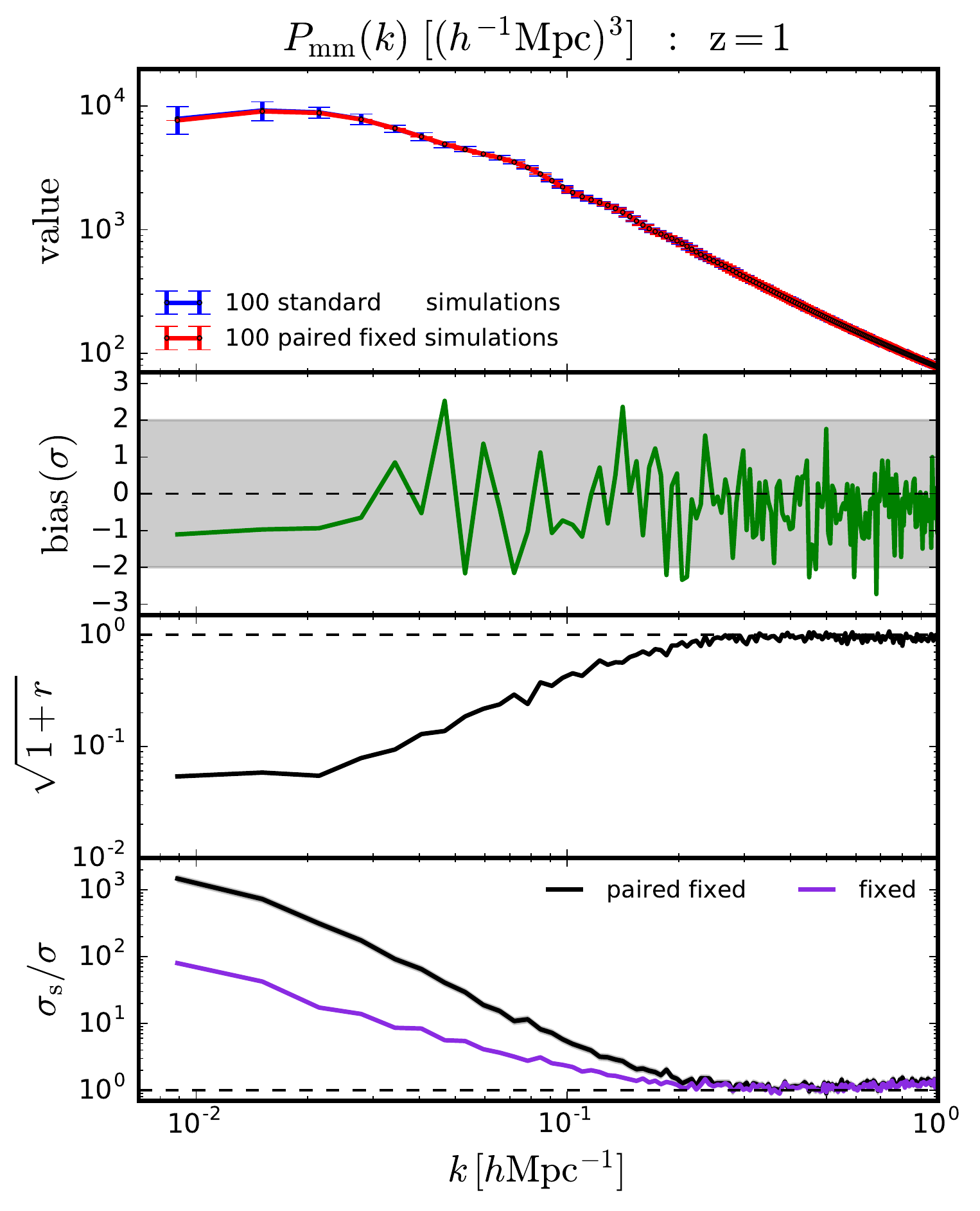}
\includegraphics[width=0.33\textwidth]{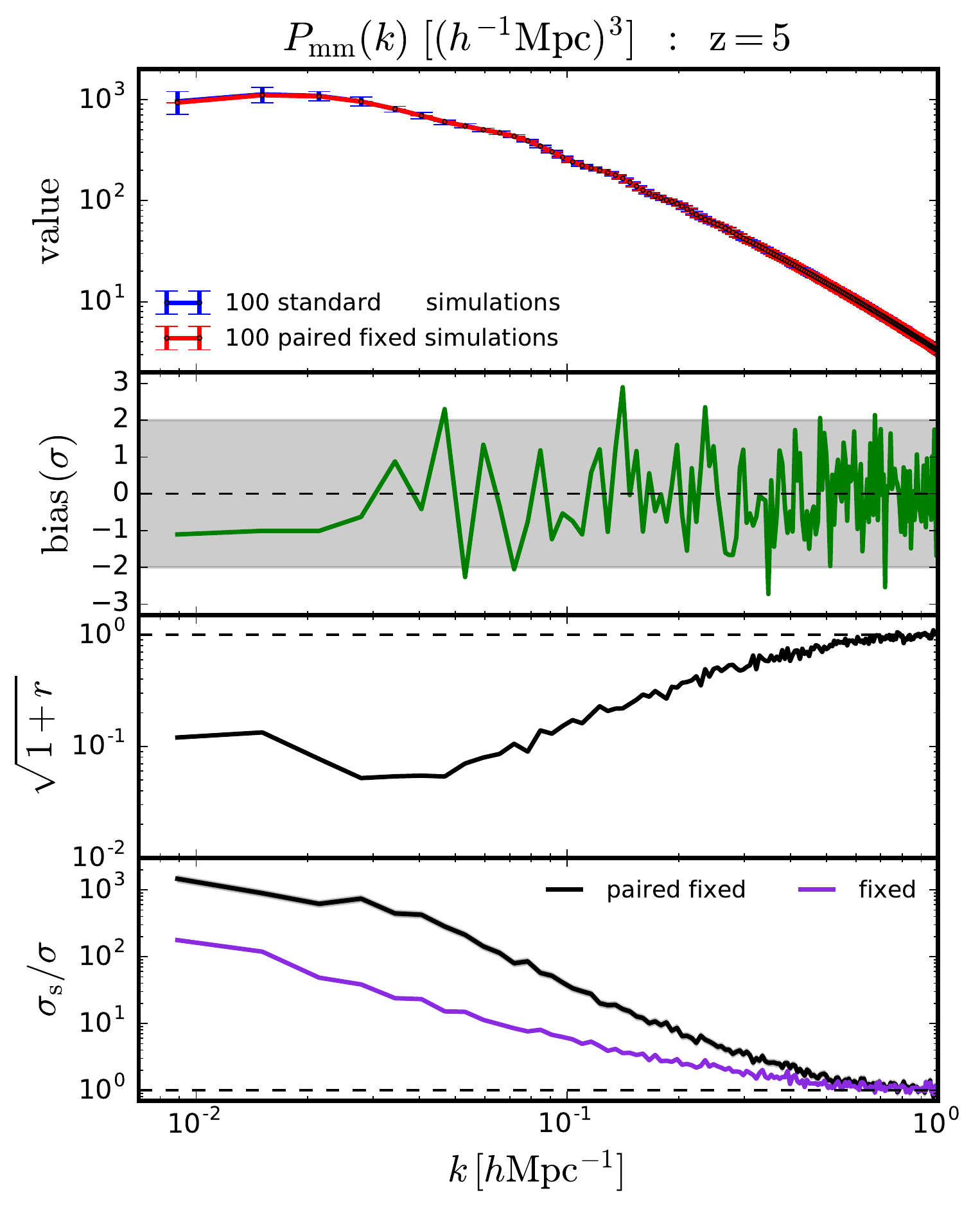}\\[2ex]
\includegraphics[width=0.33\textwidth]{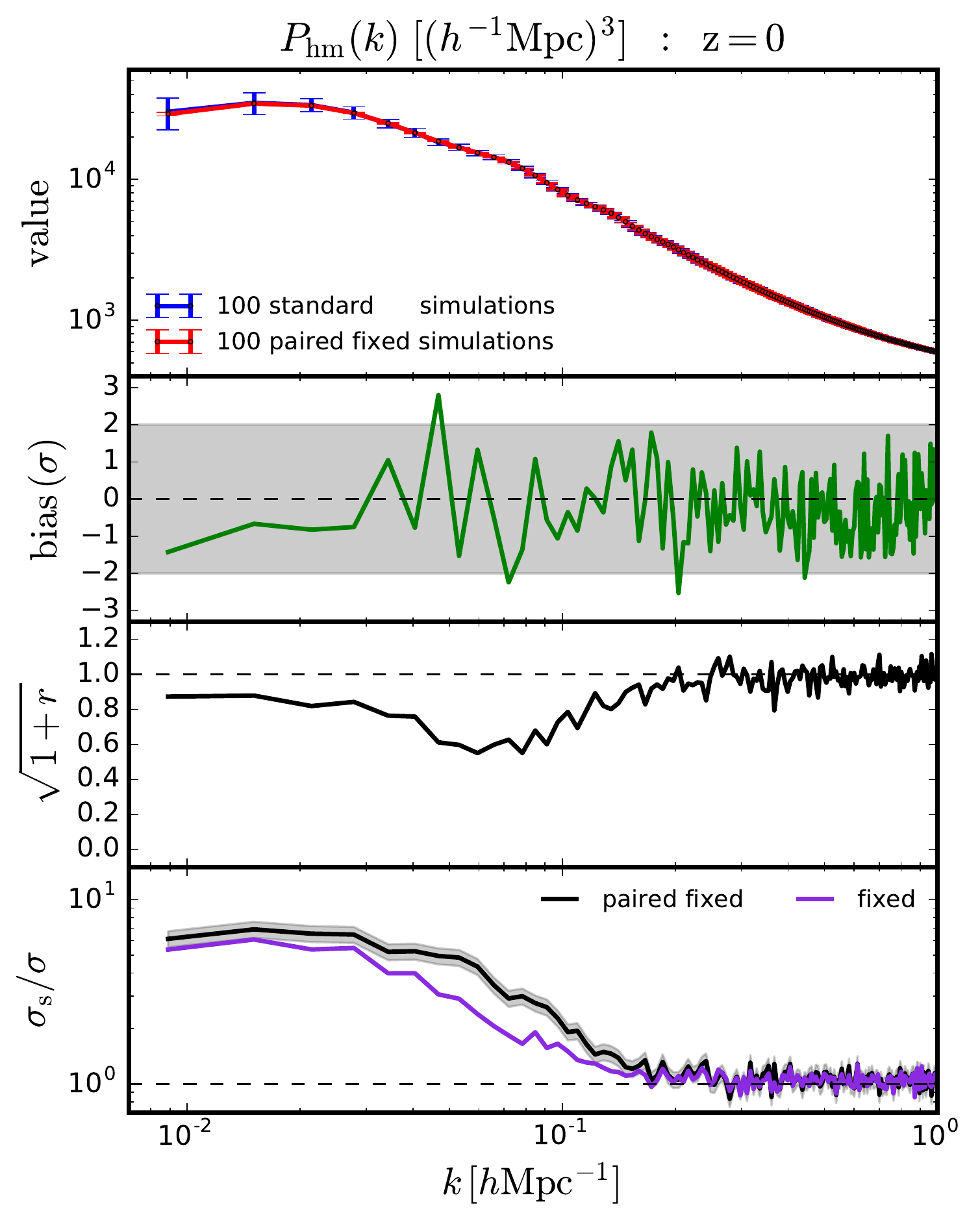}
\includegraphics[width=0.33\textwidth]{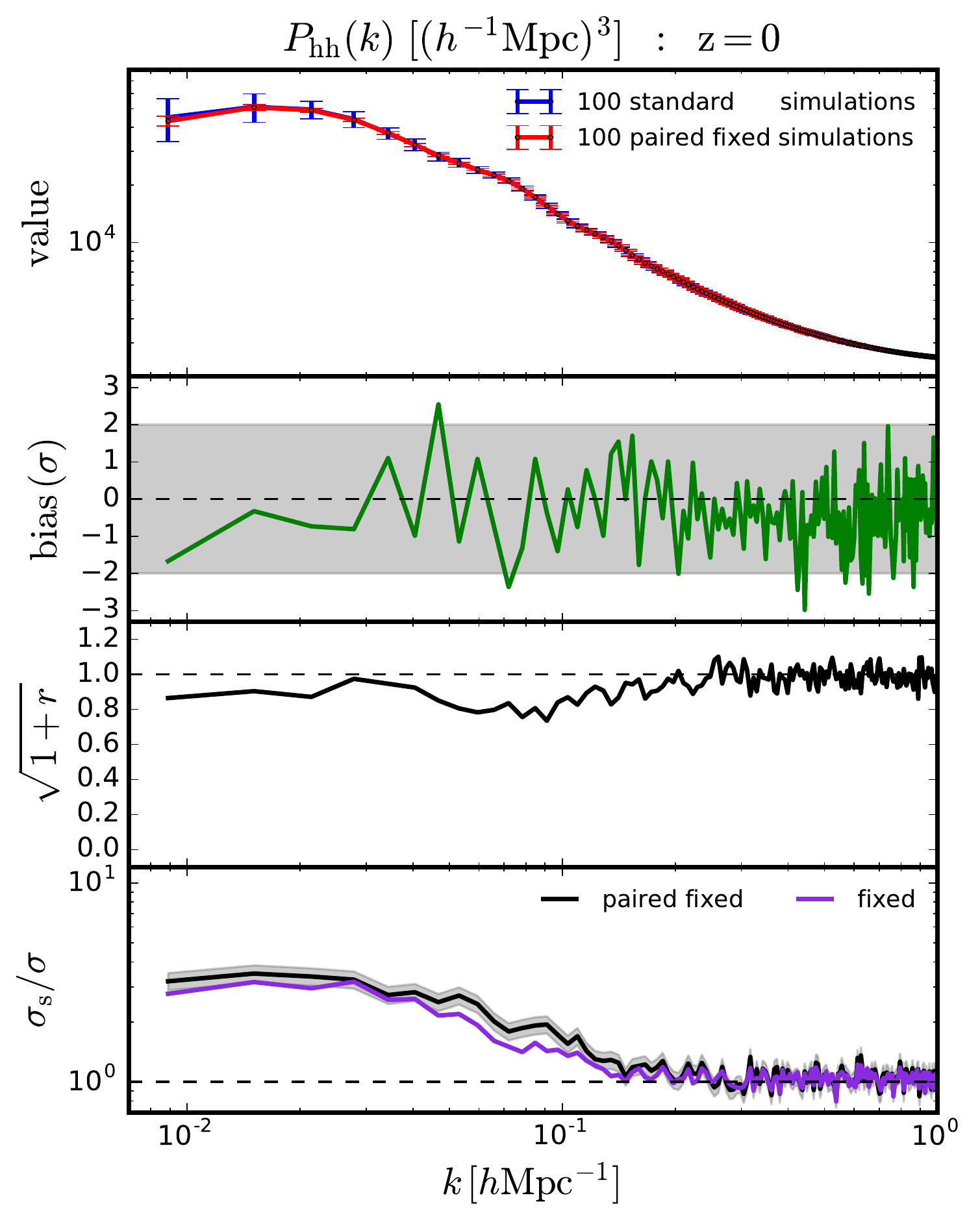}
\caption{Impact of paired fixed simulations on the clustering of matter (top row), halos (bottom-right) and on the halo-matter cross-power spectrum (bottom-left). We show results at redshifts 0 (left column), 1 (middle column) and 5 (right column) for matter, and at $z=0$ for the halo and halo-matter power spectra. paired fixed simulations can reduce the sample variance scatter on these power spectra by large quantities without introducing a bias on them.}
\label{fig:Pk_1000Mpc_Nbody}
\end{centering}
\end{figure*}

For each simulation in N1000 we have computed the matter and halo auto-power spectrum and the halo-matter cross-power spectrum. The results of our statistical analysis are displayed in Fig.~\ref{fig:Pk_1000Mpc_Nbody}. 

The upper row shows the results for the matter power spectrum at redshifts 0 (top-left), 1 (top-middle) and 5 (top-right). The bottom row displays the results for the halo-matter cross-power spectrum (bottom-left) and halo power spectrum (bottom-right). For those quantities we only show results at $z=0$, since at $z=5$ the number density of halos in our simulations is very low and results are very similar at $z=1$.

From the first panels we can see that the agreement between the results of the standard and paired fixed simulations is very good in all cases. In the second panels we quantify the bias introduced by the paired fixed simulations with respect to standard simulations, and find no evidence for a bias for any of the three quantities at the different redshifts considered. We emphasize that with a finite number of simulations, this kind of claim has to be considered as an upper bound. It may be that paired fixed simulations induce a bias on those quantities, but its magnitude is too small for detection with 100 realizations. We notice that in some cases, e.g.~the matter power spectrum at $z=0$, there seem to be a systematic bias offset on small scales. This is however due to the fact that modes on those small scales are highly correlated, through non-linear evolution, and therefore not fully independent. 

In the third panels we show the value of the cross-correlation coefficient. We find that for all the considered quantities on small scales, $k\gtrsim0.2-0.5~h{\rm Mpc}^{-1}$, its value is compatible with $\sqrt{1+r}=1$, indicating that the power spectra from the two pairs are independent. In that case, pairing does not help in reducing the statistical error due to sample variance.  We find that the value of $\sqrt{1+r}$ is smaller than $1$ on scales larger than $k\gtrsim0.2-0.5~h{\rm Mpc}^{-1}$. The scale at which $\sqrt{1+r}$ equals $1$ decreases with redshift, independently of the considered power spectrum, but the effect is more pronounced in the matter power spectrum. 

The value of $\sqrt{1+r}$ for the matter power spectrum can be as low as $\simeq4\times10^{-2}$, pointing out that pairing, once fixed, can reduce the scatter of the standard simulations by that factor. The value of $\sqrt{1+r}$ increases with scale, until reaching the value of 1. At $z=5$ however, we find a dip around $3\times10^{-2}~h{\rm Mpc}^{-1}$.  It is interesting to notice that the lowest values of $\sqrt{1+r}$ take place at $z=0$ rather than $z=5$. At present, we do not have an explanation for this.

We find much higher values of $\sqrt{1+r}$ for the matter-halo and the halo-halo power spectra than for the matter power spectrum. In those cases, we also find a dip around $4\times10^{-2}~h{\rm Mpc}^{-1}$. On large scales, the value of $\sqrt{1+r}$ barely goes below 0.7, indicating that pairing can only reduce the variance by $\simeq0.7$. We notice that halos are the main driver of the increase in the value of $\sqrt{1+r}$, as on large scales the halo power spectrum barely deviates from $1$. 

\begin{figure*}
\begin{centering}
\includegraphics[width=0.7\textwidth]{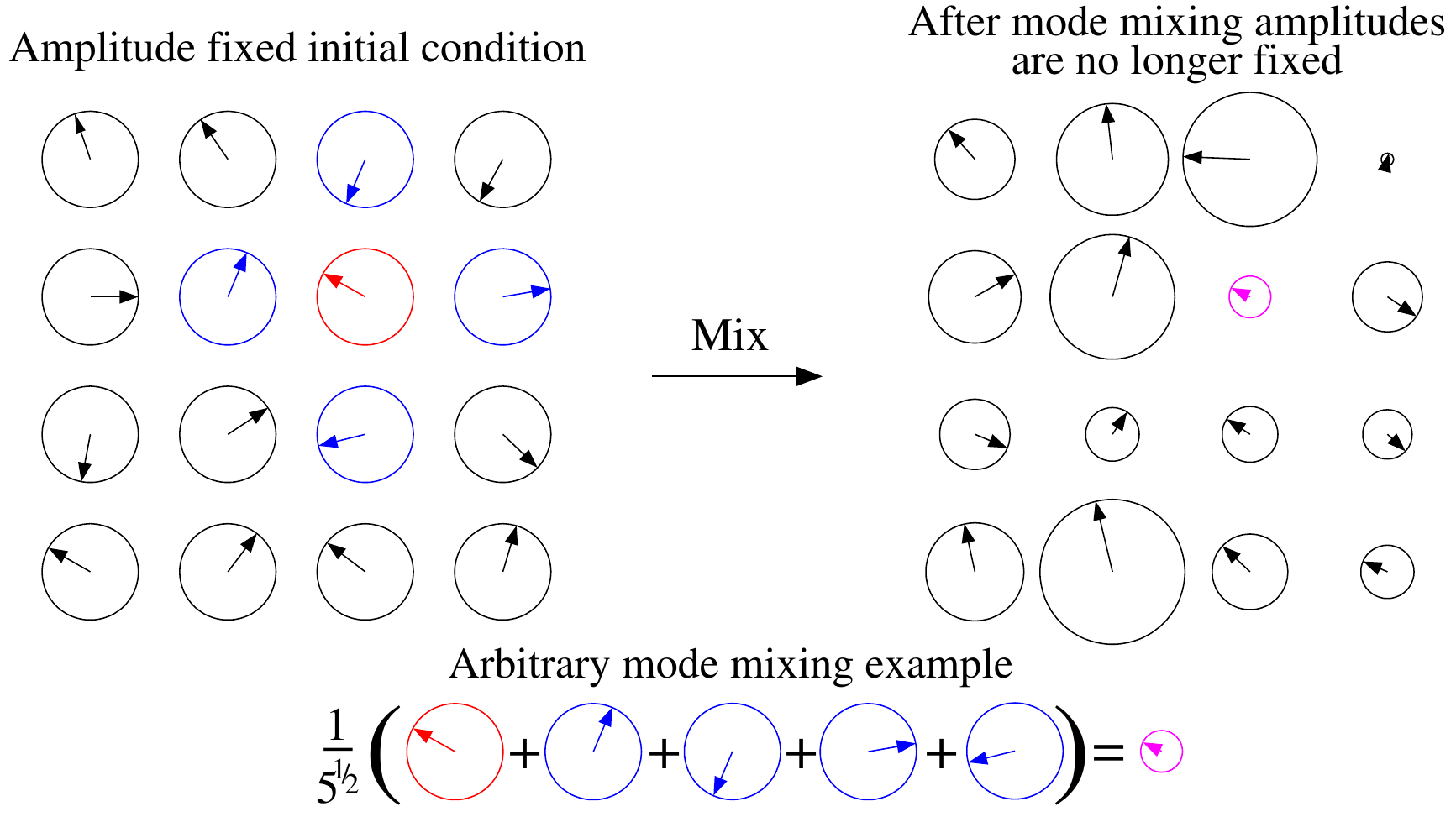}
\caption{Illustration of how mode mixing from non-linear evolution introduces sample variance in fixed simulations. \textit{Left:} In the initial conditions each cell in Fourier space has a random phase (the arrow) but a fixed amplitude (the radius of the circle). \textit{Right:} After each mode mixes with its neighbors, some amplitudes grow and others diminish, depending on how their phases align with those of their neighbors. Hence, the amplitudes are no longer fixed after mixing, reintroducing sample variance. The exact form of mixing shown here is just an example, while any kind of mixing will have similar effects.}
\label{fig:Mode_mixing}
\end{centering}
\end{figure*}

From the fourth panels of Fig.~\ref{fig:Pk_1000Mpc_Nbody} we can see that for the matter power spectrum on large scales, reductions on the standard deviation of standard simulations can be as large as $10^3$, at all redshifts considered. Since the standard deviation on the mean from standard simulations goes as $\propto1/\sqrt{N_{\rm s}}$, the above numbers can be interpreted as follows. A single paired fixed simulation can be used to evaluate the amplitude and shape of the matter power spectrum on large scales, with an error equal to that achieved by running $\sim10^6$ standard simulations. On small scales, $k\sim0.2-0.4~h{\rm Mpc}^{-1}$, the ratios tend to 1, showing that no improvement is achieved by the fixed or paired fixed simulations. The reason for why fixed and paired fixed simulations do not improve the statistics of standard simulations on small scales is that in the non-linear regime, modes get mixed in a complicated manner that affects both the amplitudes and phases and gives rises to sample variance. We show this schematically in Fig. \ref{fig:Mode_mixing}, where a set of complex numbers with fixed amplitude end up with very different amplitudes after each mode mixes with its neighbors.
This happens because whether complex numbers with the same amplitude add up or cancel depends on whether their phases align. In fixed and paired fixed simulations the phases are random\footnote{In paired fixed simulations there is a correlation between the phases  in the initial conditions of the two simulations in a pair. The argument regarding random phases applies however to the mode-mixing of each individual simulation in a pair.}.

We notice however that on very small scales and at $z=0$, the $P_{\rm mm}(k)$ results for $\sigma_s/\sigma$ are between 2 and 3. The improvement on those scales is mostly coming from fixing the amplitude rather than from pairing. We also observe this effect in the smaller box size simulations that we study in section \ref{sec:hydro_intermediate}. Thus, the above argument can explain the behavior we find in simulations only qualitatively.

The statistical improvement on large scales is much smaller for the halo-matter and halo-halo power spectra. For the halo-matter cross-power spectrum, we reach values of $\sigma_{\rm s}/\sigma_{\rm pf}\simeq6$ on large scales at both redshifts 0 and 1. For the halo auto-power spectrum those values shrink to $\sigma_{\rm s}/\sigma_{\rm pf}\simeq3$. For those two power spectra no statistical improvement is achieved by fixed or paired fixed simulations on scales smaller than $k\gtrsim0.3~h{\rm Mpc}^{-1}$. 

We conclude that while paired fixed simulations can yield very large statistical improvements, $\sigma_{\rm s}/\sigma_{\rm pf}\simeq1000$, for the matter power spectrum, for the halo-matter and halo power spectra the gain is much smaller, $\sigma_{\rm s}/\sigma_{\rm pf}\sim5$, but still valuable.

\subsection{Halo bias}

\begin{figure*}
\begin{centering}
\includegraphics[width=0.33\textwidth]{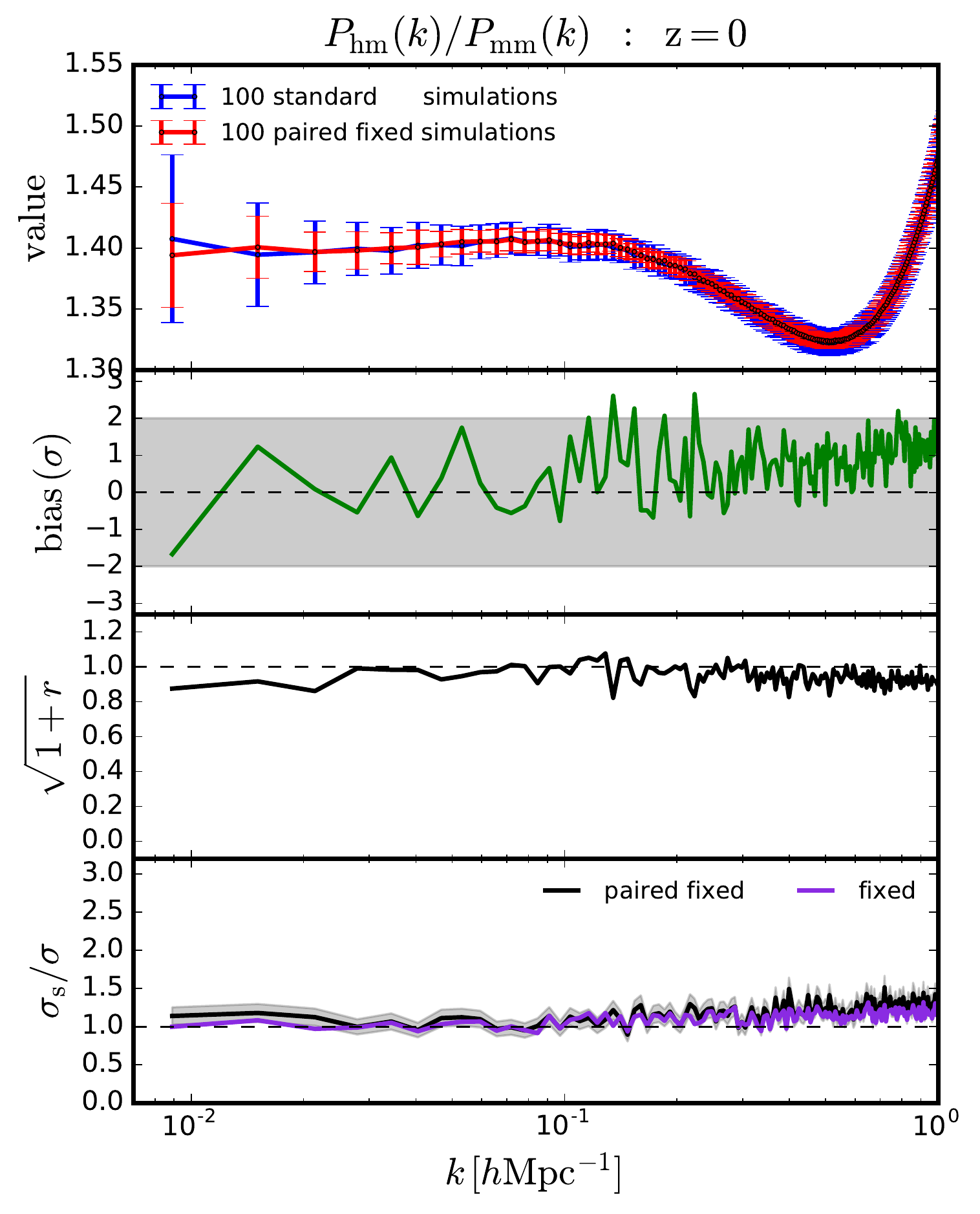}
\includegraphics[width=0.33\textwidth]{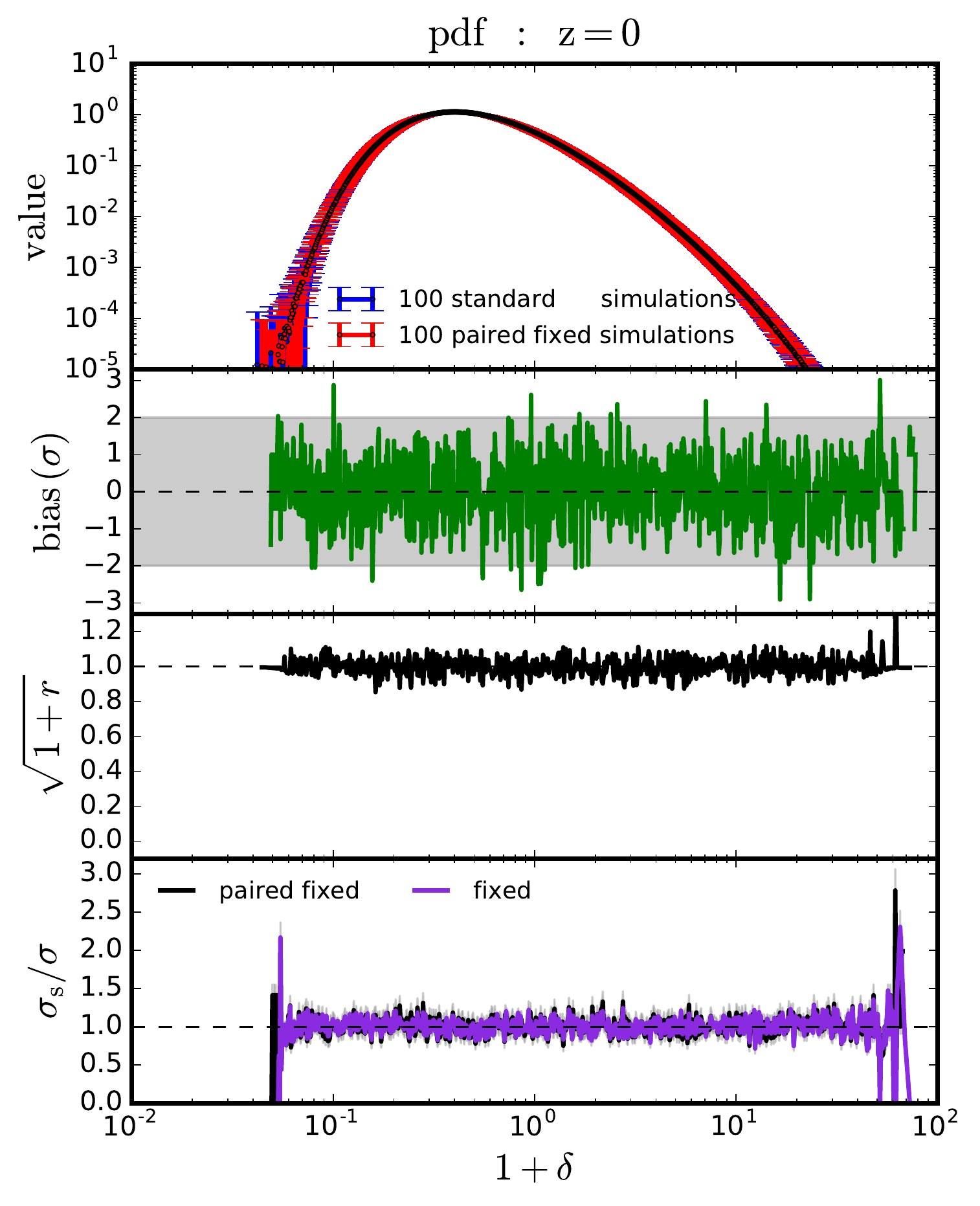}
\includegraphics[width=0.33\textwidth]{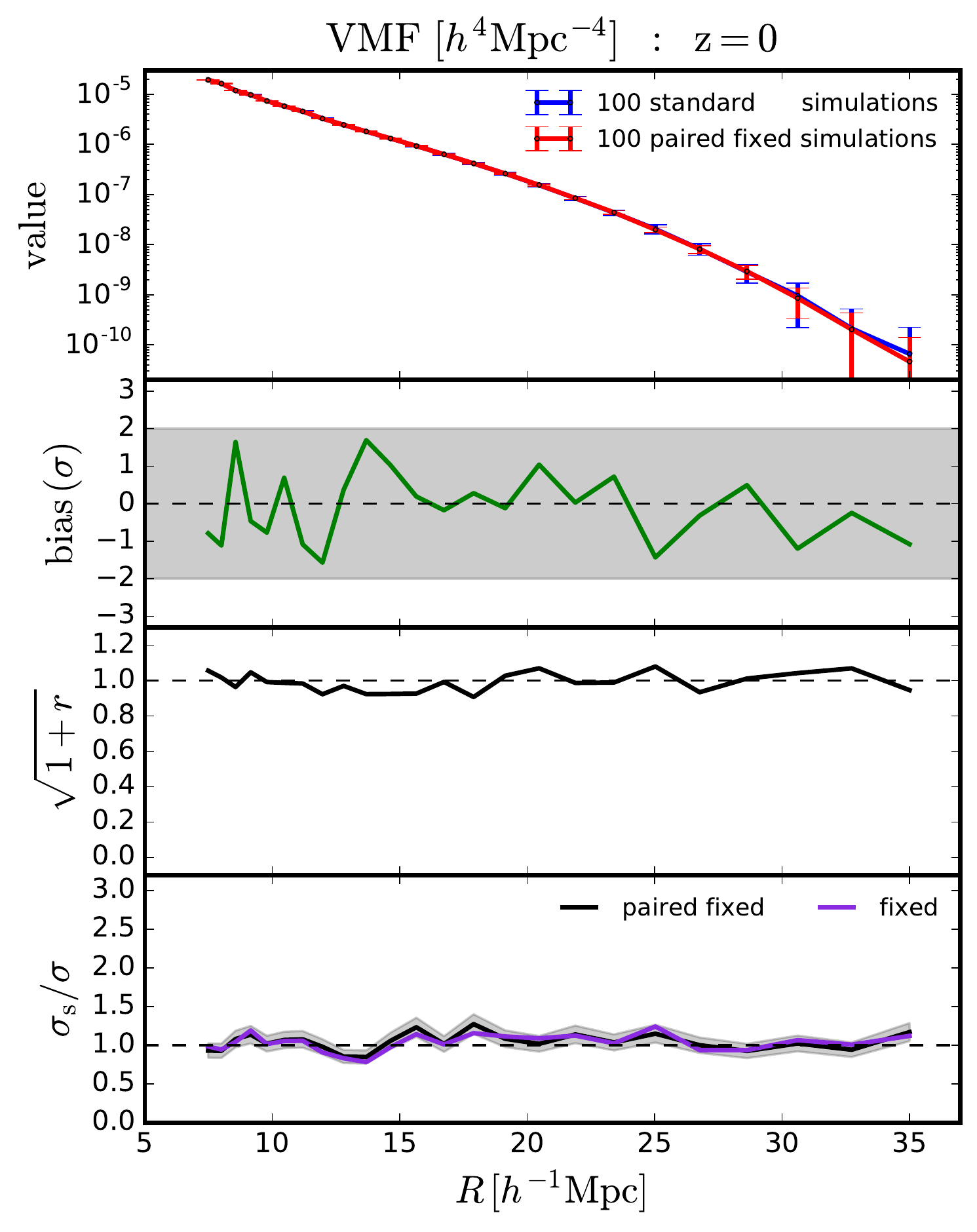}
\includegraphics[width=0.33\textwidth]{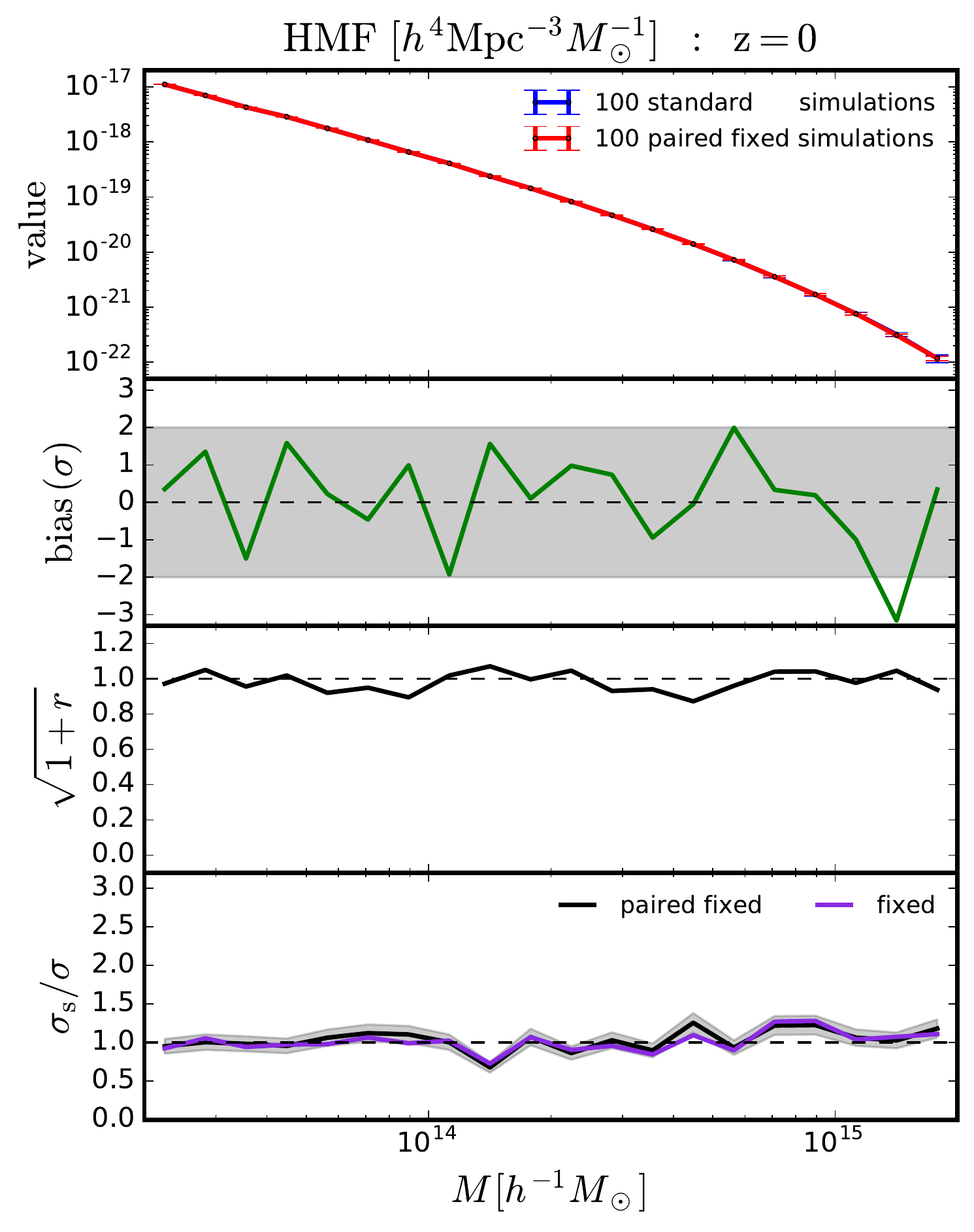}\\[2ex]
\caption{Impact of paired fixed simulations on halo bias (top-left), matter density pdf (top-right), void mass function (bottom-left) and halo mass function (bottom right) at $z=0$ from the N1000 simulation set. Paired fixed simulations do not introduce a bias on these quantities but they do not reduce their scatter neither.}
\label{fig:1000Mpc_others}
\end{centering}
\end{figure*}

We now turn our attention to the halo bias. For each standard and paired fixed realization we have computed the halo bias using the estimator
\be
b(k)=\frac{P_{\rm hm}(k)}{P_{\rm mm}(k)}
\ee
We show the results of our statistical analysis in the top-left panel of Fig.~\ref{fig:1000Mpc_others}. We only show the results at $z=0$ since at $z=1$ our conclusions are unchanged. From the first panel we see the very good agreement between the results of both simulations while in the second panel we show that paired fixed simulations do not introduce a bias on this quantity. The value of the cross-correlation coefficient is, for almost all scales, compatible with 0 ($\sqrt{1+r}=1$), implying that pairing does not help in reducing the scatter. Finally, in the fourth panel we can see how fixing the amplitude does not reduce the scatter either, and therefore, paired fixed simulations exhibit the same scatter in the halo bias as standard simulations. 

We have repeated the above analysis by computing the bias as $b(k)=\sqrt{P_{\rm hh}(k)/P_{\rm mm}(k)}$, reaching identical conclusions: fixed and paired fixed simulations exhibit the same scatter on the halo bias as standard simulations.

This result may appear surprising at first since, as we saw above, paired fixed simulations can reduce the scatter on the matter, halo-matter and halo power spectra by factors as large as $10^3$, $6$ and $3$, respectively. In order to understand the reason for this result let us write the variance of the halo bias at linear order (see appendix \ref{sec:variance_ratios} for the derivation)
\be
\sigma_{\rm b}^2 = \frac{1}{2N(k)}\left(\frac{P_{\rm hh}(k)P_{\rm mm}(k)-P_{\rm hm}^2}{P_{\rm mm}^2}\right)~,
\ee
where $N(k)$ is the number of independent modes in the considered k-interval and the halo power spectrum includes both cosmological signal and the shot-noise term, i.e.~$P_{\rm hh}(k)=P_{\rm hh}^{\rm cosmo}(k)+1/\bar{n}$. If the shot-noise amplitude were zero, $P_{\rm hh}=b^2P_{\rm mm}$, $P_{\rm hm}=bP_{\rm mm}(k)$ and the linear-order variance of the bias would be zero too. This tells us that in the absence of shot-noise, the halo-matter and matter-matter power spectra are perfectly correlated, and hence their ratio, the halo bias, has zero variance.

What breaks this perfect correlation and becomes the source of variance in our bias estimate, is the presence of shot-noise in the halo power spectrum. The amplitude of the shot-noise on large scales is the same in paired fixed and standard simulations, as it only depends on the halo number density. Thus, it should not be surprising after all that the scatter in the halo bias from paired fixed and standard simulations is the same as well.

We leave for future work a formal derivation of this result on mildly non-linear scales and a deeper understanding on why paired fixed simulations do not even reduce the scatter of the halo bias on non-linear scales.

\subsection{Matter density pdf}

We now focus our attention on the probability distribution function of the matter density field. For each realization of the standard and paired fixed simulations we have computed the matter density field on a grid with $128^3$ cells using the cloud-in-cell (CIC) mass assignment scheme. We show the results of our analysis in the top-right panel of Fig.~\ref{fig:1000Mpc_others}. We only show results at $z=0$ since results at higher redshift do not change our conclusions.

Unlike the matter power spectrum, where both pairing and amplitude fixing greatly
reduced the variance, the matter density pdf is indifferent to these techniques, at least
on this scale. There is however no harm: the bias is consistent with zero, showing full agreement between
standard and paired fixed simulations. But there is also no benefit: the value of $\sqrt{1+r}$ is consistent with 1, so pairing after fixing is of no help, and the all effects of amplitude fixing are washed out in this basis. See however Subsection \ref{subsec:1pt_20Mpc_hydro} for how this changes on smaller scales.

We can interpret these results by taking into account that paired fixed simulations do not reduce the scatter on the pdf already at the starting redshift of the simulation (see Subsection \ref{subsec:ICs}). Thus, it is unlikely that non-linear evolution would lead to different pdfs at low redshift.

\subsection{Halo and void mass functions}

Here we study the impact of paired fixed simulations on the halo and void mass functions. For each standard and paired fixed simulations we have computed the halo mass function, defined as the number density of halos per mass interval. We show the results in the bottom-right panel of Fig.~\ref{fig:1000Mpc_others}. We have also computed the void mass function, defined as the number density of voids per radius interval, for each realization of the standard and paired fixed simulations. We show the results in the bottom-left panel of Fig.~\ref{fig:1000Mpc_others}. For both cases we only show results at $z=0$, as higher redshifts lead to identical conclusions.

From the upper panels we find that the agreement between the standard and paired fixed simulations is very good for both the halo and void mass functions and in the second panels we show that no bias is introduced on these quantities by the paired fixed simulations. In the third panels we show the cross-correlation coefficient from the results of each pair. Our results are compatible with the $\sqrt{1+r}=1$, pointing out that the results of each pair are independent. From the fourth panels we find that no statistical improvement on these two quantities from fixed or paired fixed simulations.

We believe that paired fixed simulations do not improve the abundance of halos and voids statistics because the formation of those takes place on small scales, where the 1-pt properties are more relevant to determining the final outcome. As we saw in subsection \ref{subsec:ICs}, these are not affected by the fixing and pairing procedure.

In Section \ref{sec:hydro_small} we will however see that paired fixed simulations slightly reduce the scatter of the halo mass function and matter density pdf when analyzing hydrodynamic simulations with small box sizes. This may be related to non-linearities reaching the halo filtering scale but further exploration is deferred to future work.

We thus conclude that large-scale box size paired fixed simulations reduce the scatter on clustering quantities like the matter, halo-matter or halo power spectra. They however do not help in reducing the scatter of the halo bias or on 1-point statistics like the halo or void mass functions, or the matter density pdf.

\section{Intermediate scales: hydrodynamic}
\label{sec:hydro_intermediate}

In this section we investigate the statistical properties of paired fixed simulations on intermediate scales using state-of-the-art magneto-hydrodynamic simulations. We carry out the statistical analysis using the H200 simulations. Those simulations are computationally expensive, so we could only run 56 of them: 26 standard and 15 paired fixed realizations. This small number of simulations does not allow us to reach robust statistical conclusions for most of the quantities considered in this paper. For this reason we focus our analysis on clustering, where the effect is large enough to establish that paired fixed simulations do reduce the intrinsic scatter due to sample variance. 

We have also computed the matter density pdf, the halo mass function and the void mass function, and our results are in agreement with those from large scales, i.e.~paired fixed simulations do not introduce a bias but also do not reduce the intrinsic scatter. However, the associated error bars are too large to rule out a small statistical improvement such as that we observe in the H20 simulations (see section \ref{sec:hydro_small}).

\subsection{Clustering}
\label{subsec:clustering_H200}

For each standard and paired fixed simulation we have computed the matter, CDM, gas, stars and black-holes power spectrum at redshifts 0, 1 and 5. The relatively low resolution of the H200 simulation highly affects the power spectrum of stars and black-holes, due to the large amplitude of the shot-noise, on all scales we probe. Hence, we focus our analysis on the matter, CDM and gas power spectra.

\begin{figure*}
\begin{center}
\includegraphics[width=0.33\textwidth]{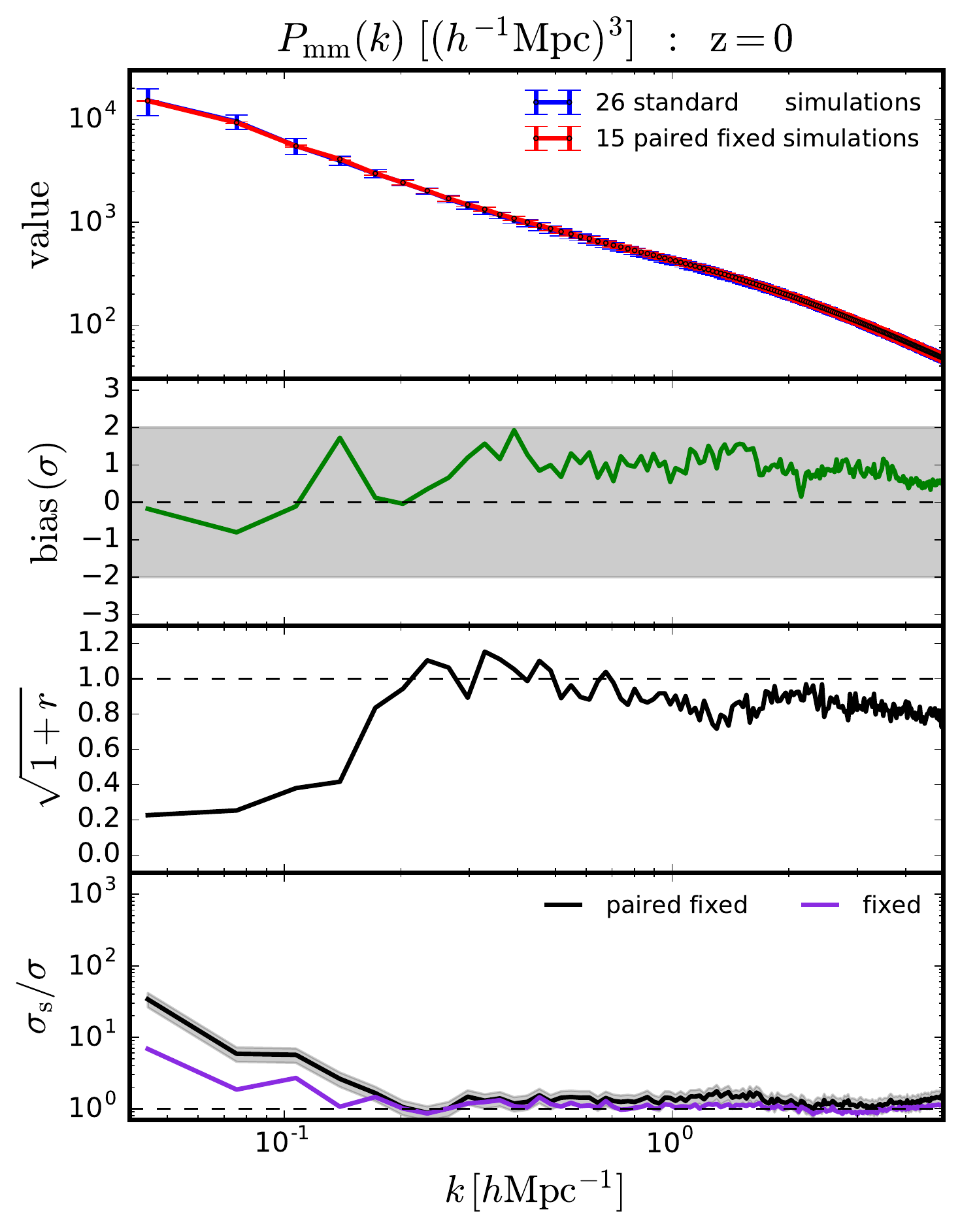}
\includegraphics[width=0.33\textwidth]{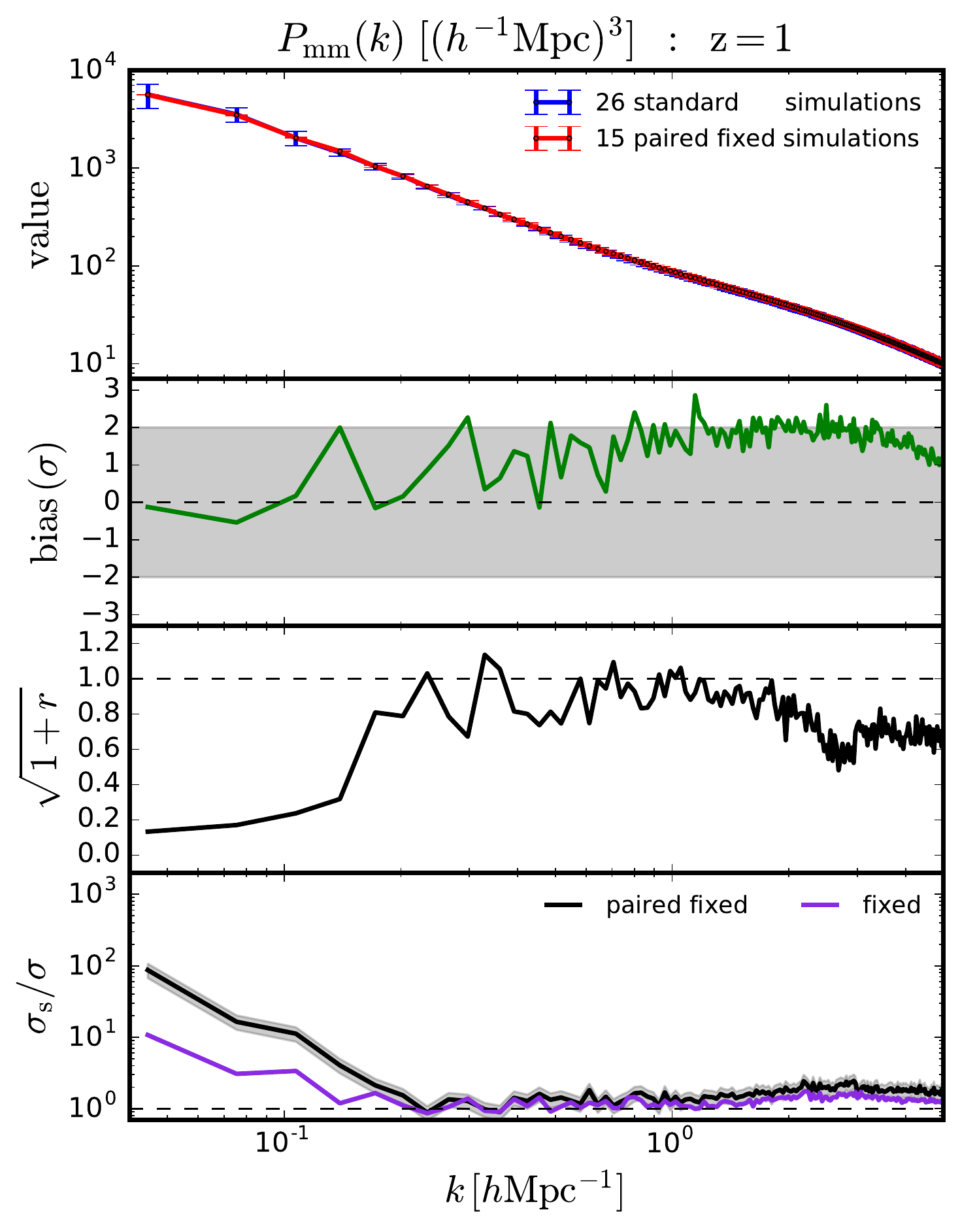}
\includegraphics[width=0.33\textwidth]{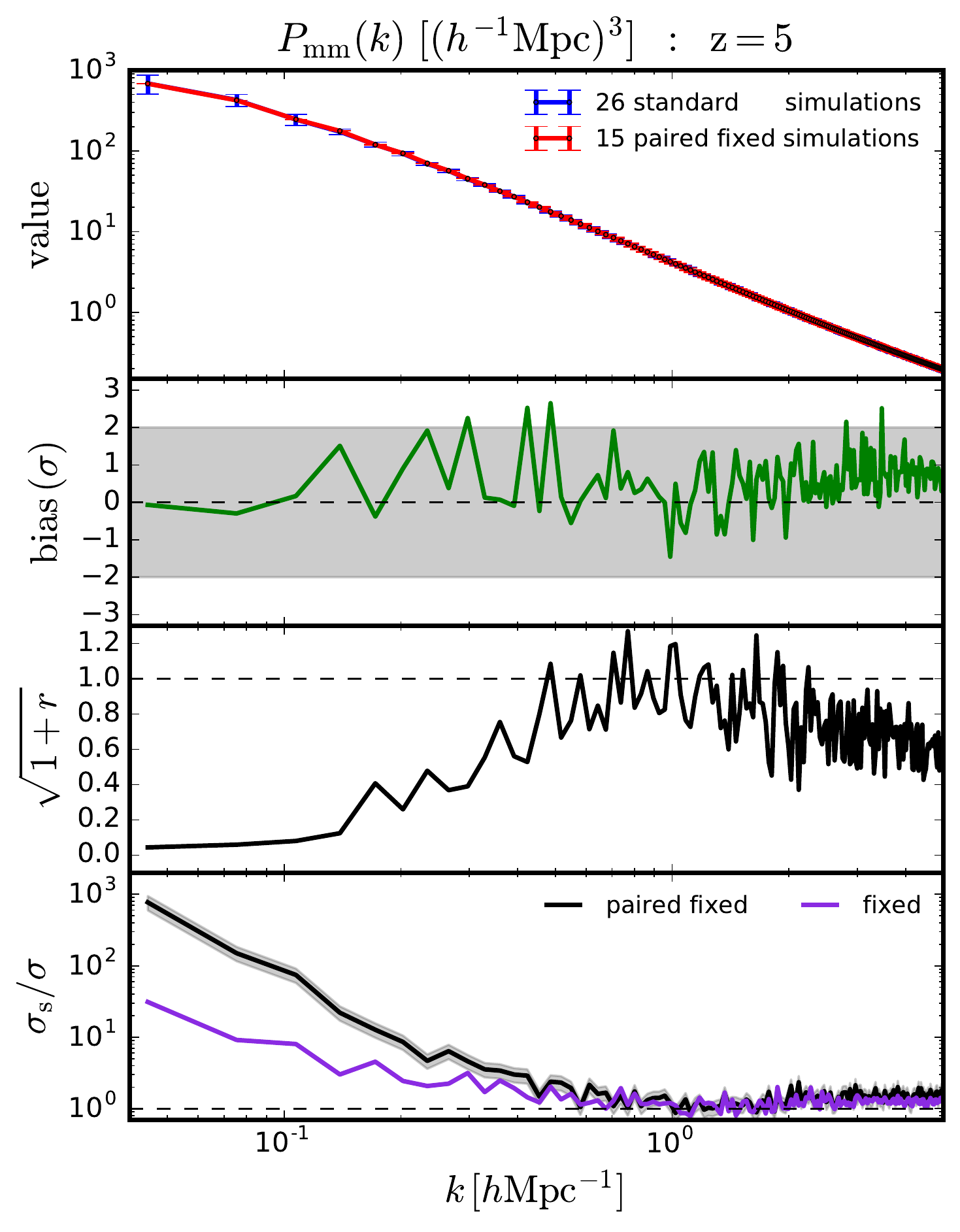}\\[2ex]
\includegraphics[width=0.33\textwidth]{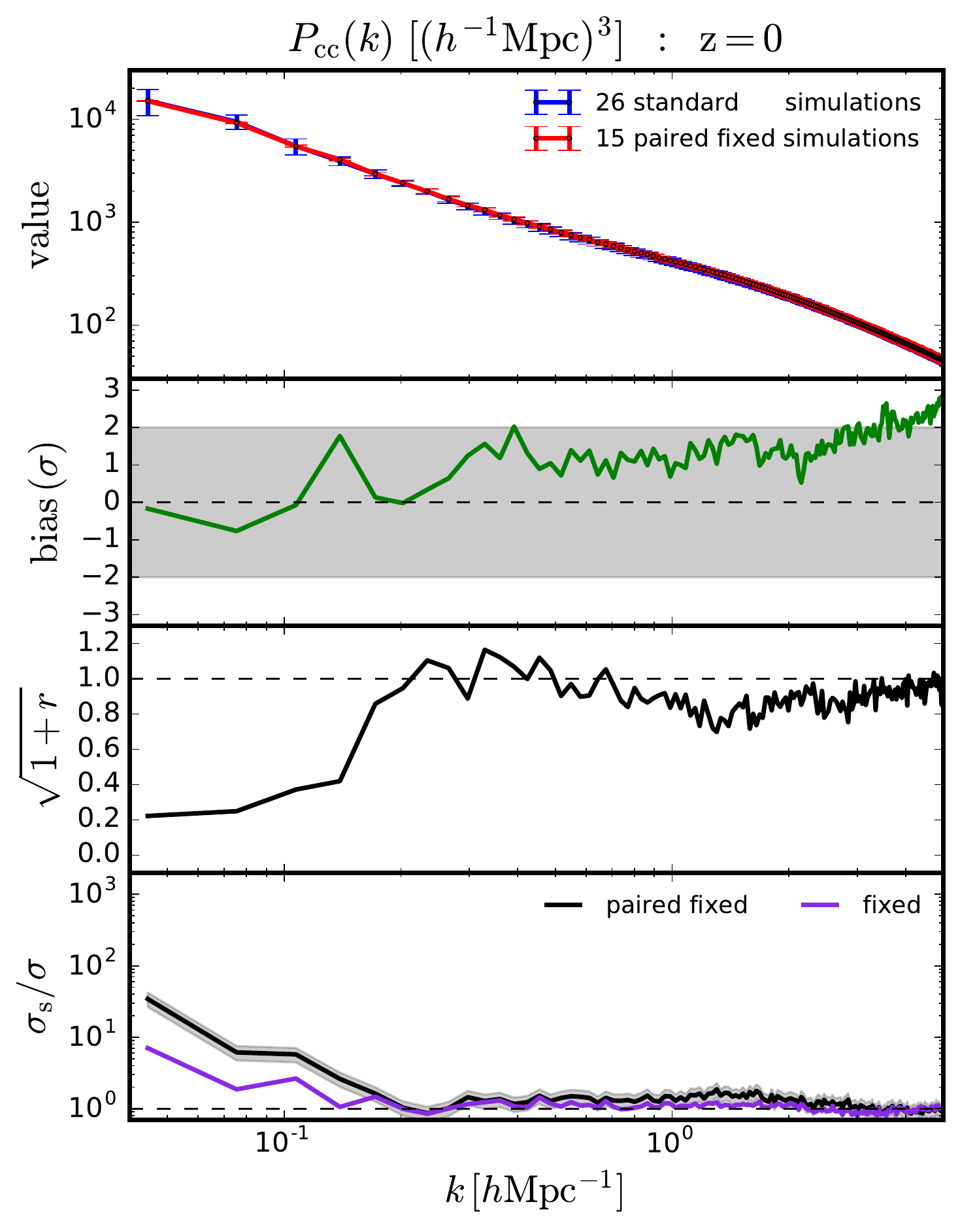}
\includegraphics[width=0.33\textwidth]{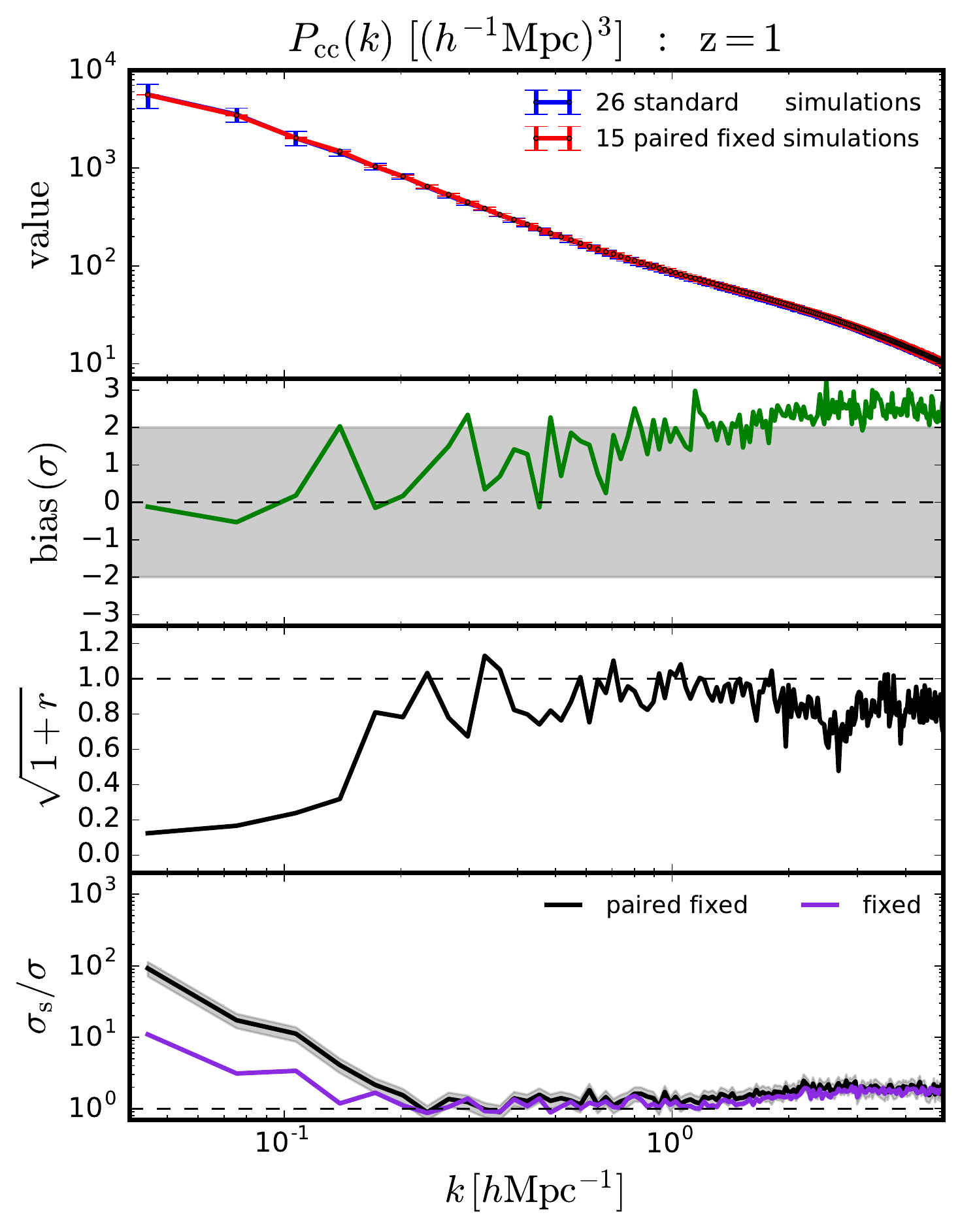}
\includegraphics[width=0.33\textwidth]{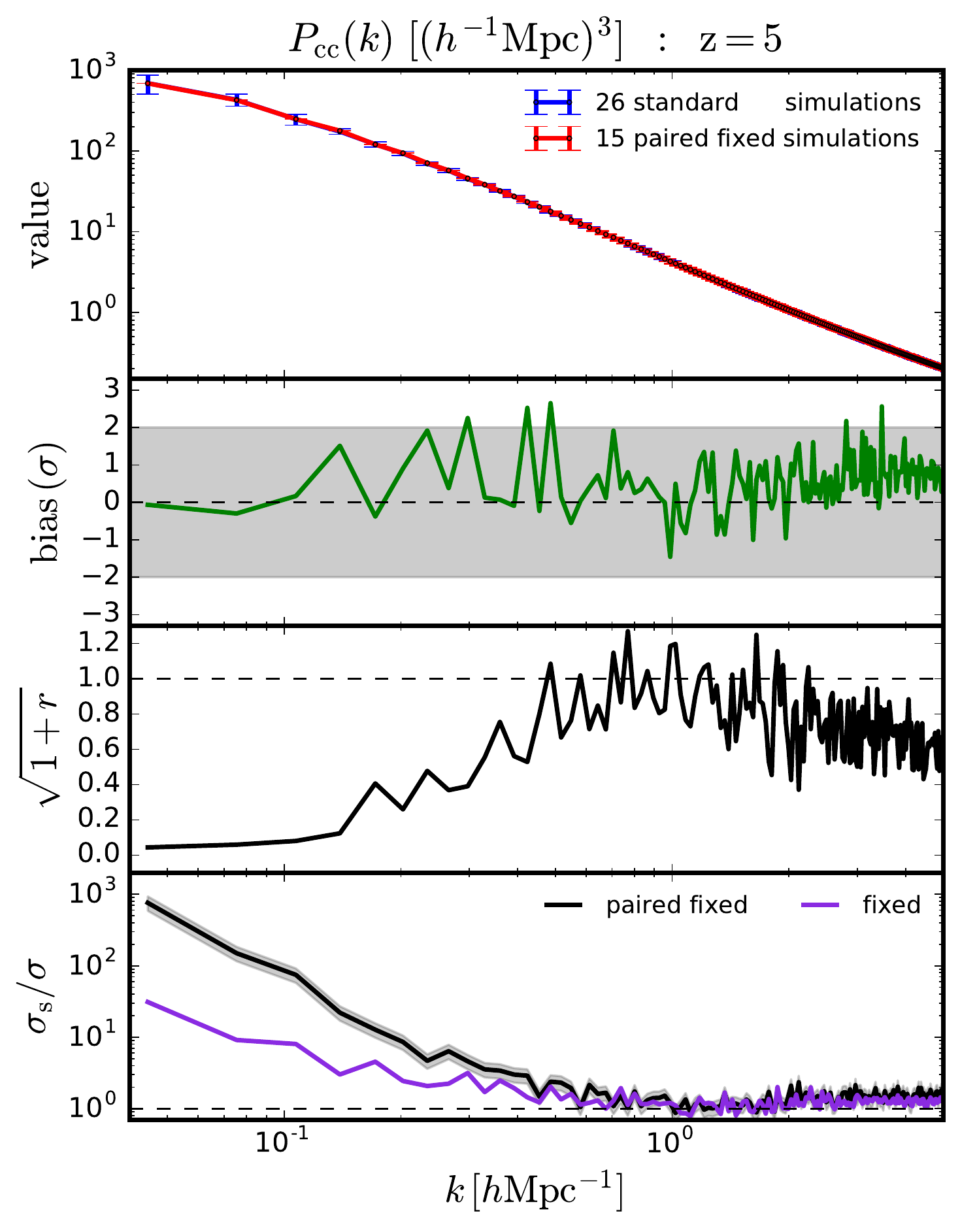}\\[2ex]
\includegraphics[width=0.33\textwidth]{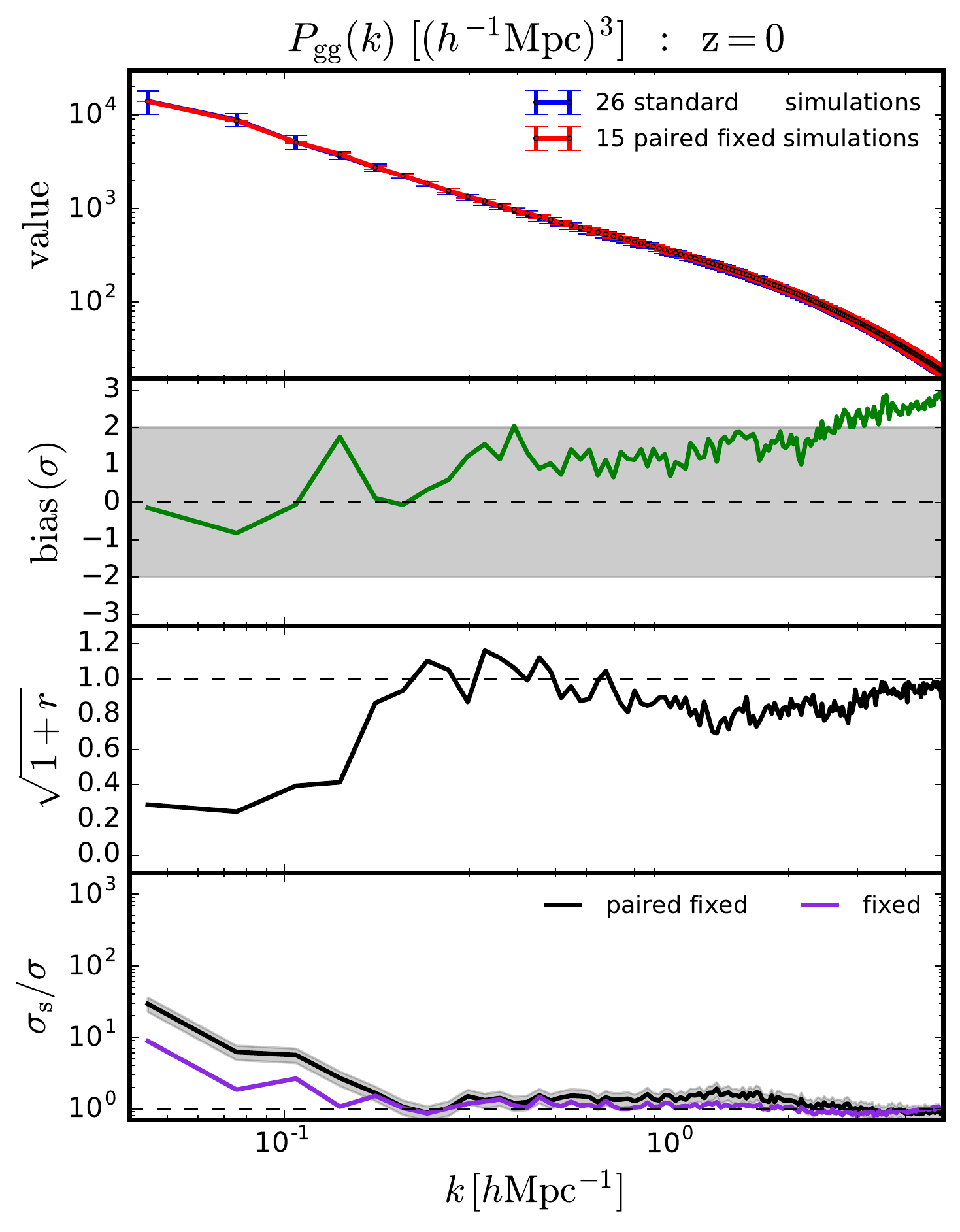}
\includegraphics[width=0.33\textwidth]{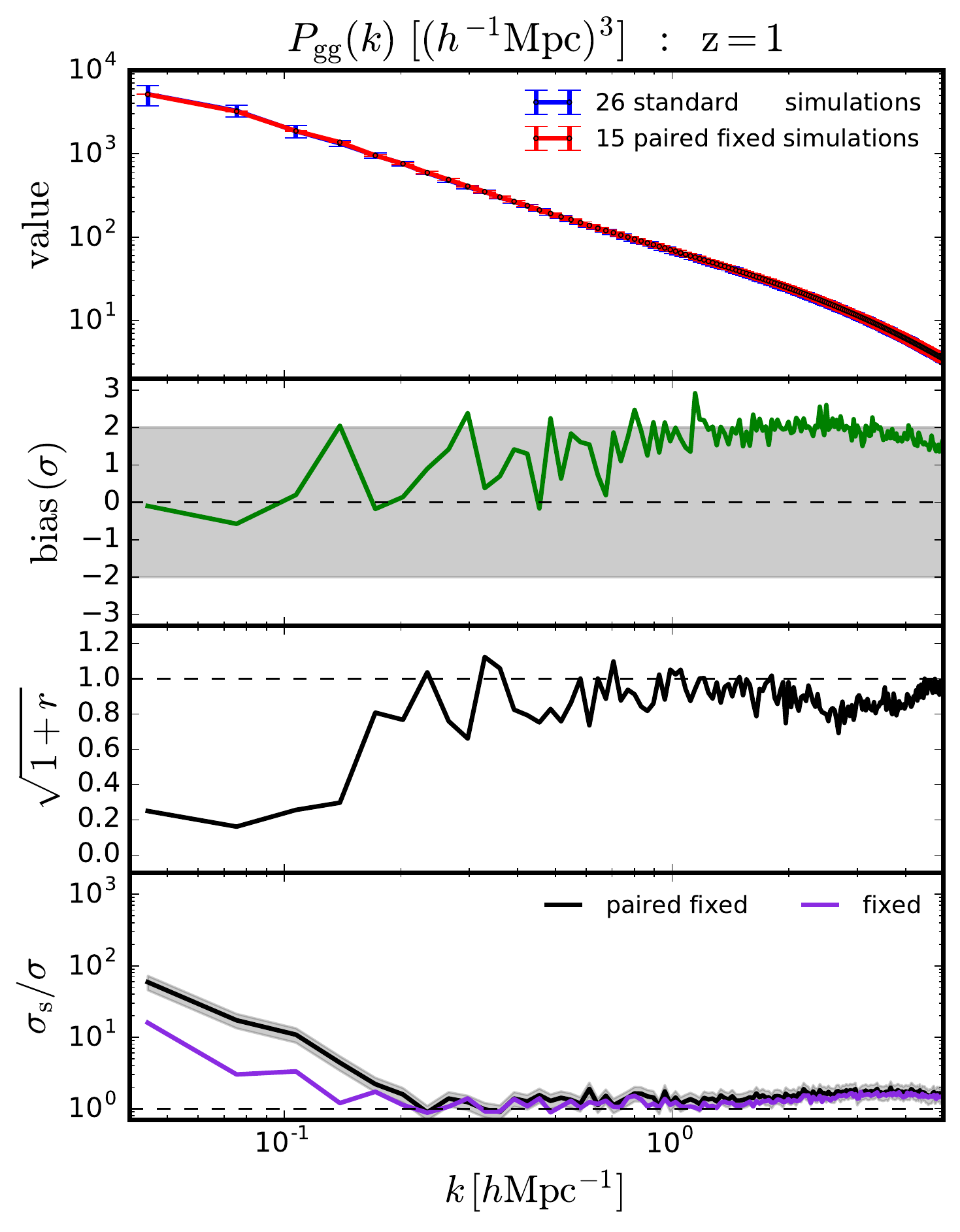}
\includegraphics[width=0.33\textwidth]{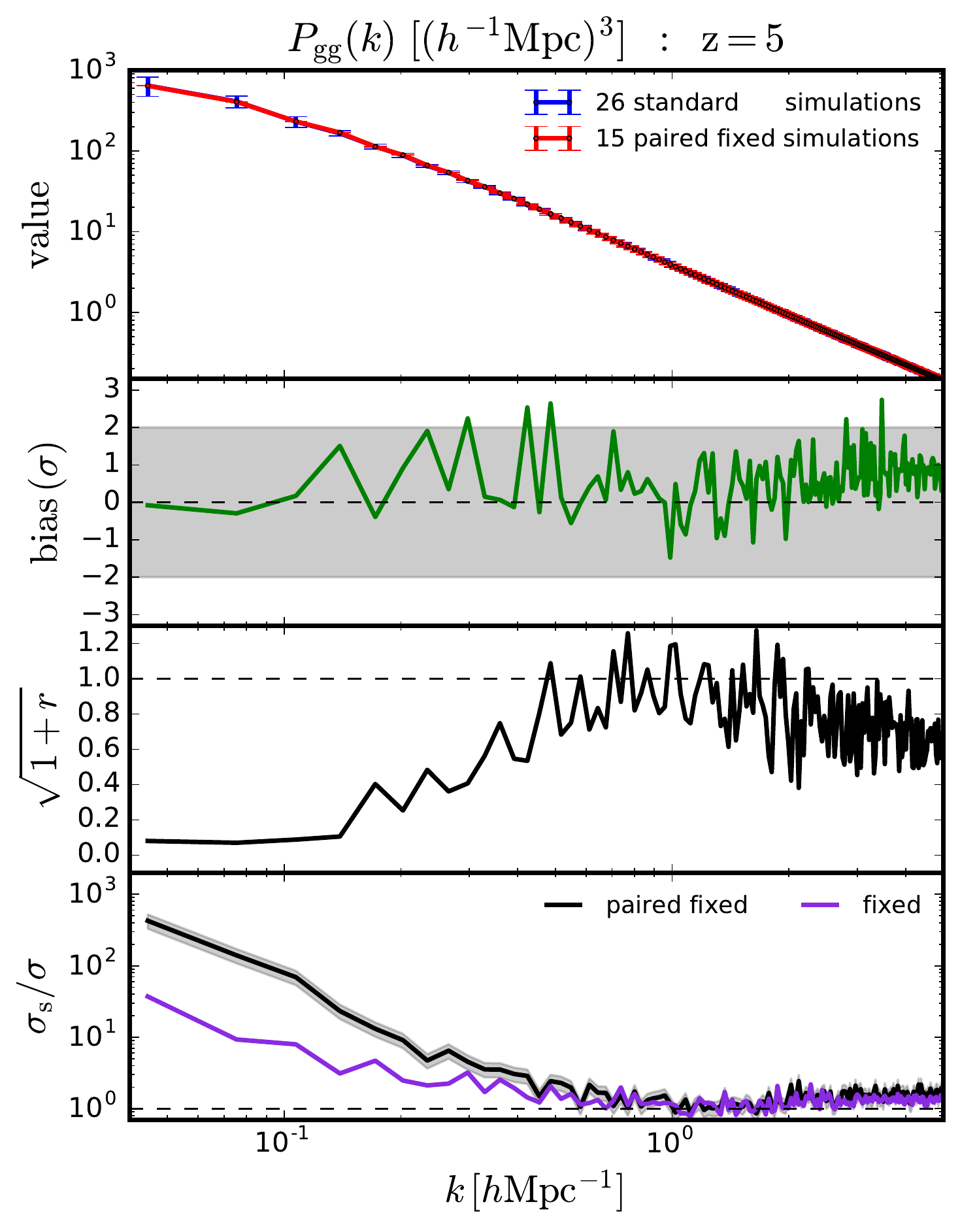}\\[2ex]
\caption{Impact of paired fixed simulations on the clustering of matter (top row), CDM (middle row) and gas (bottom row) from the magneto-hydrodynamic simulations set H200 at redshifts 0 (left column), 1 (middle column) and 5 (right column). paired fixed simulations largely reduce the sample variance errors associated to standard simulations on large-scales. We believe that some of the rather large bias values we find are not statistical significant.}
\label{fig:Pk_200Mpc_hydro}
\end{center}
\end{figure*}

We show our results in Fig.~\ref{fig:Pk_200Mpc_hydro}. From the first panels we deduce that the agreement between the different power spectra from the different simulations is very good at all redshifts. From the second panels we see that no bias is introduced by paired fixed simulations, with respect to standard simulations, on these power spectra. At redshift 5, we can see how results from the two simulation types are in agreement, within $2\sigma$, with a small fraction of points exceeding that threshold, as expected. At $z=1$, we find that many scales exhibit a discrepancy of $\simeq2\sigma$ for the three different power spectra. We emphasize that those scales are highly correlated, so it is expected that if one scale deviates, the others will exhibit the same behavior. Since the number of realizations we have in the H200 is very small, it is not unreasonable to expect mean differences of $\simeq2\sigma$. We find similar results at $z=0$, where in some cases, e.g.~gas power spectrum on very small scales, the difference between the mean of both data sets can be around $3\sigma$, but again, on highly correlated scales. In order to verify that this bias is not statistically significant we have repeated the above analysis but removing some random paired fixed or standard simulations. By doing so, we find that in most of the cases the bias between the two data sets decreases and remains below $2\sigma$. This points out that our low number of realizations may be underestimating the intrinsic scatter. Furthermore, as we will see in the next section, with a much larger number of hydrodynamic simulations covering a range of scales similar to those we explore here, we do not find a bias on any of the power spectra studied here. This reinforces our interpretation that the bias we find in the H200 simulations may be due to statistical fluctuations. More simulations are however needed to clearly disentangle this issue.

We find that the power spectra from the two pairs are strongly anti-correlated on large scales, for all the considered fields. This translates, as we shall see below, into large statistical improvements of the paired fixed simulations with respect to standard simulations. On smaller scales the value of the cross-correlation coefficient tends to zero, although usually remains smaller than zero. We note that at $z=0$ and for $k\simeq1.5~h{\rm Mpc}^{-1}$, the cross-correlation coefficient exhibits a significant dip. That dip also seems to take place at higher redshifts but on smaller scales. 

On large scales and for fixed simulations we find an improvement on the standard deviation of standard simulations that ranges from $\simeq7$ at $z=0$ to $\simeq30$ at $z=5$. The improvement for paired fixed simulations is much higher, induced by the low values of the cross-correlation coefficient. It is worth pointing out that running one paired fixed realization can be used to determine the mean of the matter, CDM or gas power spectrum with an error equal to that achieved by running $\simeq900$ standard simulations for $k\simeq0.1~h{\rm Mpc}^{-1}$, a very important scale for BAO studies. On smaller scales the statistical improvement vanishes, although we observe some residual improvement on scales where the value of the cross-correlation coefficient is below 0.

We thus conclude that paired fixed simulations bring large statistical improvements on the matter, CDM and gas power spectra on large scales from full hydrodynamic simulations.

\section{Small scales: hydrodynamic}
\label{sec:hydro_small}

We now push the limits of paired fixed simulations by studying their properties on small scales through the H20 hydrodynamic set. We focus our analysis on clustering, 1-point statistics and internal galaxy properties.

\subsection{Initial conditions}
\label{subsec:ICs_20Mpc_hydro}

We have computed the matter, CDM, and gas power spectra of each realization of the H20 simulations. The result of our statistical analysis for these quantities is similar to what we found for the N1000 simulations, i.e. a very large improvement on the largest scales of the box, while on smaller scales, the variance reduction is smaller. We thus do not show these results as they do not add much to our discussion. 

\begin{figure}
\begin{center}
\includegraphics[width=0.45\textwidth]{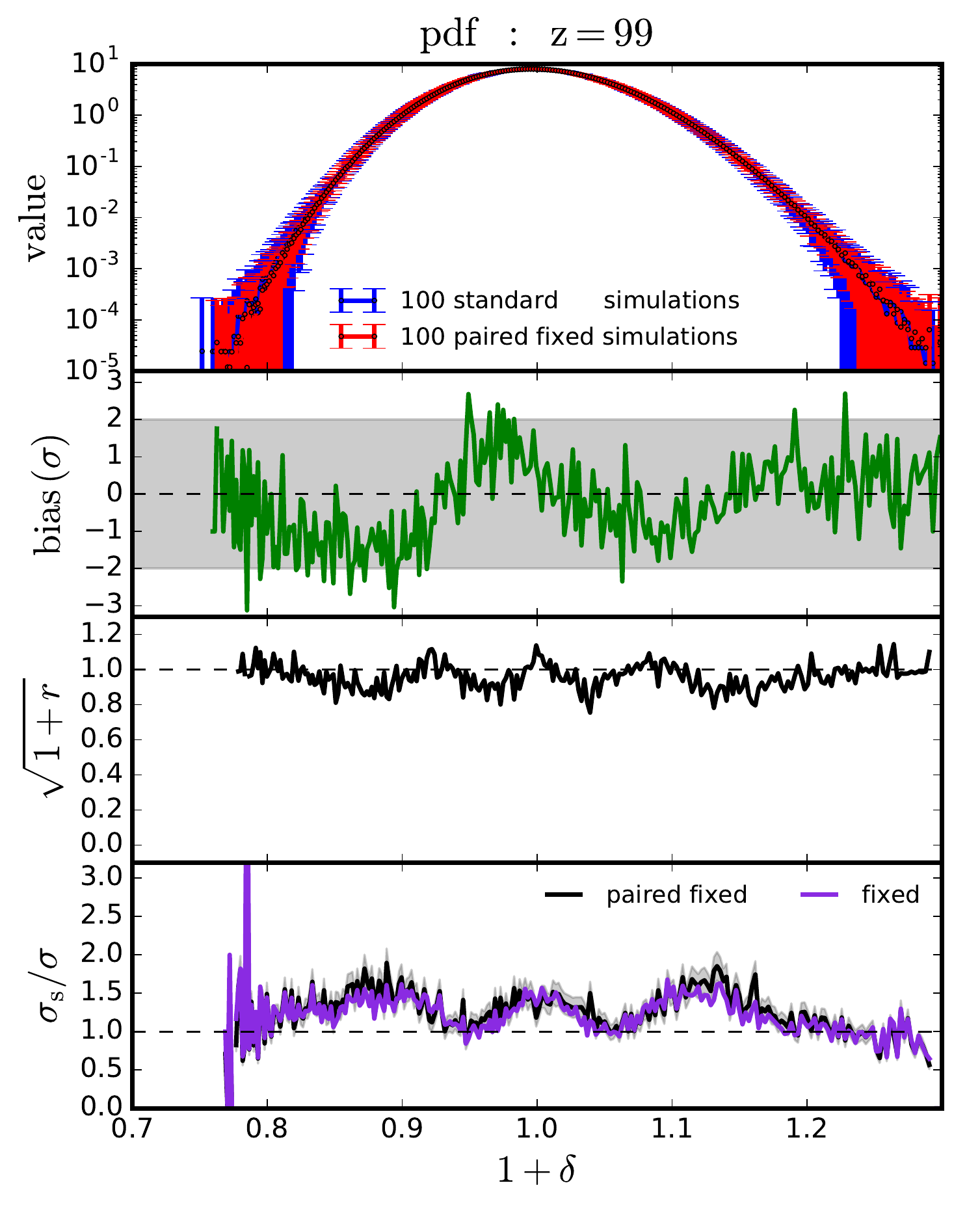}
\caption{Impact of paired fixed simulations on the matter density pdf of the initial conditions of the H20 simulation set. Contrary to what we found for the N1000 simulations, for small smoothing scales we find that fixed and paired fixed simulations can slightly reduce the scatter in the matter density pdf without introducing a bias on it. The statistical improvement takes places over different overdensities in a complicated manner.}
\label{fig:ICs_20Mpc_hydro}
\end{center}
\end{figure}

We have also computed the matter density pdf for each standard and paired fixed realization of the H20 simulations using a grid with $64^3$ cells by employing the CIC mass-assignment scheme. Fig.~\ref{fig:ICs_20Mpc_hydro} shows the result of our analysis. We find that paired fixed simulations do not introduce a bias on the matter density pdf of the standard simulations. 

The value of $\sqrt{1+r}$ is compatible with 1 for almost all overdensities, with deviations being mostly statistical fluctuations. From the fourth panel we can see that both, fixed and paired fixed simulations reduce the scatter of the matter density pdf of standard simulations in a non-trivial way. Those improvements, although small, are not statistical fluctuations. We obtain very similar results for the matter field when using the N20 simulations. We leave it for future work to understand the reason for why paired fixed simulations reduce the scatter of the matter density pdf relative to standard simulations in the way they do.

\subsection{Clustering}
\label{subsec:clustering_20Mpc_hydro}

\begin{figure*}
\begin{center}
\includegraphics[width=0.33\textwidth]{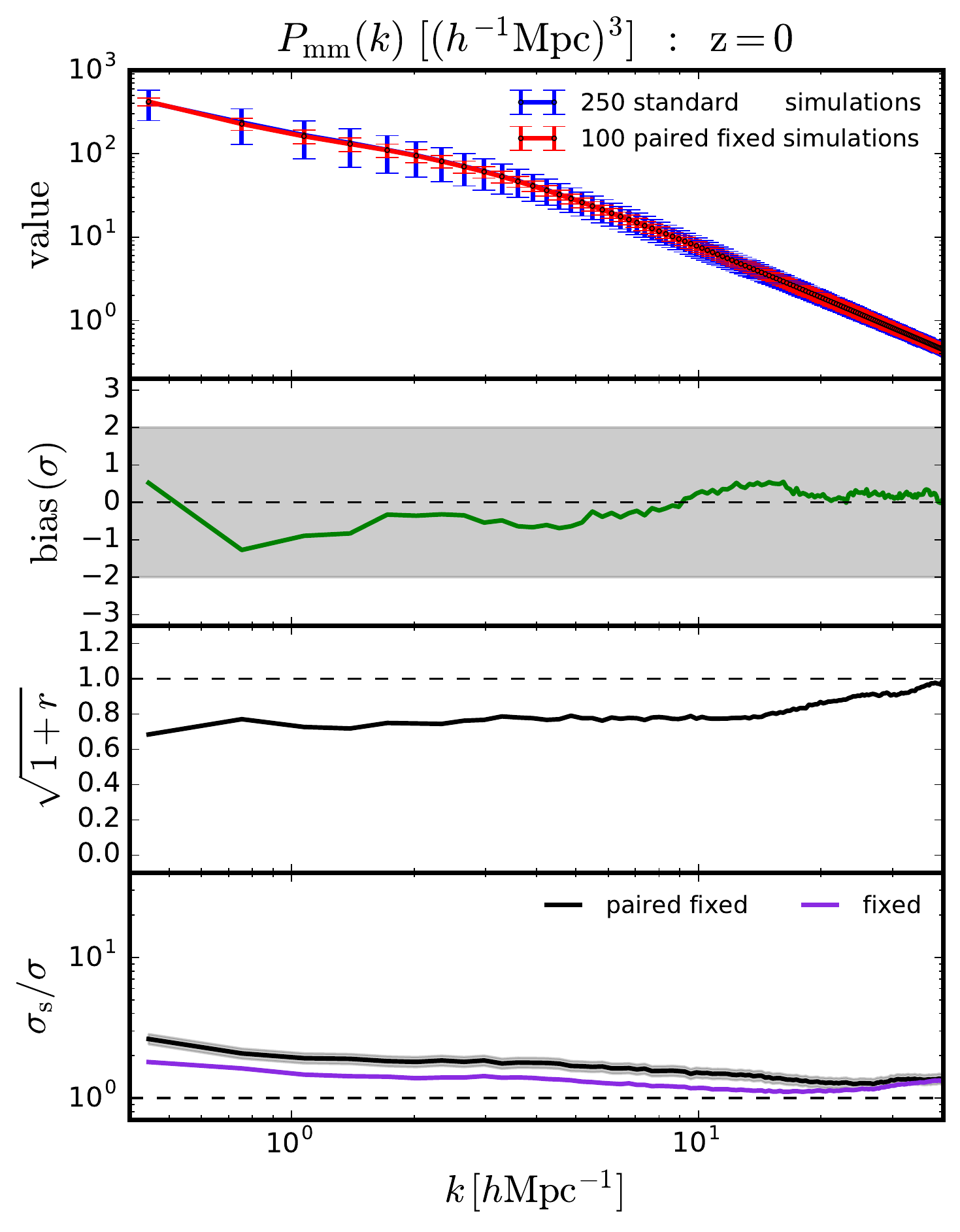}
\includegraphics[width=0.33\textwidth]{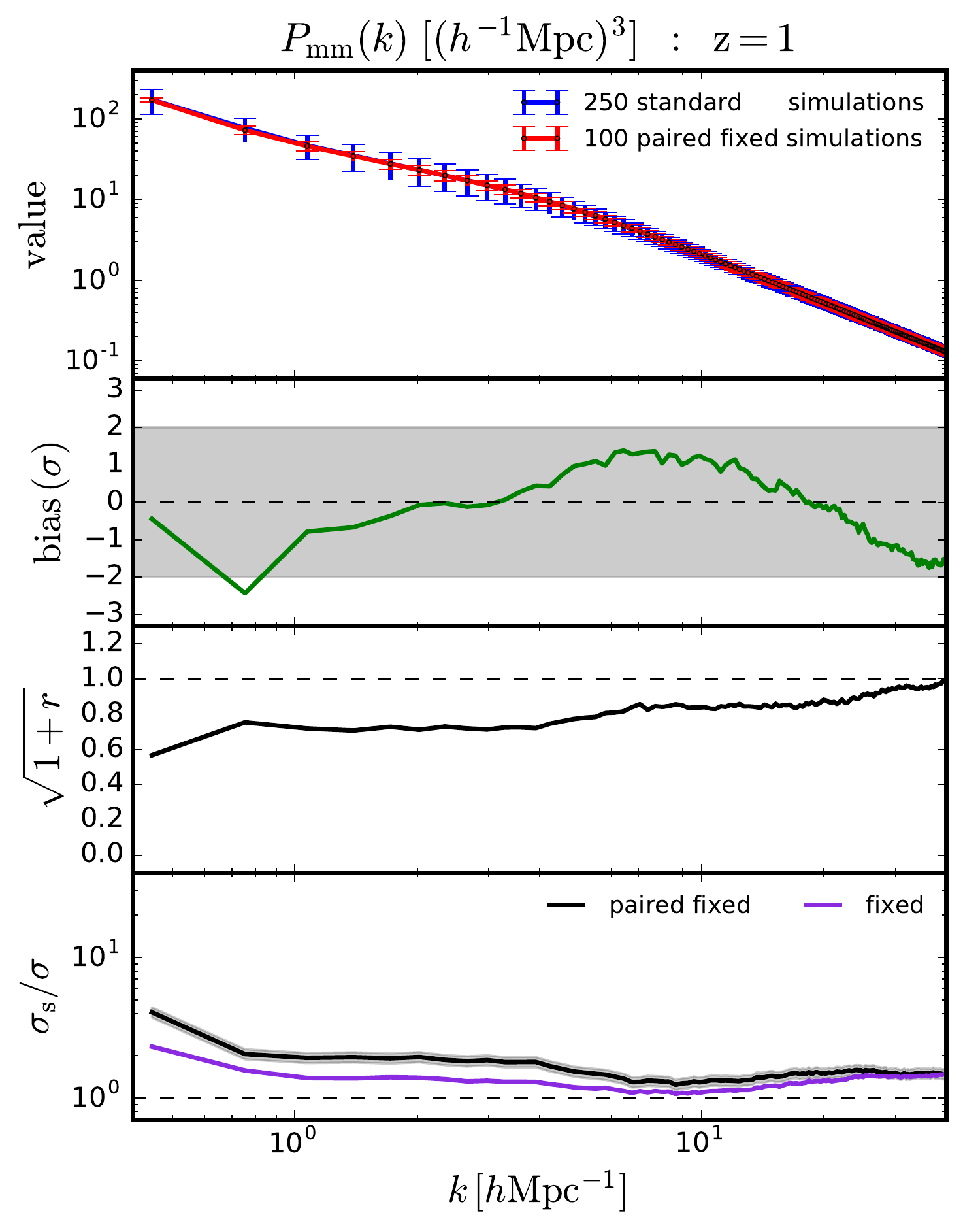}
\includegraphics[width=0.33\textwidth]{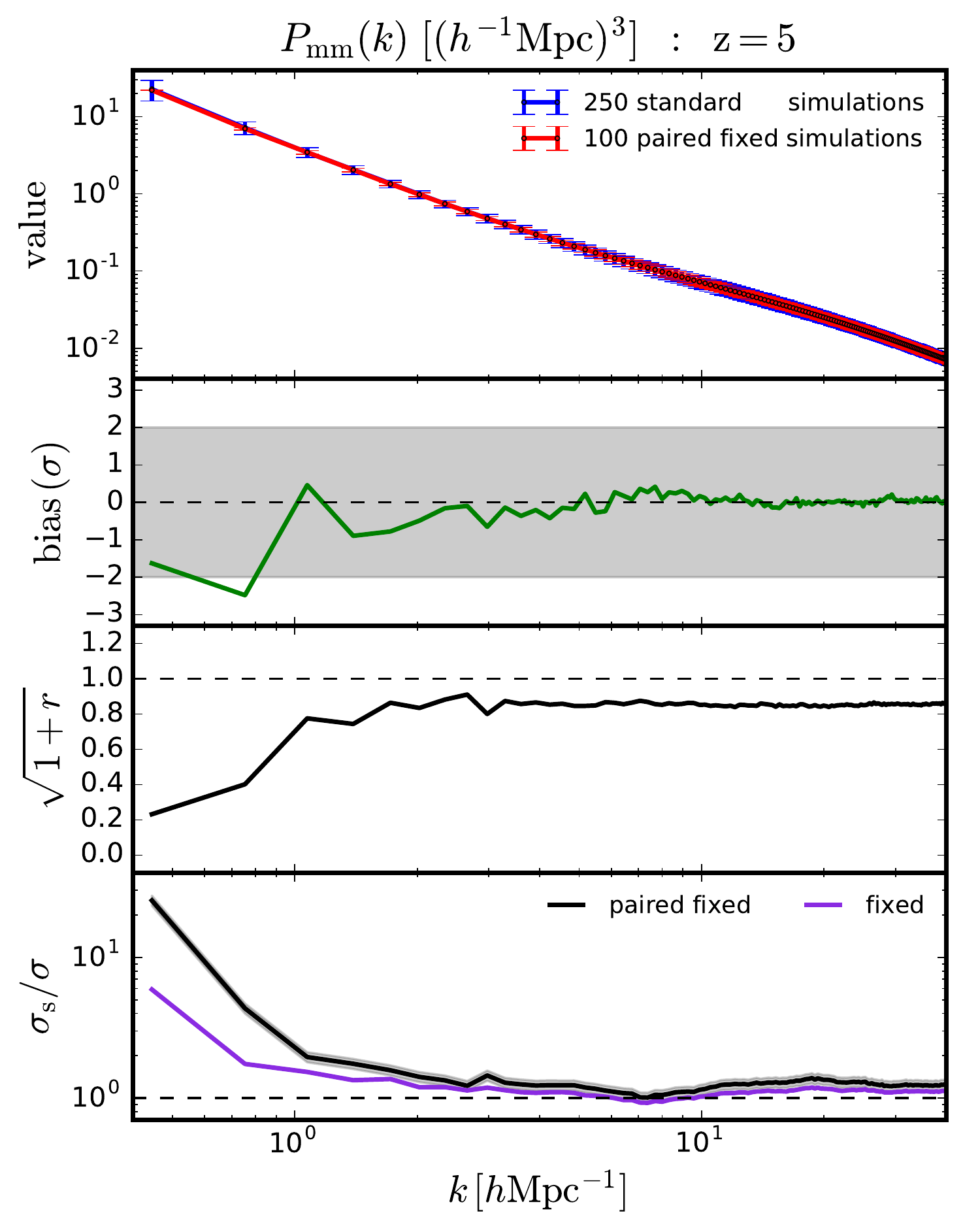}\\[2ex]
\includegraphics[width=0.33\textwidth]{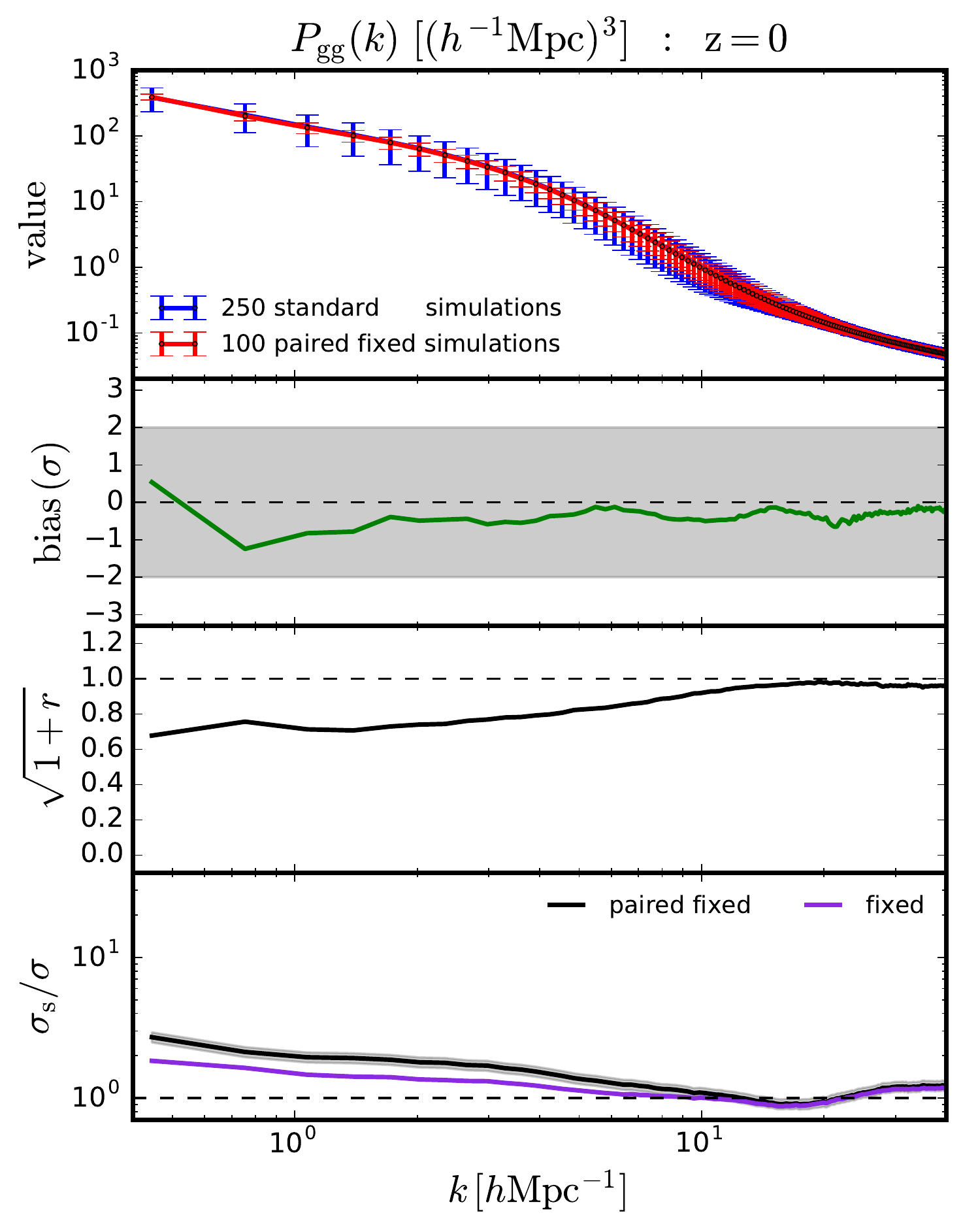}
\includegraphics[width=0.33\textwidth]{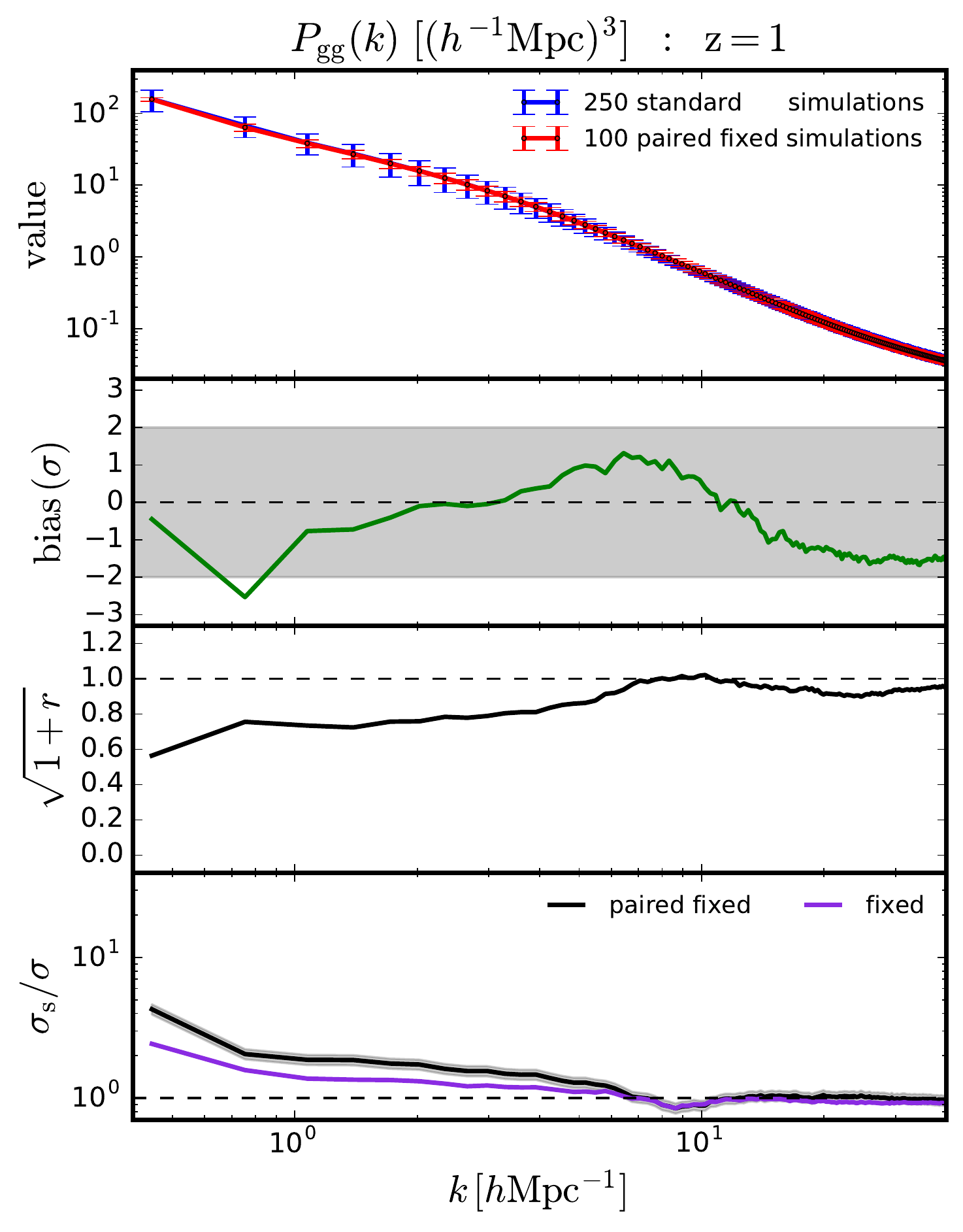}
\includegraphics[width=0.33\textwidth]{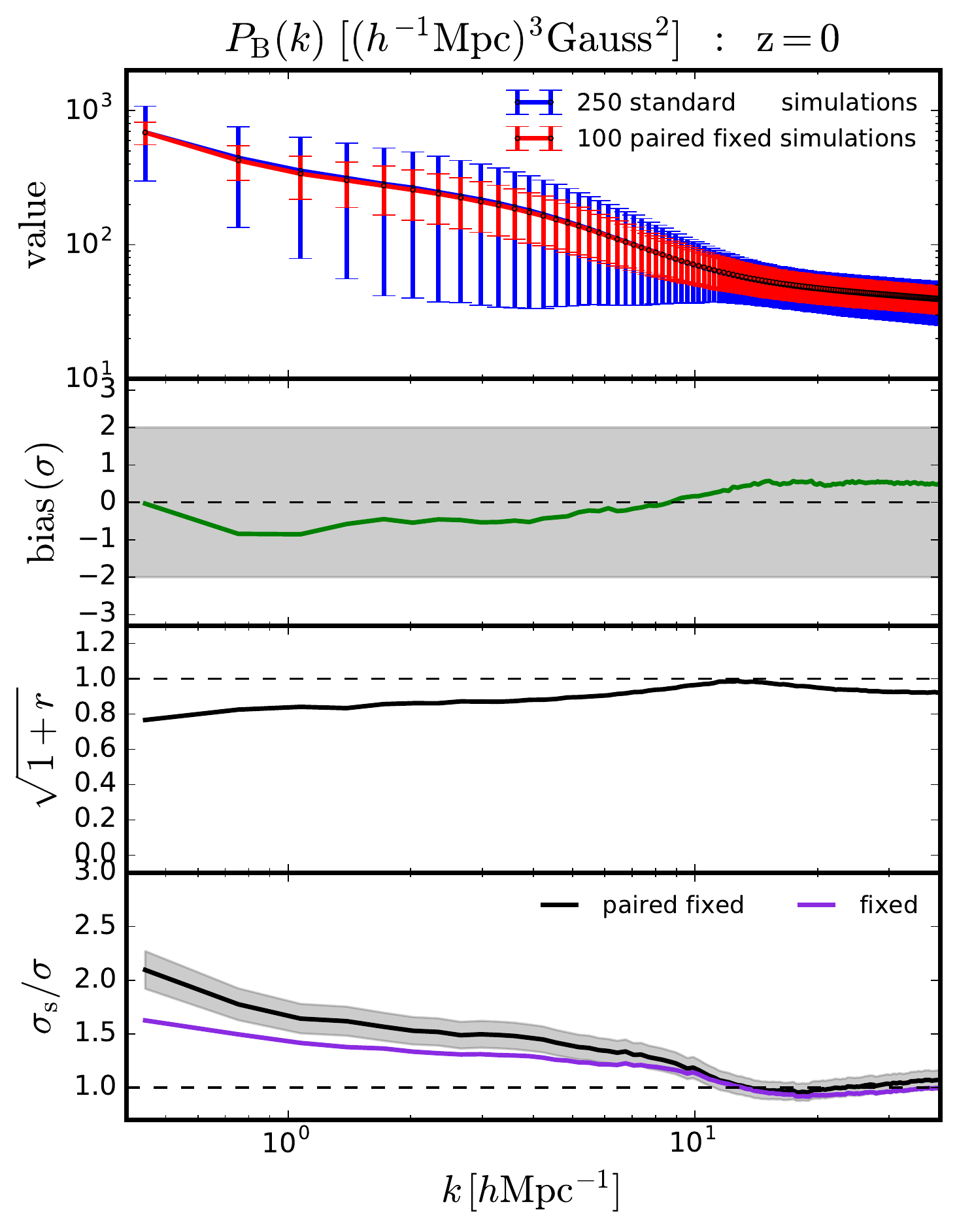}\\[2ex]
\includegraphics[width=0.33\textwidth]{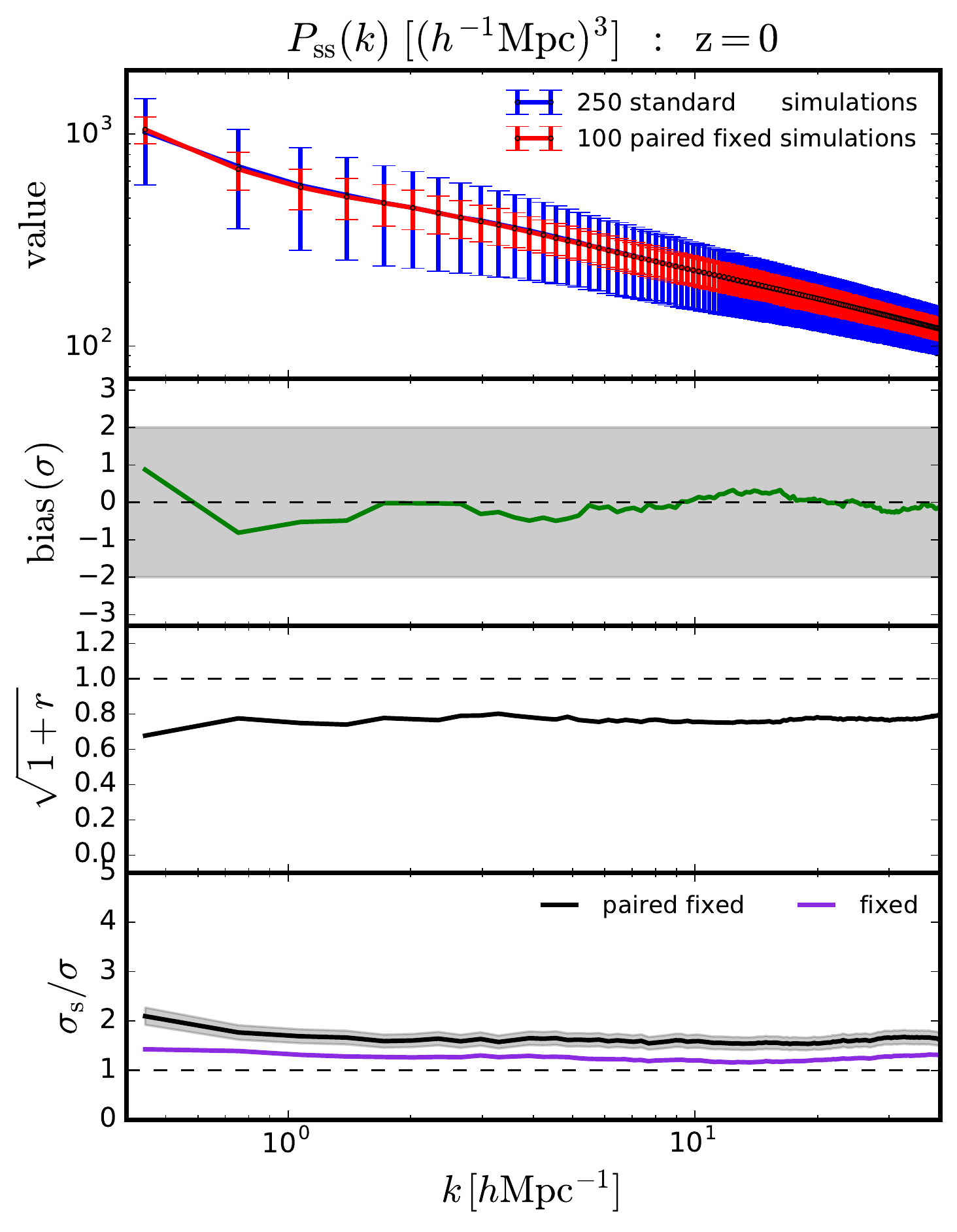}
\includegraphics[width=0.33\textwidth]{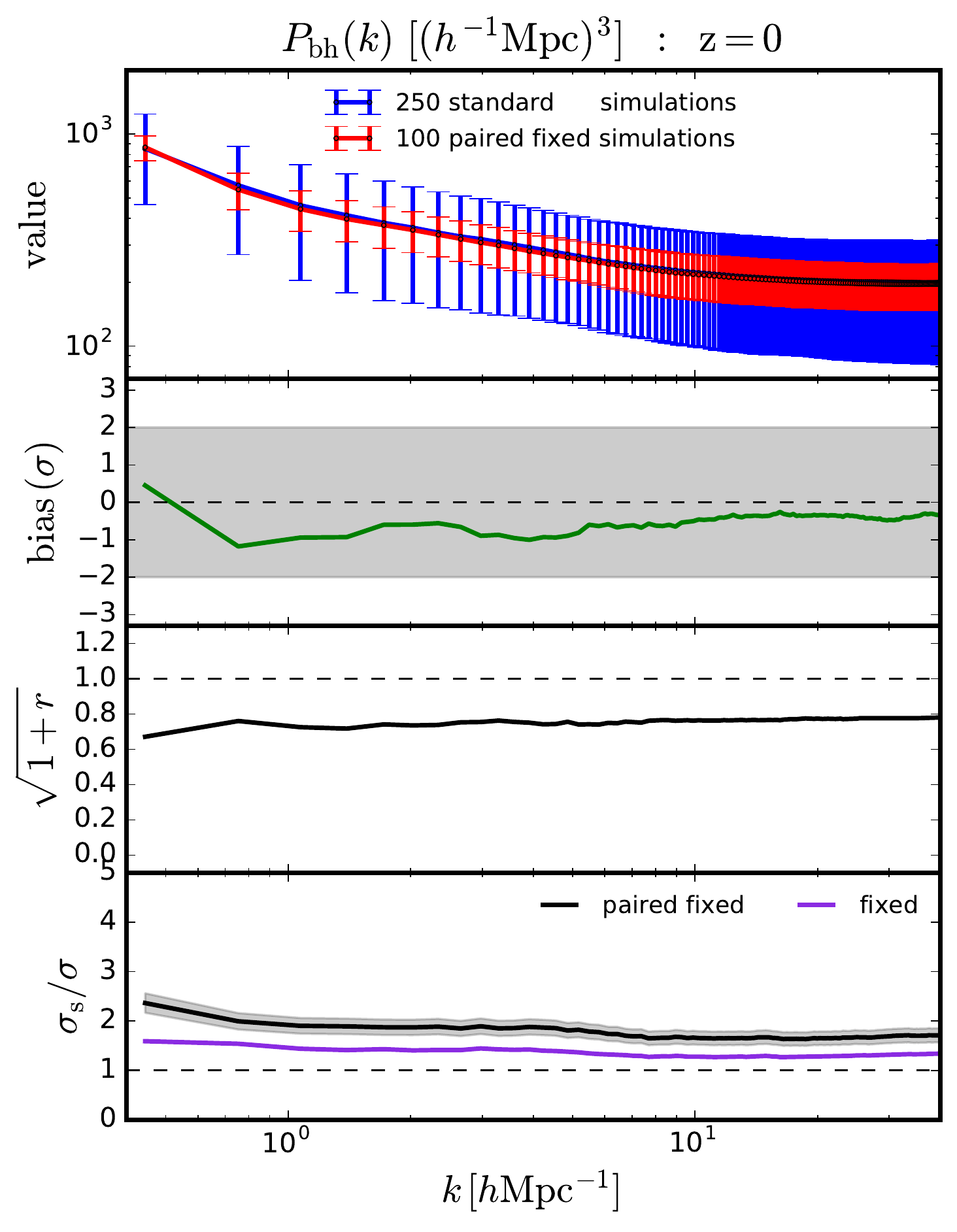}
\includegraphics[width=0.33\textwidth]{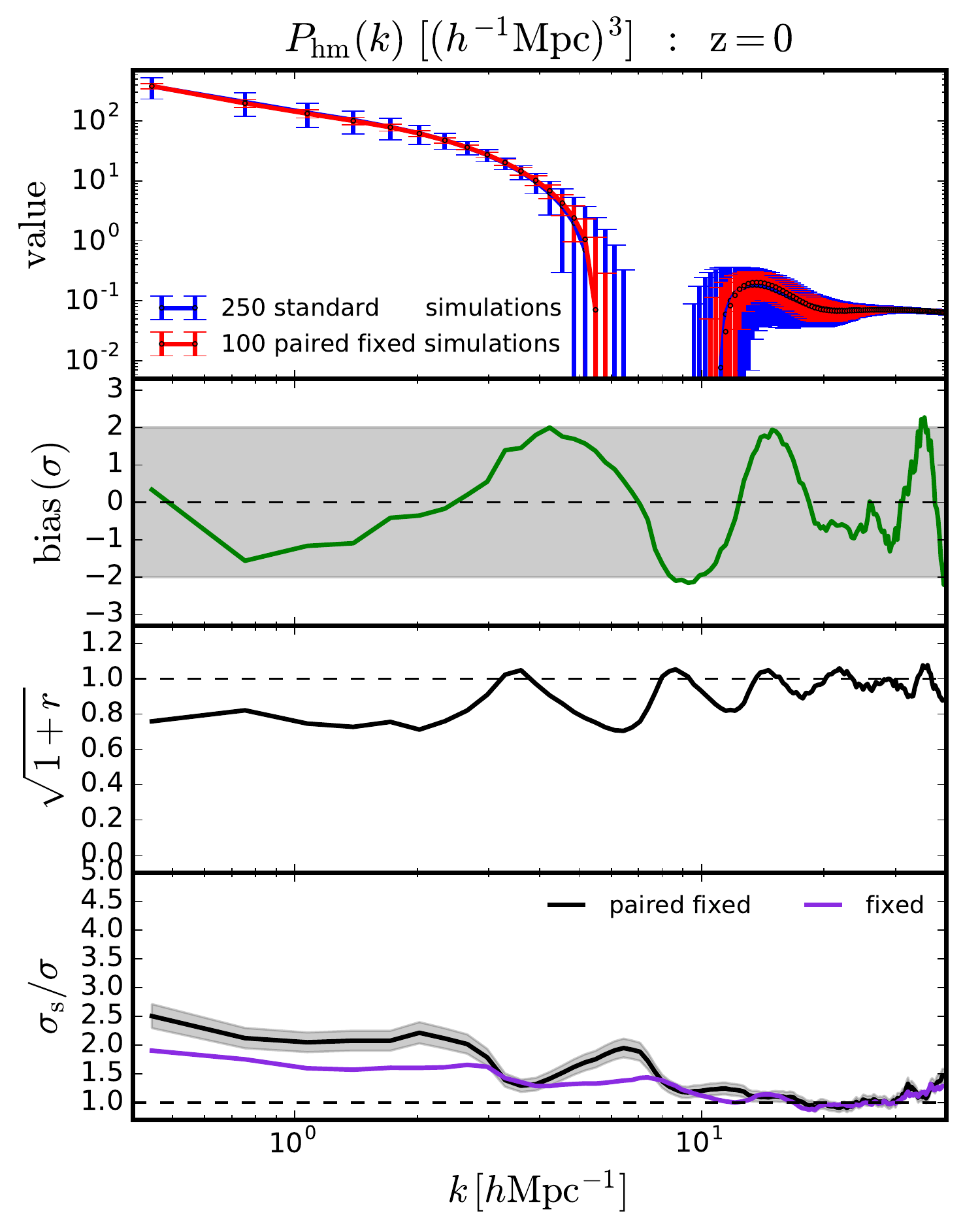}
\caption{Impact of paired fixed simulations on the power spectrum of matter (top-row), gas (left and middle panels of middle row), magnetic fields (right panel of middle row), stars (bottom-left), black-holes (bottom-middle) and halo-matter (bottom-right) from the N20 simulations. Results for matter and gas are shown are redshifts 0 (left column), 1 (middle column) and 5 (right column, only for matter), while for magnetic fields, stars, black holes and the halo-matter we only show results at $z=0$ since we observe little time evolution. Paired fixed simulations do not introduce a bias on any of these quantities and improve the statistics of standard simulations. }
\label{fig:Pk_20Mpc_hydro}
\end{center}
\end{figure*}

For each standard and paired fixed realization we have computed the power spectrum of matter, CDM, gas, magnetic fields, stars, black holes, halos and halo-matter. In Fig.~\ref{fig:Pk_20Mpc_hydro} we show the results at redshifts 0, 1 and 5 for the total matter and gas power spectra (for gas only at redshifts 0 and 1) in the top and middle rows, respectively. We do not show the results for CDM since they are pretty similar to those from total matter and gas. The results for gas at $z=5$ are also similar to those of matter at that redshift. In the middle-right panel we show the results for the magnetic field power spectrum, while in the bottom row we display our findings for the stars, black-holes and the halo-matter power spectra. Since our conclusions for those components do not change significantly with redshift, we only show those at $z=0$.

From the first panels we can see that the agreement between the results of the two simulation types is very good. In the second panels we demonstrate that for all power spectra and considered redshifts the bias introduced by paired fixed simulations, with respect to standard simulations, is compatible with 0. We emphasize that the scales we probe with the H20 simulations are highly non-linear and correlated, as expected. This is why the green curves look so smooth in comparison with those of, e.g., Fig.~\ref{fig:Pk_1000Mpc_Nbody}. 

We find that on almost all scales, for all power spectra, and at all redshifts, the value of $\sqrt{1+r}$ is lower than 1, pointing out that the power spectra from the two pairs of the paired fixed simulations exhibit a degree of anti-correlation. At $z=0$, and on the largest scales we can probe with the H20 simulations, the value of $\sqrt{1+r}$ is around 0.7. At higher redshift that value shrinks, reaching $\simeq0.2$ for matter, CDM and gas at $z=5$. As we move to smaller scales, the value of the cross-correlation coefficient increases. At low redshift and for matter, CDM and gas it tends to 1, while for stars and black-holes it remains quite constant at $\sqrt{1+r}\simeq0.8$. We observe a similar behavior at $z=5$ for matter, CDM and gas. 

In the fourth panel we see that at low redshift, the improvement on the sample variance reduction is moderate, with the standard deviation ratio reaching factors of 2 to 3 for matter, CDM and gas on the largest scales. For the magnetic fields, stars and black-holes the improvement is slightly lower, but almost scale-independent, with the exception of the magnetic field. As we move to smaller scales, the improvement decreases, although showing a non-monotonic dependence with redshift. The difference between the improvement from fixed and paired fixed is not large at low-redshifts, while at $z=5$ it can be a factor of almost 5 on the largest scales. 

We note that the power spectrum of the magnetic field, stars and black-holes is highly affected by shot-noise. It is thus interesting to see that  paired fixed simulations help to reduce the intrinsic error on it.

We find very interesting results for the halo-matter cross-power spectrum. The value of $\sqrt{1+r}$ exhibits an oscillatory behavior that is not due to statistical fluctuations and whose value is, in almost all scales, below 1. From the fourth panel we can see how on scales larger than $\simeq2~h{\rm Mpc}^{-1}$ fixed and paired fixed simulations slightly improve the statistics of the standard simulations. The oscillatory features we found in the value of the cross-correlation coefficient are reflected in the statistical improvement of paired fixed simulations, although fixed simulations also present that behavior, to a lesser extent.

We find similar oscillatory features in the halo auto-power spectrum and the halo bias. While in the former are not due to the behavior of the cross-correlation coefficient, the latter exhibit the same features as the halo-matter power spectrum. We believe that the oscillations in the standard deviation ratio of the different halo power spectra are related to the features we observe in the matter density pdf of the initial conditions, that propagate to the matter density pdf and halo mass function at lower redshift (see next Subsection). A more detailed study of this is beyond the scope of the present paper.

We thus conclude that even with small box size hydrodynamic simulations where all scales are non-linear at low-redshift, paired fixed simulations always produce power spectra with lower scatters than those from standard simulations. The statistical improvement can be pretty large at high redshift. Our results also point out that paired fixed simulations do not introduce a bias on any of the above power spectra.

\subsection{1-point statistics}
\label{subsec:1pt_20Mpc_hydro}

We now study the impact of paired fixed simulations on 1-pt statistics. We focus our analysis on the halo mass function, void mass function, the matter density pdf, the star-formation rate history and the stellar mass function. We only show results at $z=0$\footnote{For the star-formation rate history we show results between redshifts 0 and 15.} since our conclusions are unchanged at higher redshifts.

\begin{figure*}
\begin{center}
\includegraphics[width=0.33\textwidth]{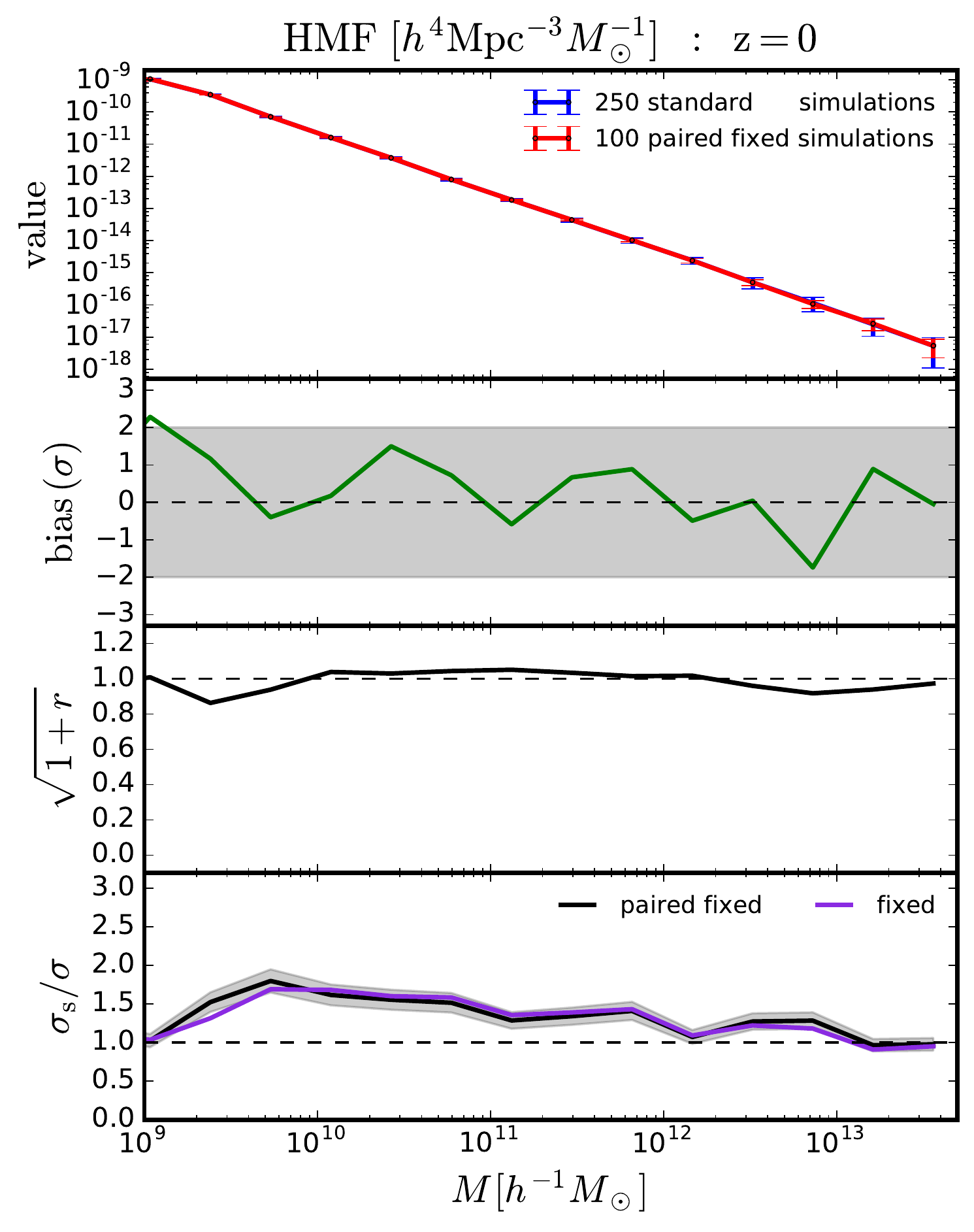}
\includegraphics[width=0.33\textwidth]{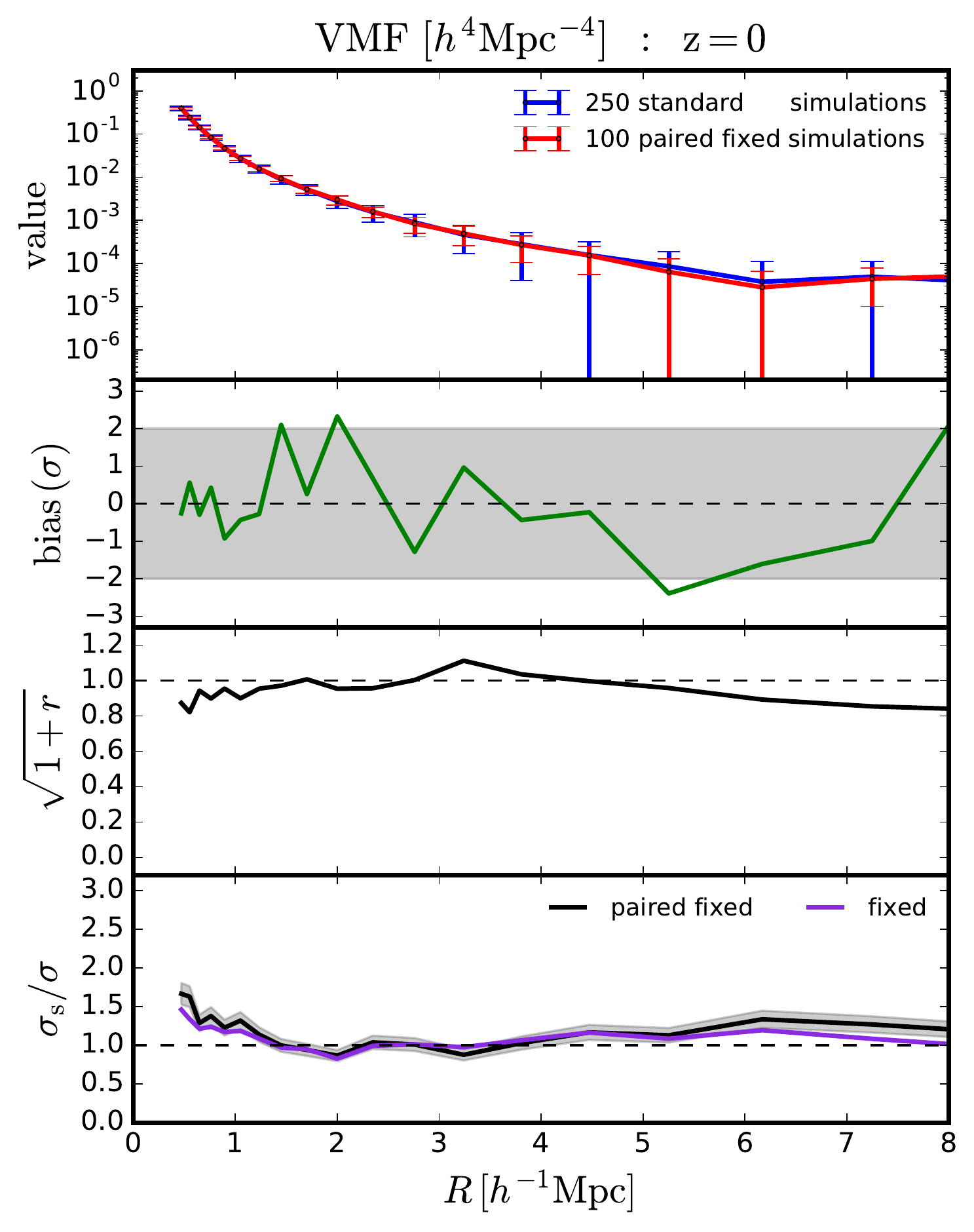}
\includegraphics[width=0.33\textwidth]{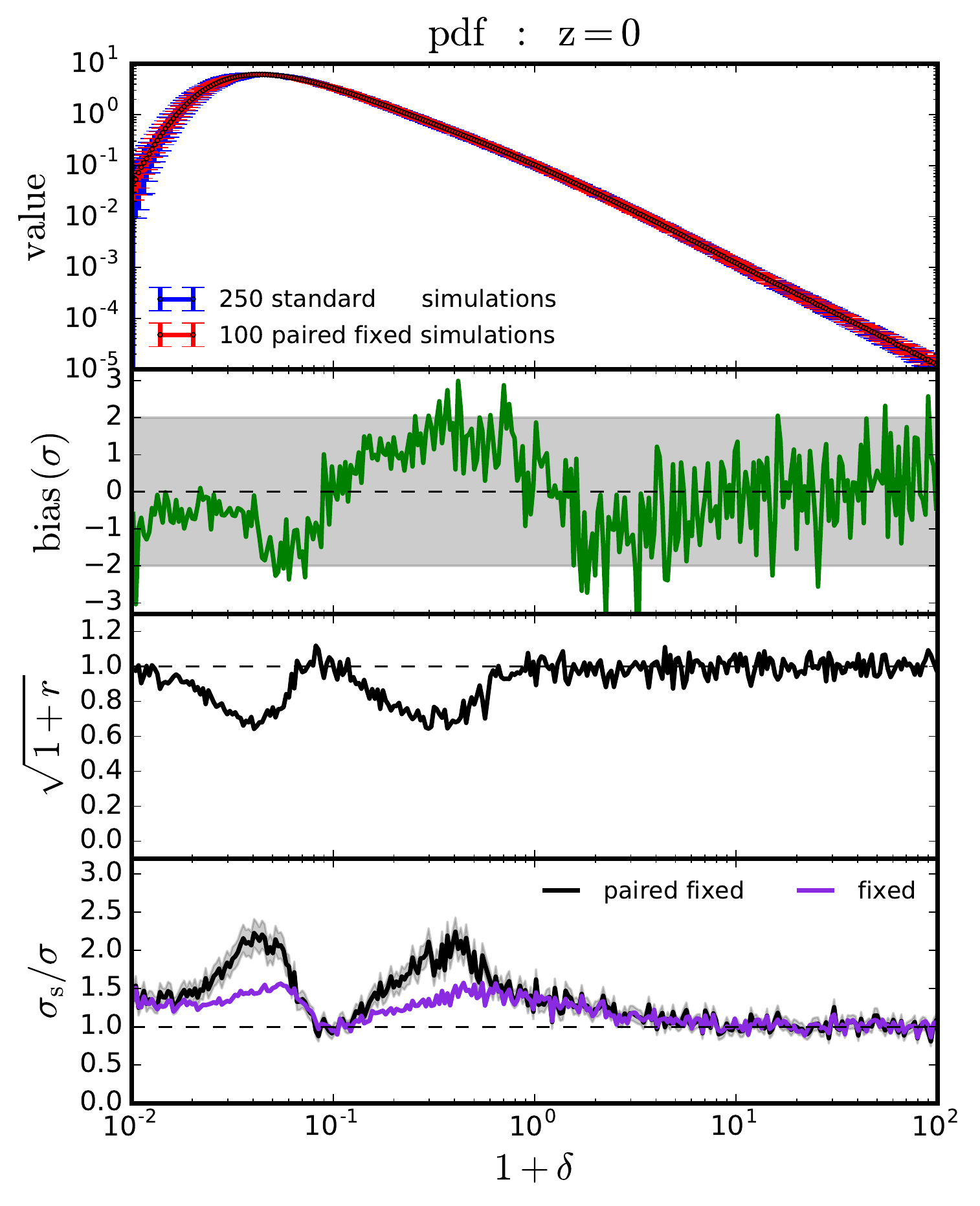}\\
\includegraphics[width=0.33\textwidth]{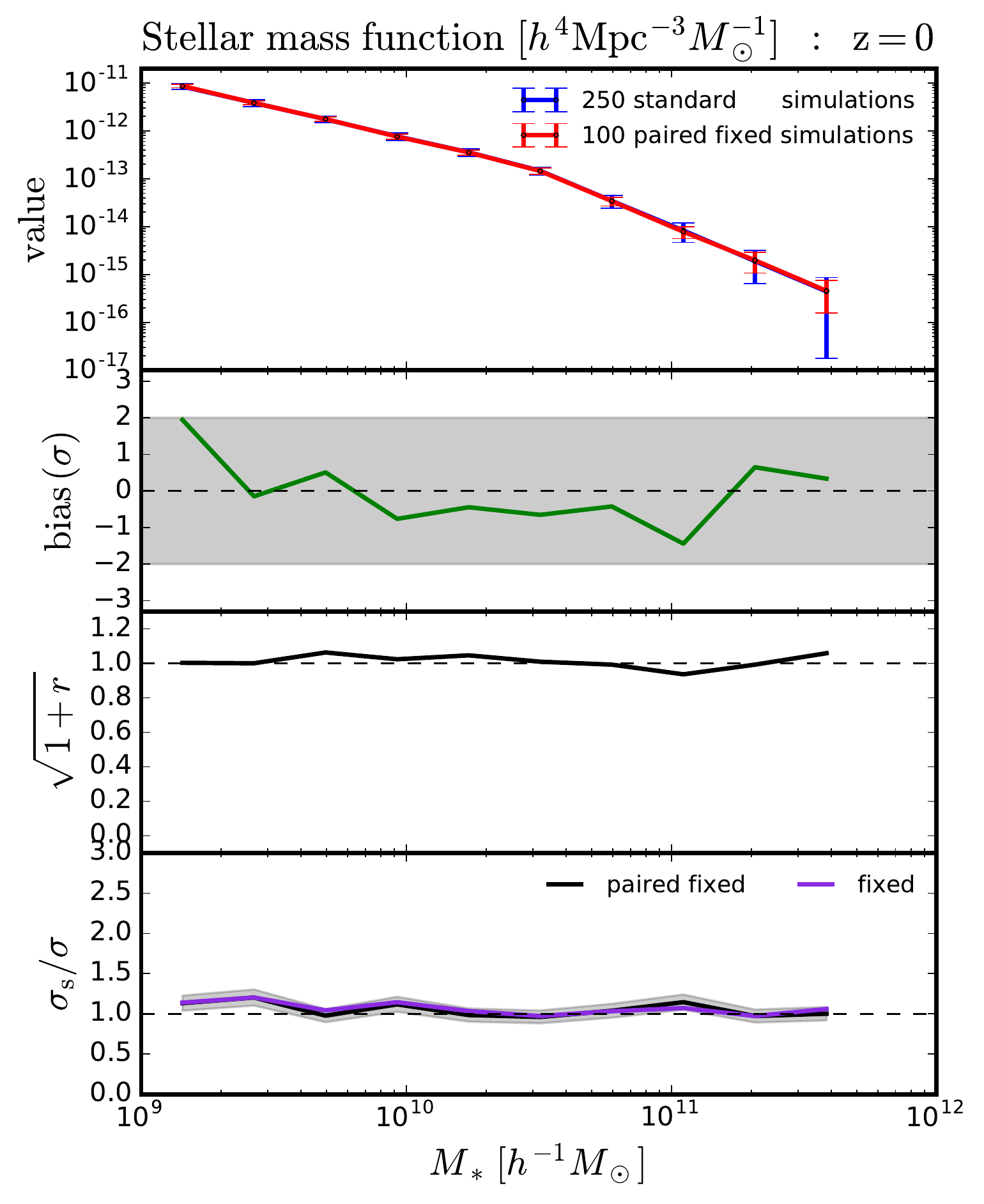}
\includegraphics[width=0.33\textwidth]{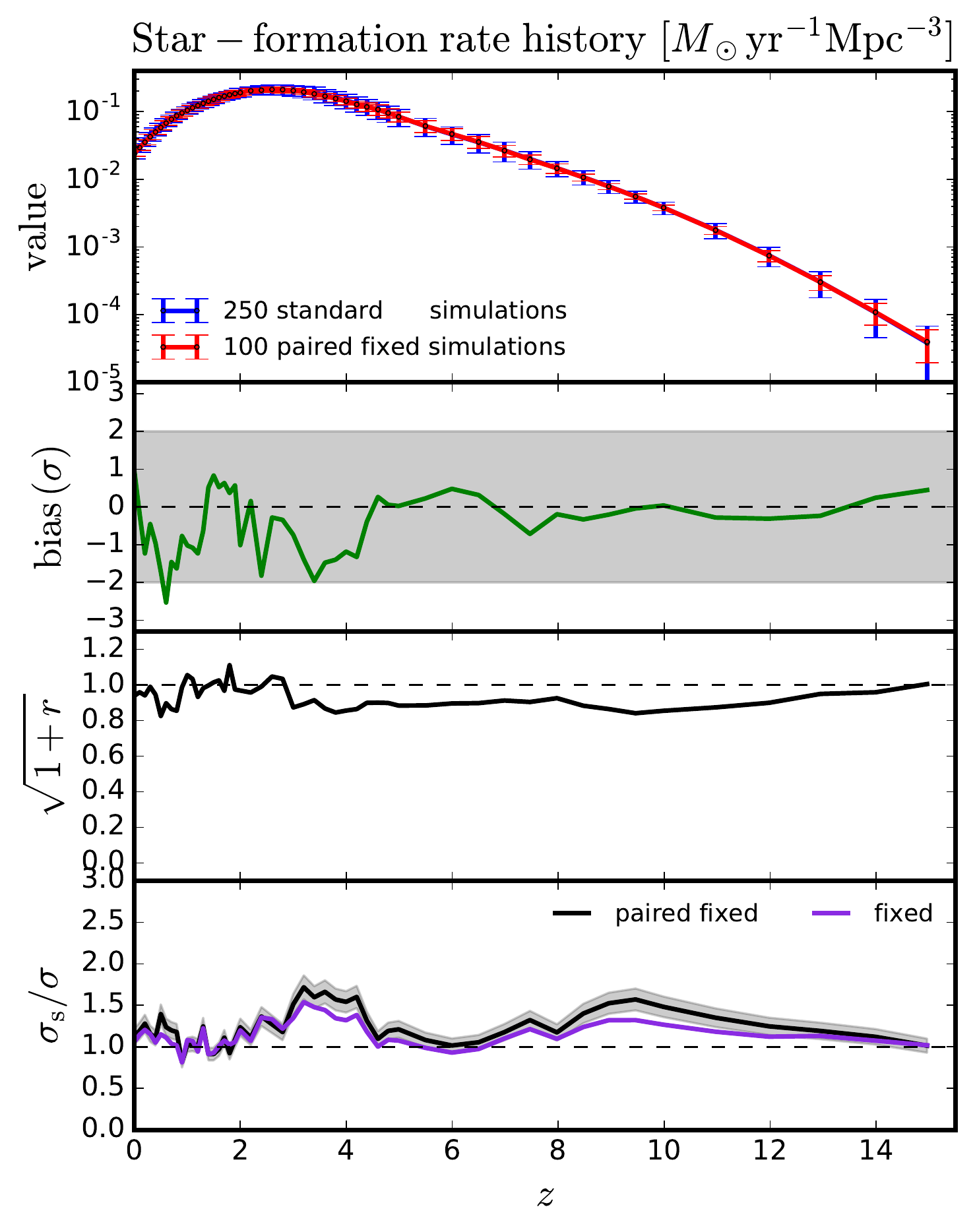}
\caption{Impact of paired fixed simulations on the halo mass function (top-left), void mass function (top-middle), matter density pdf (top-right), stellar mass function (bottom left) and star-formation rate history (bottom-right) from the N20 magneto-hydrodynamic simulations at $z=0$. We find similar results at higher redshifts. paired fixed simulations do not introduce a bias on any of these quantities and they slightly improve the statistics of some quantities in a complicated manner.}
\label{fig:1pt_20Mpc_hydro}
\end{center}
\end{figure*}

\subsubsection{Halo mass function}

For each realization of the standard and paired fixed simulations we have extracted halo catalogues by selecting all halos with masses above $\simeq9\times10^8~h^{-1}M_\odot$. We have then computed the halo mass function for each realization and show the results in the top-left panel of Fig.~\ref{fig:1pt_20Mpc_hydro}. We find an excellent agreement between the results of both simulation types and our results point out that paired fixed simulations do not introduce a bias. We can also see that the value of the $\sqrt{1+r}$ is compatible with 1 for all halo masses. 

From the fourth panel we can see how fixed and paired fixed simulations slightly reduce the scatter on the halo mass function from standard simulations for some halo masses. This contrasts with our results of section \ref{sec:Nbody}, where we found that paired fixed simulations do not reduce the scatter in the halo mass function. Note however that in section \ref{sec:Nbody} we only probed halos with masses above $\simeq10^{13}~h^{-1}M_\odot$, thus, for the halo mass range common to both simulations, our results are in agreement. 

We note that the statistical improvement is not very significant, taking into account the errorbars associated to the paired fixed simulations. In order to verify the robustness of this results we have repeated the same analysis but using the N20 simulations, which are N-body and contain a different number of paired fixed realizations. By doing so we find very similar results to what we find with the H20 simulations, implying that the improvement is not a statistical fluctuation, but a physical effect. 

Understanding the origin of this improvement on the halo mass function of small halos is beyond the scope of the current work. 

\subsubsection{Void mass function}

For each realization of the standard and paired fixed simulations we have extracted voids in the matter field. In the top-middle panel of Fig.~\ref{fig:1pt_20Mpc_hydro} we show the results for the void mass function. As always, we find a good agreement between the results of both simulations types, and a bias between the mean of both simulations that is below $\simeq2\sigma$. The value of the cross-correlation coefficient is compatible with 0 ($\sqrt{1+r}=1$) for most of the void radii. We find that fixed and paired fixed simulations do not reduce the scatter on the void mass function. This is in agreement with our findings for larger voids in section \ref{sec:Nbody}.

\subsubsection{Matter density pdf}

We have computed the matter density field on a grid with $64^3$ cells using the CIC interpolation scheme for each realization of the standard and paired fixed simulations. Our results for the pdf of the matter field are shown in the top-right panel of Fig.~\ref{fig:1pt_20Mpc_hydro}. We find good agreement among the results of both simulation types and that most of the points are below $2\sigma$. The value of $\sqrt{1+r}$ is compatible with 1 for all overdensities with the exception of two dips for values of $1+\delta$ around 0.04 and 0.4. The origin of those dips is unclear to us, but we have verified that they are not statistical fluctuations. We obtain similar results by using the N20 simulations.

The fourth panel shows the statistical improvement achieved by fixed and paired fixed simulations with respect to standard simulations. We find that for overdensities larger than $\simeq5$, fixed and paired fixed simulations do not reduce the intrinsic scatter of the standard simulations. For lower overdensities, we do however observe improvements. Those come from both the fixed and paired fixed simulations, and manifest themselves as 2 bumps for overdensity values similar to those quoted above. In paired fixed simulations the improvement is more pronounced on those bumps due to the anti-correlation of the pdfs we find in the third panel. This result is different to what we found in Section \ref{sec:Nbody}, where we concluded that paired fixed simulations do not reduce the scatter of the matter density pdf. We notice however that the scales we are probing in the two cases are very different. Besides, for these very small smoothing scales, we find that paired fixed simulations slightly improve the statistics of the matter density pdf already in the initial conditions (see Subsection \ref{subsec:ICs_20Mpc_hydro}). Future investigation of this effect will be required to disentangle whether the improvement propagates from the initial conditions or is brought by non-linear evolution.
 
\subsubsection{Stellar mass function}

The results for the stellar mass function from the H20 simulation set are shown in the bottom-left panel of Fig.~\ref{fig:1pt_20Mpc_hydro}. As expected, the results from the two simulation types show a good agreement, and we find no bias between their means within $2\sigma$ (first and second panels). The third panel shows that the value of $\sqrt{1+r}$ is compatible with 1 for all stellar masses. Finally, we find no evidence for statistical improvement of fixed and paired fixed simulations over standard simulations for the stellar mass function (fourth panel).

\subsubsection{Star-formation rate history}

We have computed the star-formation rate history of each standard and paired fixed realization as the sum of the star-formation rates of all gas particles divided by the simulation volume. That quantity informs us about the rate at which stars are being formed at a given redshift, and therefore, complements the stellar mass function when studying overall abundance. We show the results of our statistical analysis in the bottom-right panel of Fig.~\ref{fig:1pt_20Mpc_hydro}. The agreement between the results of both simulations is excellent and we find no evidence that paired fixed simulations introduce a bias on that quantity. The value of the cross-correlation coefficient is compatible with 0 ($\sqrt{1+r}=1$), although between redshifts 4 and 12 it is less than 1. 

We find that fixed simulations barely improve the statistics of standard simulations, although there exist two significant bumps at redshifts $\simeq4$ and $\simeq9$. The improvement is slightly larger in paired fixed simulations, due to the values of the cross-correlation coefficient being less than 1 at some redshifts.

\subsection{Galaxy properties}
\label{subsec:gal_properties}

The above results point out that, at least for clustering related quantities, paired fixed or fixed simulations can improve the statistics of standard simulations without introducing a bias on the results. Thus, state-of-the-art cosmological hydrodynamic simulations such as \textsc{IllustrisTNG} \citep{MarinacciF_17a,NaimanJ_17a,NelsonD_17a,PillepichA_17a,SpringelV_17a}, \textsc{Eagle} \citep{EAGLE}, \textsc{HorizonAGN} \citep{HorizonAGN}, \textsc{Magneticum} \citep{Magneticum} or \textsc{BlueTides} \citep{BlueTides} will highly benefit, for clustering analysis, by generating their initial conditions through fixed or paired fixed fields rather than standard Gaussian fields.

On the other hand, the main analysis scope of the above simulations is usually not clustering, but rather galaxy properties and evolution. It is thus very important to investigate 1) whether paired fixed simulations introduce a bias in internal galaxy properties and 2) whether the intrinsic physical scatter in their properties is changed in paired fixed simulations. The purpose of this section is to answer these two questions.

\begin{figure*}
\begin{center}
\includegraphics[width=0.33\textwidth]{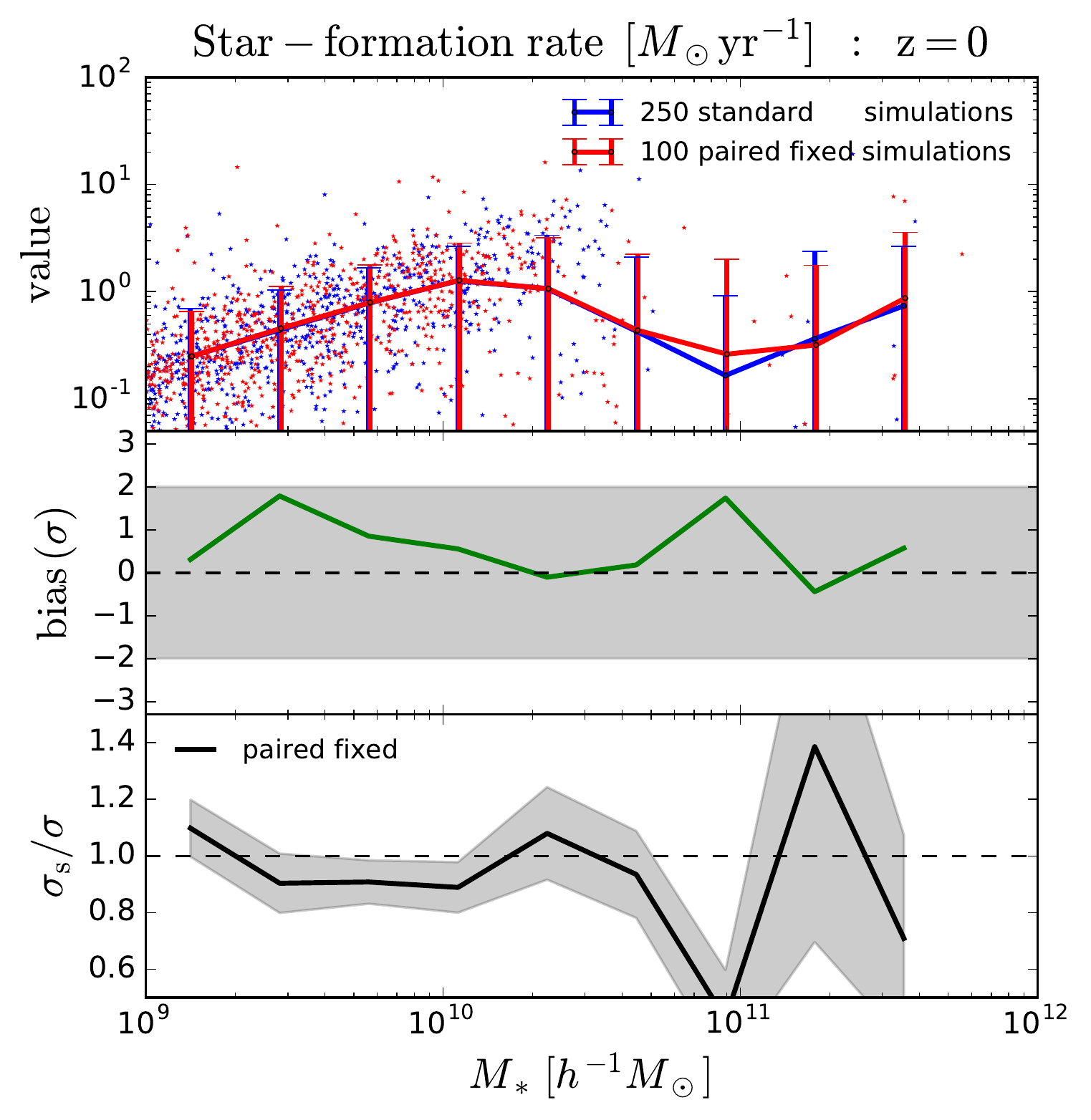}
\includegraphics[width=0.33\textwidth]{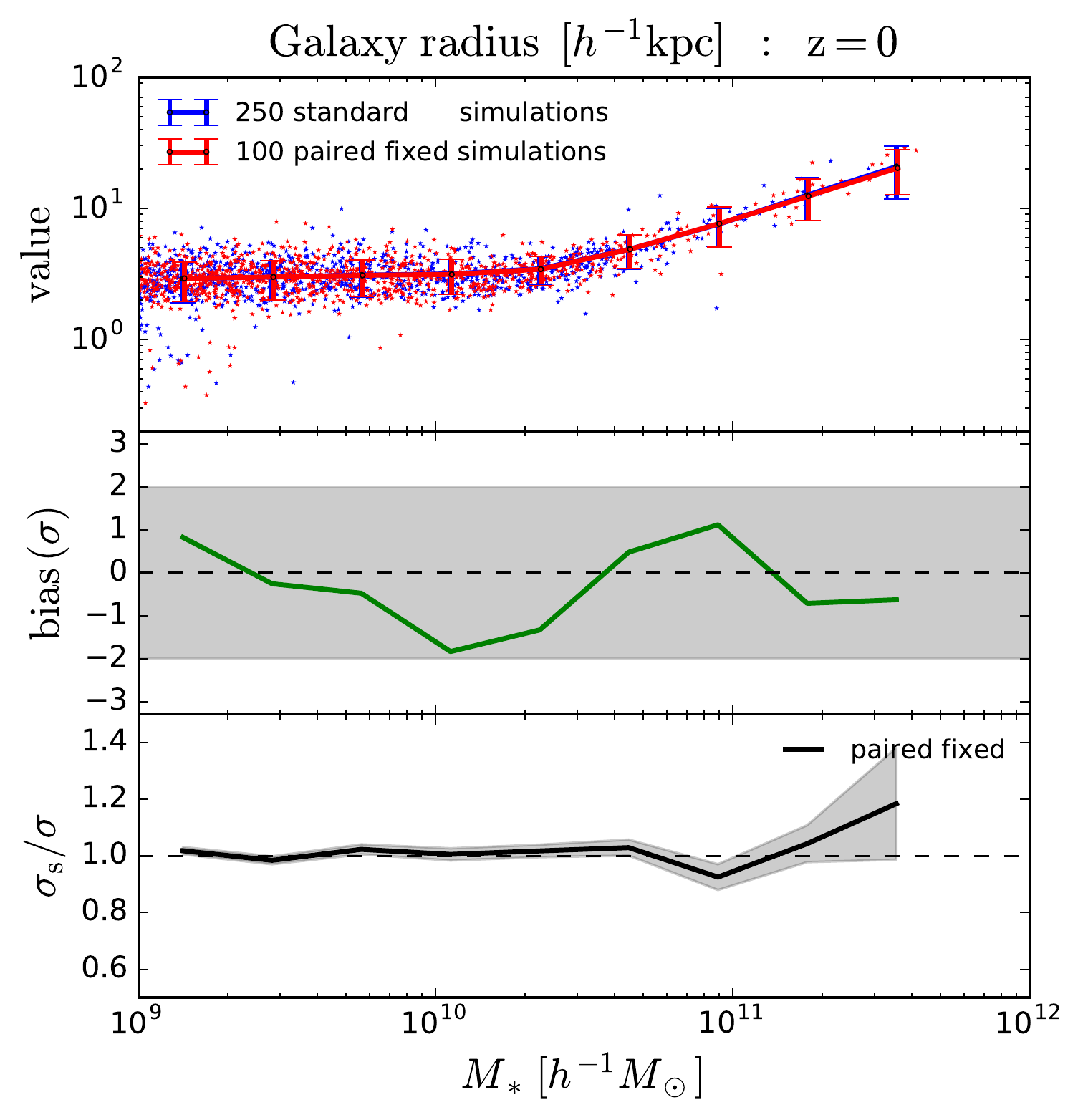}
\includegraphics[width=0.33\textwidth]{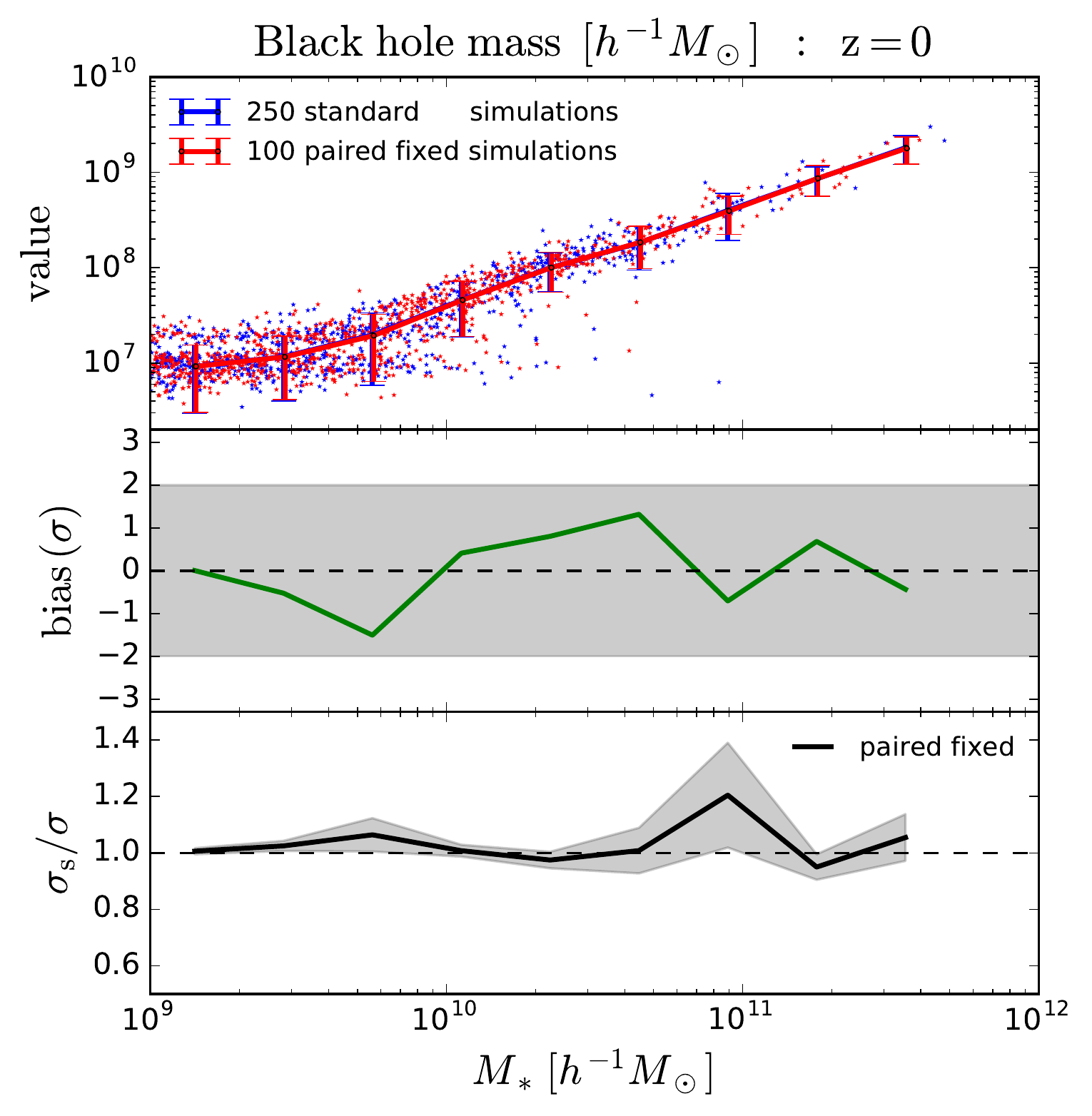}
\includegraphics[width=0.33\textwidth]{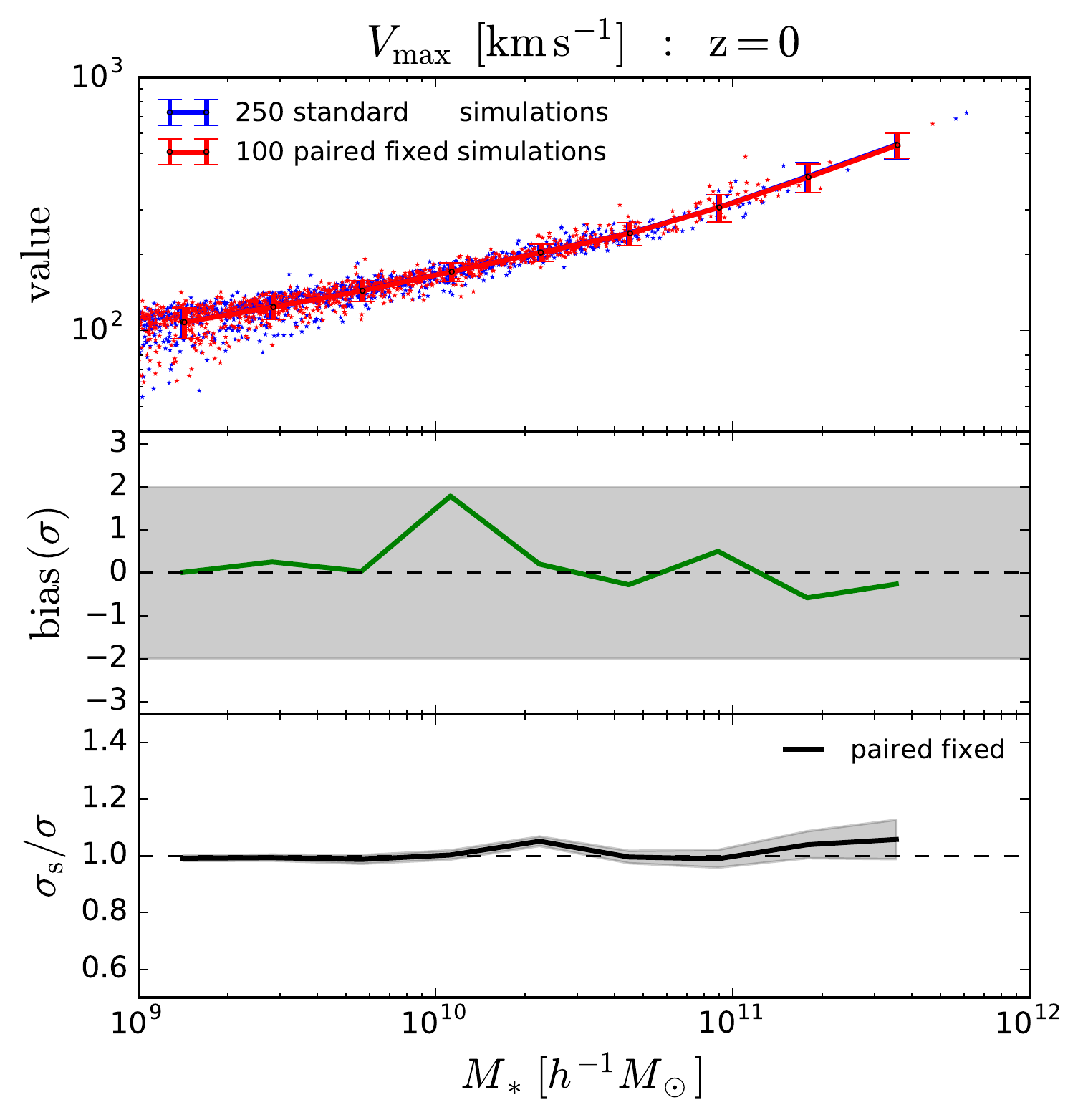}
\includegraphics[width=0.33\textwidth]{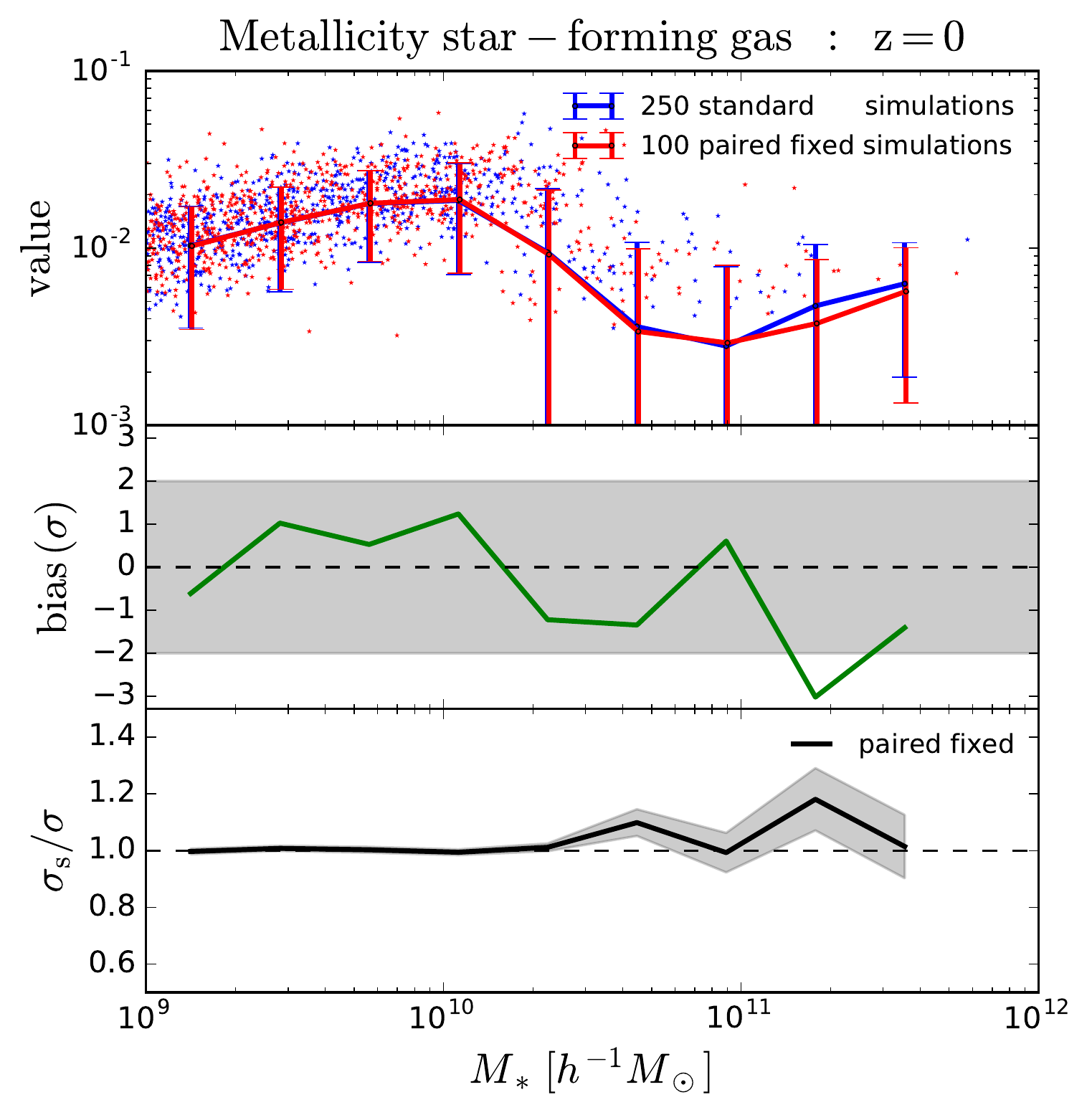}
\caption{Impact of paired fixed simulations on internal galaxy properties. We show star-formation rate (top-left), radius (top-middle), black-hole mass (top-right), maximum circular velocity (bottom-left), and metallicity of star-forming gas (bottom-right) as function of stellar mass. paired fixed simulations do not reduce the intrinsic physical scatter in these quantities and they do not introduce a bias on them. Galaxies in paired fixed simulations look thus completely normal.}
\label{fig:galaxy_properties}
\end{center}
\end{figure*}

For each galaxy in each realization of the standard and paired fixed simulations, we have computed a number of different internal quantities using the {\sc subfind} algorithm: stellar mass, star-formation rate, radius, black-hole mass, maximum circular velocity and metallicity of star forming gas. We limit our analysis to well-resolved galaxies, which we define as those with a stellar mass above $10^9~h^{-1}M_\odot$. We then make a scatter plot between the above quantities and stellar mass from the results of both simulation types. Finally, we take narrow bins in stellar mass and compute the mean and standard deviation of the results for the considered quantity. 

The above procedure is slightly different from the treatment we have been using for the paired fixed simulations. For all the quantities considered so far in this work, we have estimated the value for the paired fixed realization as the average between the results within each pair. Here, for each paired fixed realization we just create the scatter plot and compute mean and standard deviation values for all galaxies (in a mass bin) together, without separating first between each simulation in the pair. This is because there is no way to pair individual objects for taking an average; indeed, individual halos become voids in their paired partner \citep{Pontzen_2016}.

We show the results of this analysis in Fig.~\ref{fig:galaxy_properties}. From the first panels we can see that the agreement between the results of both simulation types is very good, as in all the other quantities considered in this work. From the second panels we can see that paired fixed simulations do not introduce a bias on any of the studied internal galaxy properties. We have estimated the error on the difference of the means through Eq. \ref{Eq:mean_diff_variance}, but using the number of points in standard and paired fixed simulations in each bin as the value of $N_{\rm s}$ and $N_{\rm pf}$, respectively.

The third panels show the ratio between the intrinsic scatter from each simulation type. Since the distribution of some of those properties is highly non-Gaussian, e.g. the distribution of star-formation rates at fixed stellar mass, using Eq. \ref{Eq:error_std} with $N_{\rm s}$ and $N_{\rm pf}$ being the number of standard and paired fixed points in the scatter plot will underestimate the errors on the ratio of the standard deviations. To avoid that, we have computed the errors on the ratio using bootstrap: for each studied quantity, we have created 15000 bootstrap catalogues. For each catalogue we have computed the ratio between the standard deviation of the standard and paired fixed simulations. Finally, we compute the standard deviation of the results from the previous step to get an estimate of the error on the standard deviation ratio from the whole sample. We create bootstrap catalogues by randomly subsampling, with replacement, the initial catalogues from the standard and paired fixed simulations.

The errors we obtain using this procedure are very similar to the ones we derive through Eq. \ref{Eq:error_std} for the radii, black-hole mass and maximum circular velocity versus stellar mass quantities, but very different for the star-formation rate versus stellar mass. 

We find that paired fixed simulations exhibit the same scatter on the considered quantities as standard simulations. In this case, this is precisely what we want, because the scatter on those quantities is due to internal physical processes and not to sample variance. We notice however that we find a significantly lower scatter in the standard simulations for the star-formation rate vs stellar mass of galaxies with stellar masses $\simeq10^{11}~h^{-1}M_\odot$. In that case, the ratio between the standard deviations is different from 1 at $\simeq3.5\sigma$. Although the probability of having a point with such low standard deviation ratio is pretty low (under the assumption that the variance of standard and paired fixed is the same), we believe it is not completely unreasonable given the large number of quantities considered. More simulations are however needed to disentangle whether this is a statistical fluctuation or pointing towards an increase in the scatter in paired fixed simulations.

We thus conclude that galaxies in paired fixed simulations look very much like those in standard simulations. We find no evidence that paired fixed simulations introduce a bias and they do not reduce the internal physical scatter on their internal properties.

\section{Improving 1-point statistics}
\label{sec:1pt_stats}

\begin{figure*}
	\centering
	\includegraphics[height=58mm]{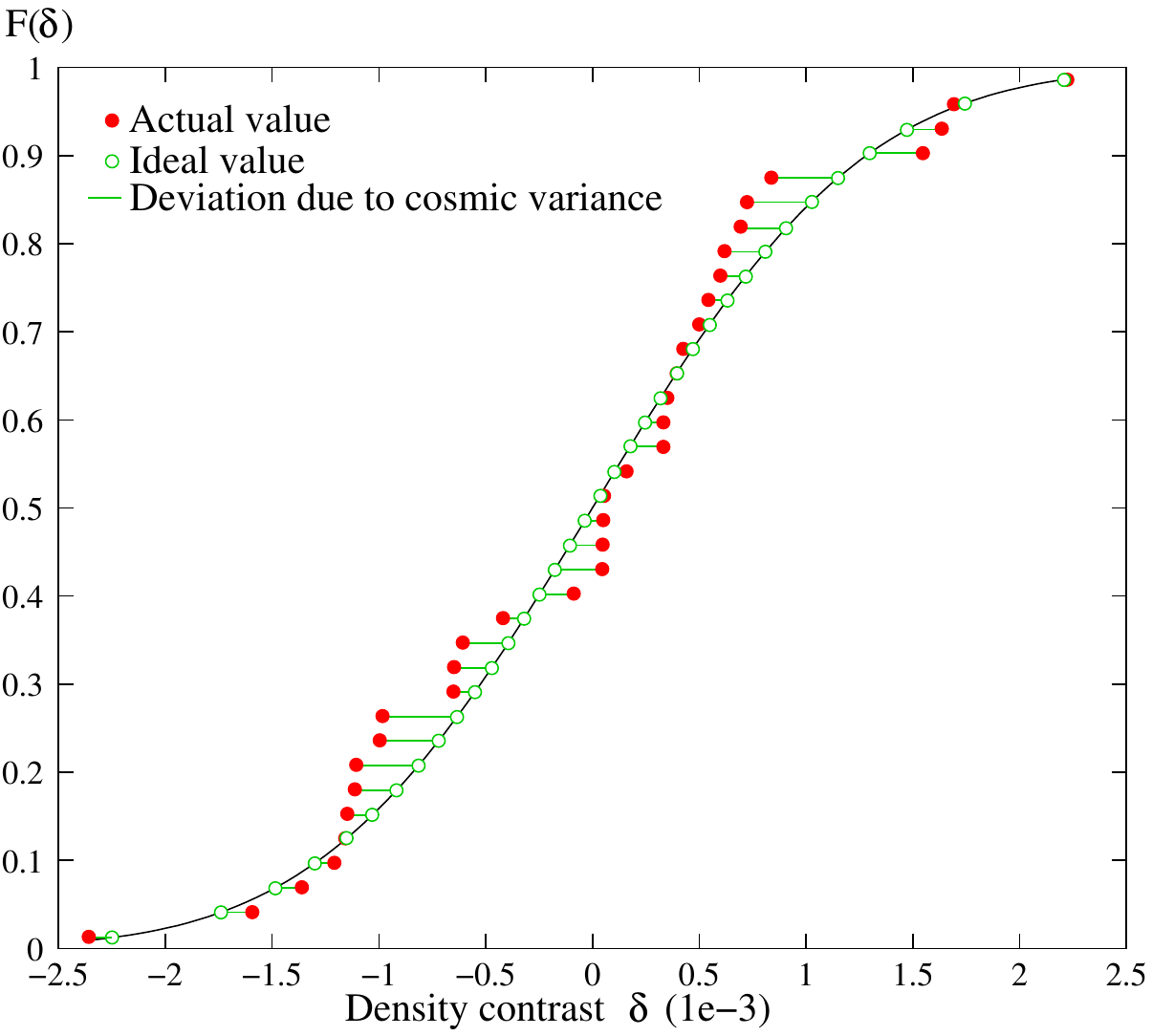}
	\includegraphics[height=56.5mm]{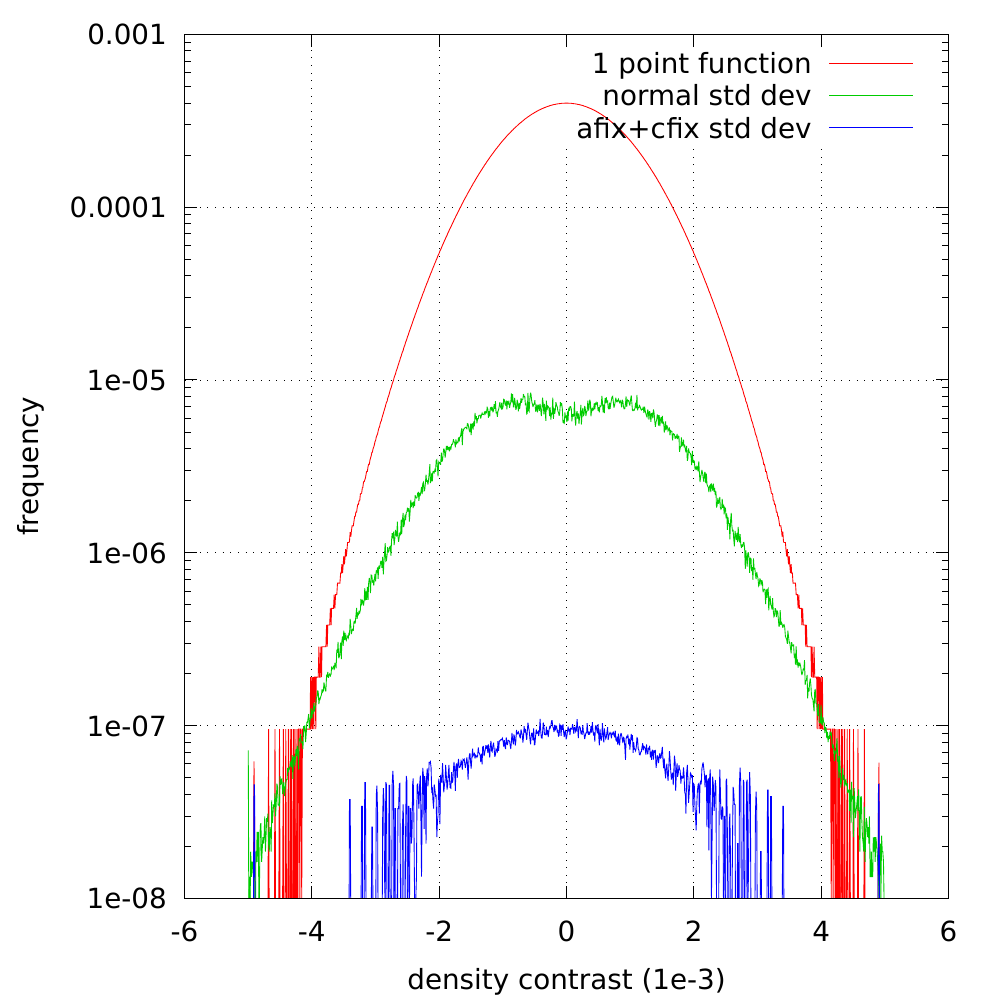}
	\includegraphics[height=56.5mm]{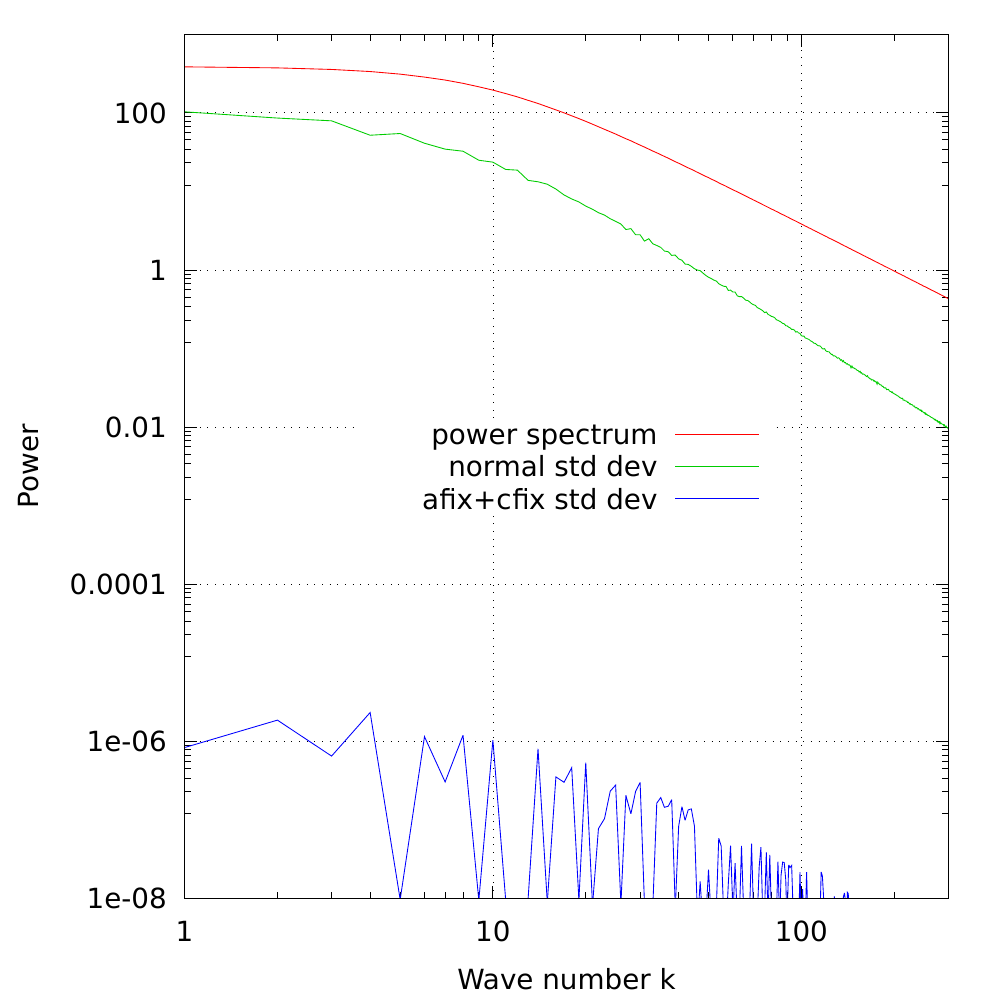}
	\caption{\emph{Left}: Sample variance in the 1-point function of a Gaussian density field 
	shows up as a random deviation of the empirical cumulative distribution function
	(red points) from the expected value (black curve). CDF-fixing the distribution
	consists of replacing each value with its corresponding point on the curve
	(green points). \emph{Middle and right}: Approximate CDF-fixing and amplitude
	fixing can be applied at the same time by iteratively CDF-fixing and amplitude
	fixing the same field, resulting in reduced sample variance in both the one point
	function and the power spectrum.}
	\label{fig:CDF-fixing}
\end{figure*}

So far we have seen that while paired fixed simulations can greatly
reduce the sample variance in the power spectra, they have little to no effect on
1-pt statistics like the matter density pdf. The fact that amplitude fixing only works for the power spectra
is not that surprising,  since that procedure was designed to carefully
tune complex amplitudes in Fourier space while letting the phases stay
random. However, this clean separation between amplitudes and phases only exists
in Fourier space. Other bases can be expressed as combinations of many different
Fourier modes, and as we have seen (Fig. \ref{fig:Mode_mixing}) mixing the amplitudes of fixed modes
undoes the fixing.

However, just like amplitude fixing is an operation designed to minimize sample
variance in the power spectrum, we could construct different operations to minimize
sample variance in other observables. For example, we could optimize for low sample
variance in the initial 1-point function of the density field by replacing the value
in each grid cell with the value from the theoretical initial 1-point function at
that cell's quantile: if there are a total of $n$ cells in the initial mass field,
then the value of the cell with the $k$'th largest value (counting from 0) would be
replaced by the theoretical cumulative distribution's $(2k+1)/(2n)$'th quantile (see the
left panel of Fig. \ref{fig:CDF-fixing}).

We can call this operation \emph{CDF-fixing}, and it does eliminate the
sample variance in the pdf of the density field in the initial conditions. Furthermore,
since amplitude fixing and CDF-fixing are defined in very different spaces, it turns out
to be possible to perform both at the same time to high accuracy. A simple algorithm
that achieves this is to iterate between fixing amplitudes in Fourier space and fixing
the CDF in real space (see the middle and right panels of Fig. \ref{fig:CDF-fixing}).

In the same way that amplitude fixing for the power spectrum works as long as Fourier
modes do not mix, CDF-fixing works as long as real-space cells do not mix.
Both conditions are fulfilled under linear evolution, but once non-linear effects appear
CDF-fixing breaks down much more quickly than amplitude fixing.
This happens because, while non-linear effects are relatively localized in Fourier space
(they are most important at small scales), they occur practically everywhere in real space.
Soon after non-linear effects become important, all cells would start mixing and
the careful tuning of quantiles needed to cancel sample variance in the pdf would be
lost.

\begin{figure*}
	\centering
	\includegraphics[height=62mm]{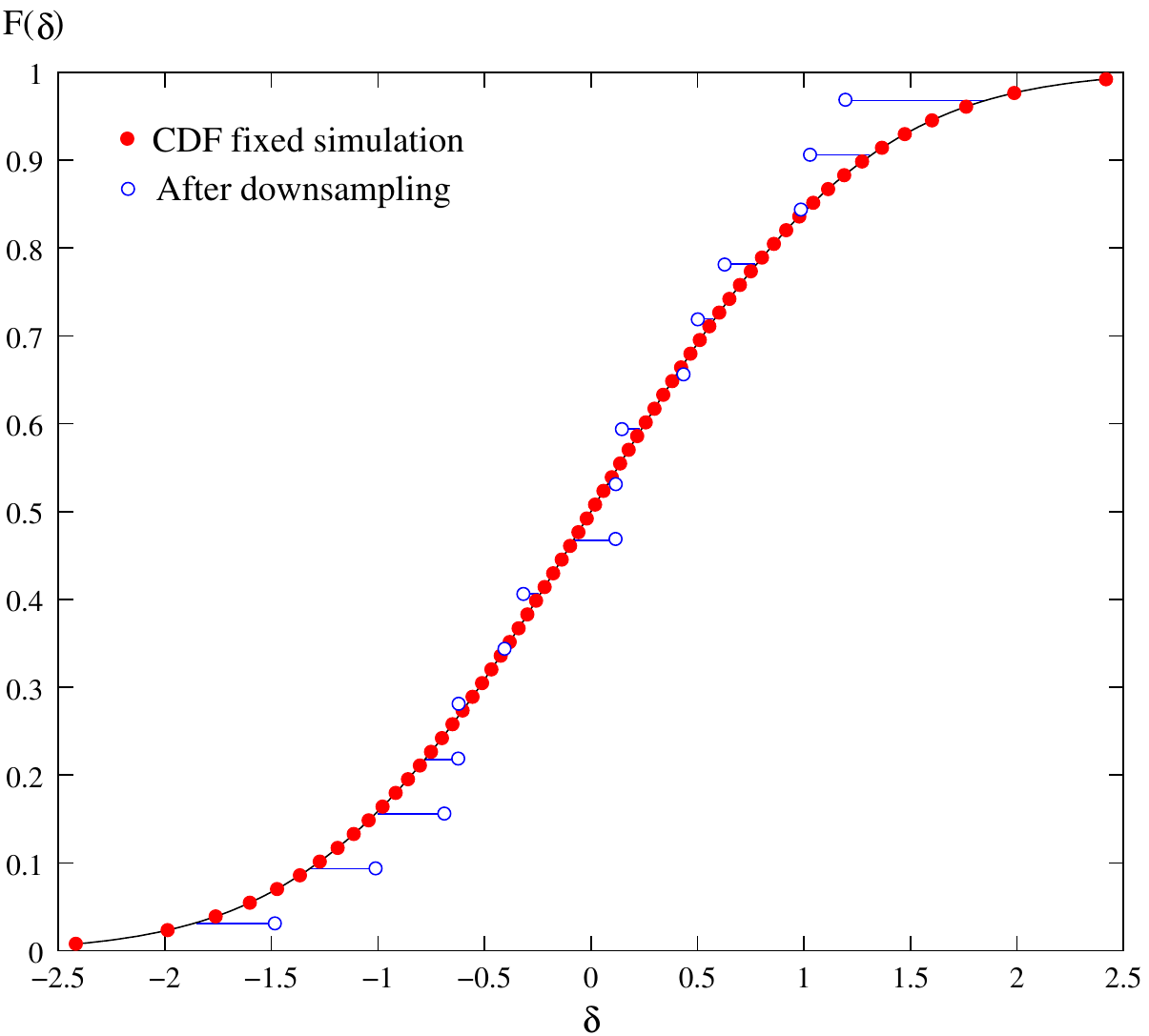}
	\includegraphics[height=60mm]{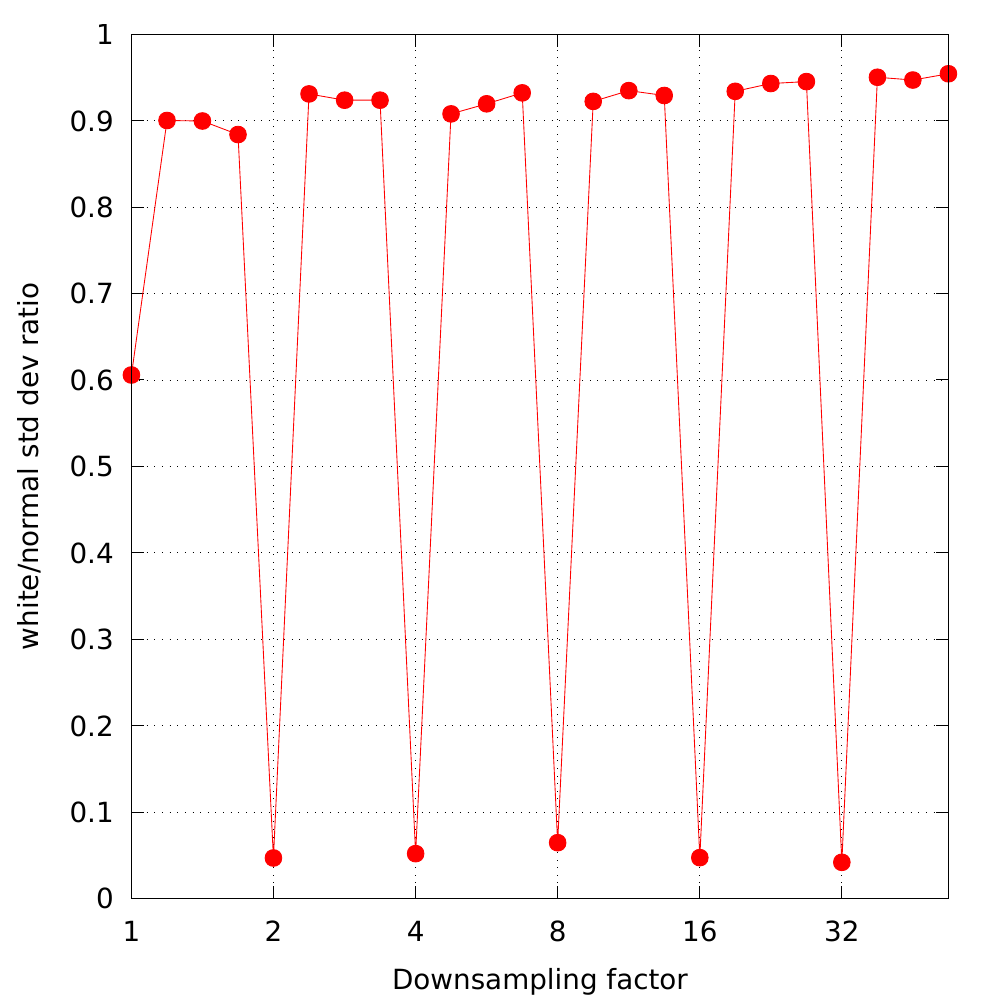}
	\caption{\emph{Left}: CDF-fixing is a nonlocal operation that only applies to
	a specific basis. This is an example of how a sample variance free CDF-fixed
	simulation (red points) completely loses its whiteness after being downsampled
	to half its original resolution. Any operation or change in basis that mixes
	voxel values will have this effect, making CDF-fixing very fragile.
	\emph{Right}:
	A field can be approximately CDF-fixed on multiple length scales at the same time
	by iteratively CDF-fixing each scale. Here a density field was CDF-fixed
	at full resolution (downsampling factor of 1) and five power-of-two reductions in resolution (2,4,8,16,32).
	The variance was then measured on these scales as well as several intermediate scales.
	The scales that were explicitly CDF-fixed have reduced standard deviation, but
	almost no benefit is seen at any other length scale.}
	\label{fig:white-flaw}
\end{figure*}

Moreover, the reduction of the 1-point function sample variance only happens for the
exact set of cells it was defined for. Changing the resolution, or even just applying
a non-integer displacement in position to the cells, will lead to destructive mixing.
For example, if we optimize the pdf at a given grid size, but measure it after downsampling
to half resolution, the pixel mixing inherent in this operation completely destroys
the sample variance cancellation. This is shown in Fig. \ref{fig:white-flaw}.

For the halo and void mass functions the problem is even worse, as the location and size
of each halo and void are not known at the outset, preventing us from tuning the volumes
that will end up as halos or voids to have reduced variance. And as we have seen, the
tuning needs to exactly match the position and size of the objects we care about for there
to be any effect. For example, simply tuning the pdf will not help as each halo and void
is a combination of multiple cells.

We conclude that while we can generate initial conditions with highly suppressed sample variance
in the power spectrum, the corresponding operation for the matter density pdf is too fragile
for practical use, and would not survive even a small amount of non-linearities and mode mixing.
Even if such an operation were possible, we believe that it would not improve any other 1-point statistics like the halo
and void mass functions due to the locality and non-linearity involved in the formation of those objects and the highly non-linear mode mixing involved thereby. \\

\section{Discussion and Conclusions}
\label{sec:conclusions}

Numerical simulations are an invaluable tool for understanding a large variety of processes such as the non-linear growth of matter perturbations, the abundance of halos and the formation and evolution of galaxies. The most powerful way to extract information from cosmological surveys will be to contrast observations versus theoretical predictions from simulations. 

The initial conditions of cosmological simulations are usually generated from Gaussian fields. The reason is that cosmic microwave background observations have shown that the temperature fluctuations in the early Universe can be very accurately described by Gaussian fields \citep{Planck_non_Gaussianity, Planck_isotropy}, whose properties are completely determined by their power spectra. The Fourier modes of a Gaussian field can be written as $\delta(\vec{k})=Ae^{i\theta}$, where $A$ follows the Rayleigh distribution of Eq. \ref{Eq:Rayleigh} and $\theta$ is a random variable with a uniform distribution between 0 and $2\pi$. 

Running simulations with initial conditions generated from Gaussian fields gives rise to \textit{sample variance}, i.e.~statistical fluctuations arising from the fact that the modes distribution is not fully sampled. That problem is particularly important on  scales approaching the box size, where only a few modes are sampled by simulations. To evaluate the likelihood and compute posteriors, the theoretical prediction should be free of statistical fluctuations. For this reason, many simulations are needed to beat down sample variance. 

Fixed fields \citep[see e.g.][]{Viel_2010} are those with $\delta(\vec{k})=Ae^{i\theta}$, where $A$ takes a fixed value as specified by Eq. \ref{Eq:fixed} and $\theta$ is a random variable with an uniform distribution between 0 and $2\pi$. The properties of those fields are that they share the same power spectrum of Gaussian fields but they do not exhibit any scatter around it. 

Paired fixed fields consists on two fixed fields $\delta_1(\vec{k})=Ae^{i\theta}$, $\delta_2(\vec{k})=Ae^{i(\theta+\pi)}=-\delta_1(\vec{k})$, where the value of $A$ and $\theta$ is the same in both. In \cite{RP_16} it was shown that if simulations are run with initial conditions generated from those fields, large reductions on the sample variance amplitude of several important quantities can be achieved. The fixing serves to prevent sample variance in the linear amplitudes, while the pairing allows us to cancel some of the leading-order effects of phase correlations on non-linear evolution in a finite box.

In this work we have further explored the properties of paired fixed fields by quantifying 1) the sample variance reduction achieved and 2) the bias introduced by paired fixed simulations with respect to standard simulations. We have carried out our analysis by using a large set of N-body (600) and state-of-the-art magneto-hydrodynamic ($506$) simulations. Our simulations cover a wide range of scales as well as mass and spatial resolutions, hence allowing us to investigate the statistical properties of paired fixed simulations in many different setups. We have analyzed the impact of paired fixed simulations in many different quantities: matter, CDM, gas, stars, black-holes, magnetic field, halos and halo-matter power spectra, matter density field pdfs, void and halo mass functions, star-formation rate history, stellar mass function and internal galaxy properties such as black-hole mass or galaxy radii.

We now enumerate the main conclusions of this work. A summary the impact of paired fixed simulations on different statistics is shown in table \ref{table:conclusions}.

\begin{itemize}

\item We find that paired fixed fields do not introduce a bias, with respect to standard Gaussian fields, on any of the quantities we have investigated in this paper. This is not an absolute statement. It may be that paired fixed simulations introduce a bias, but its magnitude has to be small since we do not find it with our rather large simulation set.

\item Paired fixed simulations reduce the scatter on the power spectrum of matter, halos, halo-matter, CDM, gas, stars, black-holes and magnetic fields. The scatter reduction depends primarily on scale, with the variance on large scales being much more suppressed than on small scales.  

\item Paired fixed simulations do not reduce the scatter on the halo bias. The linear order explanation is that the variance on the halo bias is due to the amplitude of the shot-noise, that is the same in standard and paired fixed simulations.  

\item For large box sizes paired fixed simulations do not reduce the variance of the matter density pdf
or the halo and void mass functions. For the matter density pdf, we find no improvement
already in the initial conditions. Pairing has no effect because it simply mirrors the
pdf around $\delta=0$, and there is no special connection between points with
values $\delta=+a$ and $\delta=-a$ in a simulation. Amplitude fixing has no effect either
on the pdf since it is defined in real space, where the Fourier amplitudes and phases
are scrambled. 

\item For small boxes we find a small statistically significant improvement on the matter density pdf and the halo mass function, but not on the void mass function. This may be due to a small reduction on the sample variance amplitude on the matter density pdf of those simulations that is already present in the initial conditions.

\item We find that paired fixed simulations do not reduce the scatter on the stellar mass function, while they seem to marginally improve it on the star-formation rate history. We think this follows due to the locality of the relevant physics.

\item Galaxies in paired fixed simulations look completely normal. We do not find any bias among the several intrinsic quantities, such as radii, black hole mass, star-formation rate, metallicity, maximum circular velocity and stellar mass, that we have investigated. The intrinsic, physical, scatter on those quantities, is not reduced by paired fixed simulations. 

\item We have shown that procedures aiming at fixing the matter density pdf in the initial conditions are very fragile, and it seems almost impossible to fix the pdf on all possible scales. We thus conclude that it is unlikely that general operations performed in the initial conditions can be used to reduce the sample variance associated to statistics like the matter density pdf, the halo or the void mass functions.

\end{itemize}

\begin{table*}
\begin{center}
\renewcommand{\arraystretch}{1.5}
\resizebox{1\textwidth}{!}{\begin{tabular}{| c | c | c | c | c | c | r | r | c |}
\hline
Simulation & $N_{\rm s}$ & $N_{\rm pf}$ & Statistics & redshift & Bias? & $\max(\sigma_{\rm s}/\sigma_{\rm f})$ & $\max(\sigma_{\rm s}/\sigma_{\rm pf})$ & Corresponding\\
set & & & & & & & & figure \\
\hline \hline

\multirow{11}{*}{N1000} & \multirow{11}{*}{100} & \multirow{11}{*}{100} & \multirow{4}{*}{$P_{\rm mm}(k)$} & 99 & no & 1,169.0 & 136,821.0 & \ref{fig:Pk_ICs} \\
\cline{5-9}
& & & & 5 & no & $176.9$ & $1,464.1$ & \ref{fig:Pk_1000Mpc_Nbody} \\
\cline{5-9}
&& && 1 & no & $80.0$ & $1,472.4$ & \ref{fig:Pk_1000Mpc_Nbody} \\
\cline{5-9}
& && & 0 & no & $49.6$ & $1,187.4$ & \ref{fig:Pk_1000Mpc_Nbody} \\
\cline{4-9}
 & & & $P_{\rm hm}(k)$ & 0 & no & $6.3$ & $7.3$ & \ref{fig:Pk_1000Mpc_Nbody} \\
\cline{4-9}
 & & & $P_{\rm hh}(k)$ & 0 & no & $3.2$ & $3.5$ & \ref{fig:Pk_1000Mpc_Nbody} \\
\cline{4-9}
 & & & $P_{\rm hm}(k)/P_{\rm mm}(k)$ & 0 & no & $1.3$ & $1.6$ & \ref{fig:1000Mpc_others} \\
\cline{4-9}
 & & & \multirow{2}{*}{${\rm matter\,\,pdf}$} & 99 & no & 1.4 & 1.4 & \ref{fig:Pk_ICs} \\
\cline{5-9}
 & & &  & 0 & no & $1.3$ & $1.3$ & \ref{fig:1000Mpc_others} \\
\cline{4-9}
 & & & ${\rm halo\,\,mass\,\,function}$ & 0 & no & $1.3$ & $1.3$ & \ref{fig:1000Mpc_others} \\
\cline{4-9}
 & & & ${\rm void\,\,mass\,\,function}$ & 0 & no & $1.2$ & $1.3$ & \ref{fig:1000Mpc_others} \\
\hline

\multirow{9}{*}{H200} & \multirow{9}{*}{26} & \multirow{9}{*}{15} & \multirow{3}{*}{$P_{\rm mm}(k)$} & 5 & no & $31.2$ & $771.9$ & \ref{fig:Pk_200Mpc_hydro}\\
\cline{5-9}
 & & &  & 1 & no & $10.8$ & $87.1$ & \ref{fig:Pk_200Mpc_hydro}\\
\cline{5-9}
 & & &  & 0 & no & $6.9$ & $34.0$ & \ref{fig:Pk_200Mpc_hydro}\\
\cline{4-9}
 & & & \multirow{3}{*}{$P_{\rm cc}(k)$} & 5 & no? & $31.3$ & $761.0$ & \ref{fig:Pk_200Mpc_hydro}\\
\cline{5-9}
 & & & & 1 & no? & $10.9$ & $92.0$ & \ref{fig:Pk_200Mpc_hydro}\\
\cline{5-9}
 & & & & 0 & no & $7.0$ & $34.3$ & \ref{fig:Pk_200Mpc_hydro}\\
\cline{4-9}
 & & & \multirow{3}{*}{$P_{\rm gg}(k)$} & 5 & no? & $37.2$ & $426.3$ & \ref{fig:Pk_200Mpc_hydro}\\
\cline{5-9}
 & & & & 1 & no & $16.0$ & $59.0$ & \ref{fig:Pk_200Mpc_hydro}\\
\cline{5-9}
 & & & & 0 & no & $8.8$ & $29.3$ & \ref{fig:Pk_200Mpc_hydro}\\
\hline

\multirow{18}{*}{H20} & \multirow{18}{*}{250} & \multirow{18}{*}{100} & \multirow{3}{*}{$P_{\rm mm}(k)$} & 5 & no & $5.9$ & $25.4$ & \ref{fig:Pk_20Mpc_hydro}\\
\cline{5-9}
 & & & & 1 & no & $2.3$ & $4.1$ & \ref{fig:Pk_20Mpc_hydro}\\
\cline{5-9}
 & & &  & 0 & no & $1.8$ & $2.6$ & \ref{fig:Pk_20Mpc_hydro}\\
\cline{4-9}
 & & & \multirow{3}{*}{$P_{\rm gg}(k)$} & 5 & no & $6.3$ & $23.9$ & \ref{fig:Pk_20Mpc_hydro}\\
\cline{5-9}
 & & &  & 1 & no & $2.4$ & $4.3$ & \ref{fig:Pk_20Mpc_hydro}\\
\cline{5-9}
 & & &  & 0 & no & $1.8$ & $2.7$ & \ref{fig:Pk_20Mpc_hydro}\\
\cline{4-9}
 & & & $P_{\rm ss}(k)$ & 0 & no & $1.8$ & $2.1$ & \ref{fig:Pk_20Mpc_hydro}\\
\cline{4-9}
 & & & $P_{\rm bh}(k)$ & 0 & no & $1.6$ & $2.4$ & \ref{fig:Pk_20Mpc_hydro}\\
\cline{4-9}
 & & & $P_{\rm B}(k)$ & 0 & no & $1.6$ & $2.1$ & \ref{fig:Pk_20Mpc_hydro}\\
\cline{4-9}
 & & & $P_{\rm hm}(k)$ & 0 & no & $1.9$ & $2.5$ & \ref{fig:Pk_20Mpc_hydro}\\
\cline{4-9}
 & & & ${\rm halo\,\,mass\,\,function}$ & 0 & no & $1.7$ & $1.8$ & \ref{fig:1pt_20Mpc_hydro}\\
\cline{4-9}
 & & & ${\rm void\,\,mass\,\,function}$ & 0 & no & $1.4$ & $1.6$ & \ref{fig:1pt_20Mpc_hydro}\\
\cline{4-9}
 & & & \multirow{2}{*}{${\rm matter\,\,pdf}$} & 99 & no & $1.5$ & $1.6$ & \ref{fig:ICs_20Mpc_hydro}\\
\cline{5-9}
 & & &  & 0 & no & $1.6$ & $2.2$ & \ref{fig:1pt_20Mpc_hydro}\\
\cline{4-9}
 & & & ${\rm stellar\,\,mass\,\,function}$ & 0 & no & $1.2$ & $1.2$ & \ref{fig:1pt_20Mpc_hydro}\\
\cline{4-9}
 & & & ${\rm star\,\,formation\,\,rate\,\,history}$ & [0-15] & no & $1.5$ & $1.7$ & \ref{fig:1pt_20Mpc_hydro}\\
\cline{4-9}

 & & & ${\rm star\,\,formation\,\,rate\,\,vs\,\,stellar\,\,mass}$ & 0 & no & $-$ & $1.4$ & \ref{fig:galaxy_properties}\\
\cline{4-9}
 & & & ${\rm radius\,\,vs\,\,stellar\,\,mass}$ & 0 & no & $-$ & $1.2$ & \ref{fig:galaxy_properties}\\
\cline{4-9}
 && & ${\rm black\,\,hole\,\,mass\,\,vs\,\,stellar\,\,mass}$ & 0 & no & $-$ & $1.2$ & \ref{fig:galaxy_properties}\\
\cline{4-9}
 & & & ${\rm V_{\rm max}\,\,vs\,\,stellar\,\,mass}$ & 0 & no & $-$ & $1.1$ & \ref{fig:galaxy_properties}\\
\cline{4-9}
 & & & ${\rm Metallicity\,\,vs\,\,stellar\,\,mass}$ & 0 & no & $-$ & $1.2$ & \ref{fig:galaxy_properties}\\
\hline
\end{tabular}}
\end{center}
\caption{This table summarizes the main findings of this work. The first column indicates the simulation set used to carry out the analysis. The first letter indicates whether it is from N-body (N) or hydrodynamic (N) simulations, while the following number represents the simulation box size in $h^{-1}{\rm Mpc}$. The second and third columns show the numbers of standard and pairs of fixed simulations comprising each set, respectively. The fourth column represents the statistic considered and its redshift is shown in the fifth column. The sixth column indicates whether we find that paired fixed simulations introduce a bias on the considered quantity with respect to standard simulations. The maximum reduction on the standard deviation from standard simulations achieved by fixed and paired fixed simulations is shown in the seventh and eighth columns. The relative error on those values is given by $0.5\sqrt{2/N_{\rm s}+2/N_{\rm pf}}$. For galaxy properties (last five rows) we estimated the errors through bootstrap, finding that paired fixed simulations do not reduce the intrinsic scatter on galaxy properties. Finally, the ninth column shows the corresponding figure where we plot our results for the considered quantity. We note that for the matter density pdf, we find larger reductions on the scatter of standard simulations, but we do not quote them as they are due to statistical fluctuations.}
\label{table:conclusions}
\end{table*}

From the above results we can derive two further conclusions. First, let us recall that the value of parts of the trispectrum and higher order moments are expected to be different in standard and fixed simulations. But we do not see any biases in any of our measured quantities.  This suggests the perturbation theory argument of \cite{RP_16} - that the modifications do not propagate to observables except in a very specific, measure-zero subset - seems to hold even in highly non-linear regimes.  

The second conclusion is that, since paired fixed simulations help in reducing the scatter of clustering-related quantities while they do not improve the statistics of 1-pt quantities (or improve them marginally), we believe that the two quantities cannot be very correlated. If they were, we would have expected that as we reduce the scatter in one, the other should also be affected by it. We thus believe that the information embedded in clustering and 1-pt statistics should be highly complementary. While this is not surprising \citep[see e.g.][]{Schaan_2014}, our conclusions arise from a completely different methodology than more traditional methods.

This paper constitutes an empirical confirmation of the benefits brought about by paired fixed simulations. We believe that upcoming large box size hydrodynamic simulations can highly benefit by being run with initial conditions from paired fixed fields. If running two simulations is computationally expensive, a fixed field can be used to generate the initial conditions.

\section*{ACKNOWLEDGEMENTS} 
This work has made extensive use of the python \textsc{pylians} libraries, publicly available at \url{https://github.com/franciscovillaescusa/Pylians}. The simulations have been run in the Gordon cluster at the San Diego Supercomputer Center. The work of FVN, SN, SG, LA, NB and DNS is supported by the Simons Foundation. AP is funded by the Royal Society. This work was partially enabled by funding from the UCL Cosmoparticle Initiative.

\begin{appendix}

\section{A. Variance of paired fixed simulations}
\label{sec:analytics}

In this appendix we derive Eq. \ref{eq:sigma_pf} and discuss the different origins of the statistical improvement of paired fixed simulations over traditional simulations for any generic quantity.

Suppose we are considering a quantity, $X_{\rm s}$, e.g.~the amplitude of the power spectrum at a given wavenumber $k$, or the halo mass function at mass $M$, from standard simulations, with a variance given by
\be
\sigma_{\rm s}^2 = \langle (X_{\rm s} -\bar{X}_{\rm s})^2 \rangle = \langle X_{\rm s}^2 \rangle - \bar{X}_{\rm s}^2~,
\ee
where $\bar{X}_{\rm s}=\langle X_{\rm s} \rangle$. Now consider the same quantity but estimated through the paired fixed simulations
\be
X_{\rm pf}=\frac{1}{2}(X_{\rm pf,1}+X_{\rm pf,2})~,
\ee
where $X_{\rm pf,1}$ and $X_{\rm pf,2}$ are the value of $X$ from the two pairs of a paired fixed simulation. The variance of $X_{\rm pf}$ is given by
\begin{eqnarray}
\sigma_{\rm pf}^2 &=& \langle (X_{\rm pf} - \bar{X}_{\rm pf})^2\rangle=\langle X_{\rm pf}^2 \rangle - \bar{X}_{\rm pf}^2\\
 &=& \frac{1}{4}\left( \langle X_{\rm pf,1}^2\rangle + \langle X_{\rm pf,2}^2\rangle +  2\langle X_{\rm pf,1}X_{\rm pf,2}\rangle -\bar{X}_{\rm pf,1}^2 -\bar{X}_{\rm pf,2}^2 -2\bar{X}_{\rm pf,1}\bar{X}_{\rm pf,2} \right)\\
 &=&\frac{1}{4}\left( \sigma_{\rm pf,1}^2 +\sigma_{\rm pf,2}^2 + 2{\rm cov}_{12}\right)
\end{eqnarray}
where ${\rm cov_{12}}=\langle (X_{\rm pf,1}-\bar{X}_{\rm pf,1})(X_{\rm pf,2}-\bar{X}_{\rm pf,2})\rangle$. Finally, since the variance of the two pairs from the paired fixed simulations is the same, $\sigma_{\rm pf,1}=\sigma_{\rm pf,2}=\sigma_{\rm f}$, we obtain
\be
\sigma_{\rm pf}^2=\sigma_{\rm f}^2\left(\frac{1 + r}{2}\right)
\ee
where the cross-correlation coefficient $r$ is defined as $r={\rm cov}_{12}/\sigma_{\rm f}^2$, and it satisfies $-1\leqslant r \leqslant 1$. We note that the variance of each individual pair within paired fixed simulations is, by definition, equivalent to the variance of individual fixed simulations. This is why we write $\sigma_{\rm f}$ above. It is interesting to consider some limiting situations:

\begin{itemize}

\item The two sets of simulations of paired fixed simulations are independent, $r=0$, and their variance is the same as in standard simulations, $\sigma_{\rm s}=\sigma_{\rm f}$. In this case, statistical improvement of the paired fixed Gaussian simulations will be just $\sigma_{\rm pf}=\sigma_{\rm s}/\sqrt{2}$. In this situation, the variance reduction arises simply because in the paired fixed simulations the quantity considered is estimated using two independent realizations instead of  one. 

\item The two sets of simulations of paired fixed simulations are completely correlated, $r=1$ and the variance of each set is the same as in standard simulations, $\sigma_{\rm s}=\sigma_{\rm f}$. In this case no improvement is achieved by the paired fixed simulations: $\sigma_{\rm pf}=\sigma_{\rm s}$. This corresponds to a situation where the two paired fixed simulations are equivalent to one, e.g.~the second is the same as the first, and therefore no improvement can be achieved.

\item The two sets of simulations of paired fixed simulations are completely anti-correlated, $r=-1$. In this case, the variance of the paired fixed simulations will be 0, independently of the variance of each pair, $\sigma_{\rm f}$. The interpretation of this situation is that since the two simulations in each pair are completely anti-correlated, if $X_{\rm pf,1}$ increases its value $X_{\rm pf,2}$ will decrease, such as $X_{\rm pf,1}+X_{\rm pf,2}$ will be kept constant. 

\item The variance of each set of paired fixed simulations is lower than the variance of the standard simulations, $\sigma_{\rm f}^2<\sigma_{\rm s}^2$. In this case, even if the two paired fixed simulations are completely correlated, there will be a statistical improvement. This happens simply because even if the two pairs are completely correlated, i.e.~only one independent realization is available, its variance is lower than that of a standard simulation. We notice that this case applies to fixed simulations as well.

\end{itemize}

From the above arguments we see that, in most situations, paired fixed simulations will perform better than standard simulations by a factor of at least $1/\sqrt{2}$. This arises because each paired fixed realization contains two simulations while fixed or standard does only one. In order to avoid that artificial improvement, and to be able to carry out a fair comparison, in this paper we work with the \textit{normalized variance}, defined as the variance per number of simulations. In that case, we can express the normalized variance of paired fixed simulations as
\be
\sigma_{\rm pf}^2=\sigma_{\rm f}^2\left(1 + r\right).
\label{Eq:variance_paired_lazy}
\ee
This is the expression we have used along the text. It is interesting to relate the different pieces of Eq. \ref{Eq:variance_paired_lazy} with the properties of the paired fixed fields. On the one hand, a fixed field is expected to have different variance from a standard Gaussian field. Thus, the improvement of the fixed fields will arise from $\sigma_{\rm f}$ in Eq. \ref{Eq:variance_paired_lazy}. On the other hand, the two simulations in a pair, independently on whether they are from pair simulations or paired fixed simulations, will contribute to the variance through $r$. We however emphasize that the value of $r$ will, in general, be different for paired and paired fixed simulations. Thus, the correct interpretation of the $(1+r)$ factor is the statistical improvement brought by pairing (for paired simulations) or by pairing once the amplitude is fixed (for paired fixed simulations).

In other words, for paired simulations $\sigma_{\rm f}=\sigma_{\rm s}$ and any statistical improvement arises solely from $r$. For fixed simulations the statistical improvement comes through $\sigma_{\rm f}$, while for paired fixed simulations the improvement comes from both, by fixing the amplitude through $\sigma_{\rm f}$ and by pairing, once fixed, through the value of $r$.

\section{B. Variance of the ratio}
\label{sec:variance_ratios}

Here we derive the expression we use to compute the variance of the ratio of two quantities. In general, given two random variables $X$ and $Y$ the distribution of their ratio $Z=Y/X$ cannot be expressed analytically. We now derive a well-known expression for the variance of the ratio making the assumption that the variances of both $X$ and $Y$ are smaller than their mean values. Given two random variables, $X$ and $Y$, with means and variances given by
\begin{align*}
\bar{X}&=\langle X \rangle & \sigma_x^2 &= \langle (X-\bar{X})^2\rangle\\
\bar{Y}&=\langle Y \rangle & \sigma_y^2 &= \langle (Y-\bar{Y})^2\rangle
\end{align*}
we can Taylor expand any function of them, $Z=f(X,Y)$, around the mean as
\be
\left. Z\simeq f(\bar{X},\bar{Y})+\frac{\partial f}{\partial X}\right|_{\bar{X},\bar{Y}}(X-\bar{X})+ \left. \frac{\partial f}{\partial Y}\right|_{\bar{X},\bar{Y}}(Y-\bar{Y})+...
\ee
At leading order, the mean of $Z$ will be given by $\bar{Z}=f(\bar{X},\bar{Y})$ while its variance
\begin{eqnarray}
\sigma_z^2&=&\langle (Z-\bar{Z})^2\rangle \simeq \left. \left\langle \left(\frac{\partial f}{\partial X}\right|_{\bar{X},\bar{Y}}(X-\bar{X})+ \left. \frac{\partial f}{\partial Y}\right|_{\bar{X},\bar{Y}}(Y-\bar{Y})\right)^2 \right\rangle+...\\
&=&\left.\left(\frac{\partial f}{\partial X}\right|_{\bar{X},\bar{Y}}\right)^2\sigma_x^2 + \left.\left(\frac{\partial f}{\partial Y}\right|_{\bar{X},\bar{Y}}\right)^2\sigma_y^2 + 2\left.\left(\frac{\partial f}{\partial X}\right|_{\bar{X},\bar{Y}}\right) \left. \left(\frac{\partial f}{\partial Y}\right|_{\bar{X},\bar{Y}}\right)\sigma_x\sigma_y r + ...
\end{eqnarray}
where $r$ is the cross-correlation coefficient between $X$ and $Y$. In the case where $Z=Y/X$ we obtain
\be
\sigma_z^2 \simeq \frac{\bar{Y}^2}{\bar{X}^4}\sigma_x^2 + \frac{\sigma_y^2}{\bar{X}^2}-2\frac{\bar{Y}}{\bar{X}^3}\sigma_x\sigma_y r
\label{Eq:ratio_variance}
\ee
We can finally express the above quantity as
\be
\sigma_z^2\simeq\frac{\bar{Y}^2}{\bar{X}^2}\left(\frac{\sigma_x^2}{\bar{X}^2} +  \frac{\sigma_y^2}{\bar{Y}^2} -2r\frac{\sigma_x}{\bar{X}}\frac{\sigma_y}{\bar{Y}}\right)~.
\label{Eq:ratio_error}
\ee

For $b=P_{\rm hm}(k)/P_{\rm mm}(k)$ the above expression reduces to
\be
\sigma_{\rm b}^2 \simeq \frac{1}{2N(k)}\left(\frac{P_{\rm hh}(k)P_{\rm mm}(k)-P_{\rm hm}^2}{P_{\rm mm}^2}\right)
\ee
where we have used the fact that at linear order \citep[see e.g.][]{Smith_2009}
\begin{eqnarray}
\sigma^2(P_{\rm mm}(k))&=&\frac{P_{\rm mm}^2(k)}{N(k)}\\
\sigma^2(P_{\rm hm}(k))&=&\frac{P_{\rm hm}^2(k)+P_{\rm hh}(k)P_{\rm mm}(k)}{2N(k)}\\
r&=&P_{\rm hm}(k)P_{\rm mm}(k)
\end{eqnarray}
where $N(k)$ is the number of independent modes in the interval $[k,k+dk]$ where the different power spectra are measured and $P_{\rm hh}(k)$ is the halo power spectrum, which includes both the cosmological signal and the shot-noise term
\be
P_{\rm hh}(k)=P_{\rm hh}^{\rm cosmo}(k)+\bar{n}^{-1}
\ee
where $\bar{n}$ is the mean number density of halos.

We can also use Eq. \ref{Eq:ratio_error} to compute the error on the ratio between the standard deviation of standard and paired fixed simulations. Let us first compute the variance of $r^2=\sigma_{\rm s}^2/\sigma_{\rm pf}^2$
\be
\sigma^2_{r^2} = \frac{\sigma_{\rm s}^4}{\sigma_{\rm pf}^4}\left( \frac{\sigma^2_{\sigma^2_{\rm s}}}{\sigma^4_{\rm s}} +  \frac{\sigma^2_{\sigma^2_{\rm pf}}}{\sigma^4_{\rm pf}}\right)
\ee
where $\sigma^2_{\sigma^2_{\rm s}}$ and $\sigma^2_{\sigma^2_{\rm pf}}$ denote the variance on the variance of standard and paired fixed simulations, respectively. Under the assumption that data is Gaussian distributed, the quantity $\sum_{i=1}^N(X-\bar{X})^2$ follows a $\chi^2$ distribution with with $N$ degrees of freedom. Thus, the variance of the variance is given by $\sigma^2_{\sigma^2}=2\sigma^4/N$ and we obtain
\be
\sigma^2_{r^2} = \frac{\sigma_{\rm s}^4}{\sigma_{\rm pf}^4}\left( \frac{2}{N_{\rm s}} +  \frac{2}{N_{\rm pf}} \right)
\ee
We are however interested in the variance of the standard deviations, i.e. $\sigma^2_r$. By using the above Taylor expansion we obtain $\sigma^2_{r^2}=4r^2\sigma^2_r$, thus, the standard deviation of the standard deviation ratio is given by
\be
\sigma_r=\frac{1}{2}\left(\frac{\sigma_{\rm s}}{\sigma_{\rm pf}}\right)\sqrt{\frac{2}{N_{\rm s}} +  \frac{2}{N_{\rm pf}}}~.
\ee

\end{appendix}


\bibliography{references}{}

\begin{thebibliography}{47}
\expandafter\ifx\csname natexlab\endcsname\relax\def\natexlab#1{#1}\fi

\bibitem[{{Angulo} \& {Pontzen}(2016)}]{RP_16}
{Angulo}, R.~E., \& {Pontzen}, A. 2016, \mnras, 462, L1, [arXiv:1603.05253]

\bibitem[{{Banerjee} \& {Dalal}(2016)}]{Arka_2016}
{Banerjee}, A., \& {Dalal}, N. 2016, \jcap, 11, 015, [arXiv:1606.06167]

\bibitem[{{Bernardeau} {et~al.}(2002){Bernardeau}, {Colombi}, {Gazta{\~n}aga},
  \& {Scoccimarro}}]{Bernardeau_2002}
{Bernardeau}, F., {Colombi}, S., {Gazta{\~n}aga}, E., \& {Scoccimarro}, R.
  2002, \physrep, 367, 1, [arXiv:astro-ph/0112551]

\bibitem[{{Chuang} {et~al.}(2015{\natexlab{a}}){Chuang}, {Kitaura}, {Prada},
  {Zhao}, \& {Yepes}}]{EZmocks}
{Chuang}, C.-H., {Kitaura}, F.-S., {Prada}, F., {Zhao}, C., \& {Yepes}, G.
  2015{\natexlab{a}}, \mnras, 446, 2621, [arXiv:1409.1124]

\bibitem[{{Chuang} {et~al.}(2015{\natexlab{b}}){Chuang}, {Zhao}, {Prada},
  {Munari}, {Avila}, {Izard}, {Kitaura}, {Manera}, {Monaco}, {Murray}, {Knebe},
  {Sc{\'o}ccola}, {Yepes}, {Garcia-Bellido}, {Mar{\'{\i}}n}, {M{\"u}ller},
  {Skibba}, {Crocce}, {Fosalba}, {Gottl{\"o}ber}, {Klypin}, {Power}, {Tao}, \&
  {Turchaninov}}]{Chuang_2015}
{Chuang}, C.-H. {et~al.} 2015{\natexlab{b}}, \mnras, 452, 686,
  [arXiv:1412.7729]

\bibitem[{{Davis} {et~al.}(1985){Davis}, {Efstathiou}, {Frenk}, \&
  {White}}]{FoF}
{Davis}, M., {Efstathiou}, G., {Frenk}, C.~S., \& {White}, S.~D.~M. 1985, \apj,
  292, 371

\bibitem[{{Dolag} {et~al.}(2017){Dolag}, {Mevius}, \& {Remus}}]{Magneticum}
{Dolag}, K., {Mevius}, E., \& {Remus}, R.-S. 2017, Galaxies, 5, 35,
  [arXiv:1708.00027]

\bibitem[{{Dubois} {et~al.}(2014){Dubois}, {Pichon}, {Welker}, {Le Borgne},
  {Devriendt}, {Laigle}, {Codis}, {Pogosyan}, {Arnouts}, {Benabed}, {Bertin},
  {Blaizot}, {Bouchet}, {Cardoso}, {Colombi}, {de Lapparent}, {Desjacques},
  {Gavazzi}, {Kassin}, {Kimm}, {McCracken}, {Milliard}, {Peirani}, {Prunet},
  {Rouberol}, {Silk}, {Slyz}, {Sousbie}, {Teyssier}, {Tresse}, {Treyer},
  {Vibert}, \& {Volonteri}}]{HorizonAGN}
{Dubois}, Y. {et~al.} 2014, \mnras, 444, 1453, [arXiv:1402.1165]

\bibitem[{{Feng} {et~al.}(2016{\natexlab{a}}){Feng}, {Chu}, \&
  {Seljak}}]{FastPM}
{Feng}, Y., {Chu}, M.-Y., \& {Seljak}, U. 2016{\natexlab{a}}, ArXiv e-prints,
  [arXiv:1603.00476]

\bibitem[{{Feng} {et~al.}(2016{\natexlab{b}}){Feng}, {Di-Matteo}, {Croft},
  {Bird}, {Battaglia}, \& {Wilkins}}]{BlueTides}
{Feng}, Y., {Di-Matteo}, T., {Croft}, R.~A., {Bird}, S., {Battaglia}, N., \&
  {Wilkins}, S. 2016{\natexlab{b}}, \mnras, 455, 2778, [arXiv:1504.06619]

\bibitem[{{Genel} {et~al.}(2014){Genel}, {Vogelsberger}, {Springel}, {Sijacki},
  {Nelson}, {Snyder}, {Rodriguez-Gomez}, {Torrey}, \& {Hernquist}}]{Genel_2014}
{Genel}, S. {et~al.} 2014, \mnras, 445, 175, [arXiv:1405.3749]

\bibitem[{{Heitmann} {et~al.}(2009){Heitmann}, {Higdon}, {White}, {Habib},
  {Williams}, {Lawrence}, \& {Wagner}}]{Coyote}
{Heitmann}, K., {Higdon}, D., {White}, M., {Habib}, S., {Williams}, B.~J.,
  {Lawrence}, E., \& {Wagner}, C. 2009, \apj, 705, 156, [arXiv:0902.0429]

\bibitem[{{Howlett} {et~al.}(2015){Howlett}, {Manera}, \&
  {Percival}}]{L-PICOLA}
{Howlett}, C., {Manera}, M., \& {Percival}, W.~J. 2015, Astronomy and
  Computing, 12, 109, [arXiv:1506.03737]

\bibitem[{{Kitaura} \& {He{\ss}}(2013)}]{Kitaura_2013}
{Kitaura}, F.-S., \& {He{\ss}}, S. 2013, \mnras, 435, L78, [arXiv:1212.3514]

\bibitem[{{Kitaura} {et~al.}(2014){Kitaura}, {Yepes}, \& {Prada}}]{PATCHY}
{Kitaura}, F.-S., {Yepes}, G., \& {Prada}, F. 2014, \mnras, 439, L21,
  [arXiv:1307.3285]

\bibitem[{{Marinacci} {et~al.}(2017){Marinacci}, {Vogelsberger}, {Pakmor},
  {Torrey}, {Springel}, {Hernquist}, {Nelson}, {Weinberger}, {Pillepich},
  {Naiman}, \& {Genel}}]{MarinacciF_17a}
{Marinacci}, F. {et~al.} 2017, ArXiv e-prints, 1707.03396, [arXiv:1707.03396]

\bibitem[{{Monaco} {et~al.}(2013){Monaco}, {Sefusatti}, {Borgani}, {Crocce},
  {Fosalba}, {Sheth}, \& {Theuns}}]{Monaco_2013}
{Monaco}, P., {Sefusatti}, E., {Borgani}, S., {Crocce}, M., {Fosalba}, P.,
  {Sheth}, R.~K., \& {Theuns}, T. 2013, \mnras, 433, 2389, [arXiv:1305.1505]

\bibitem[{{Monaco} {et~al.}(2002{\natexlab{a}}){Monaco}, {Theuns}, \&
  {Taffoni}}]{Monaco_2002}
{Monaco}, P., {Theuns}, T., \& {Taffoni}, G. 2002{\natexlab{a}}, \mnras, 331,
  587, [arXiv:astro-ph/0109323]

\bibitem[{{Monaco} {et~al.}(2002{\natexlab{b}}){Monaco}, {Theuns}, {Taffoni},
  {Governato}, {Quinn}, \& {Stadel}}]{Monaco_2002b}
{Monaco}, P., {Theuns}, T., {Taffoni}, G., {Governato}, F., {Quinn}, T., \&
  {Stadel}, J. 2002{\natexlab{b}}, \apj, 564, 8, [arXiv:astro-ph/0109322]

\bibitem[{{Naiman} {et~al.}(2017){Naiman}, {Pillepich}, {Springel Enrico
  Ramirez-Ruiz}, {Torrey}, {Vogelsberger}, {Pakmor}, {Nelson}, {Marinacci},
  {Hernquist}, {Weinberger}, \& {Genel}}]{NaimanJ_17a}
{Naiman}, J.~P. {et~al.} 2017, ArXiv e-prints, 1707.03401, [arXiv:1707.03401]

\bibitem[{{Nelson} {et~al.}(2018){Nelson}, {Pillepich}, {Springel},
  {Weinberger}, {Hernquist}, {Pakmor}, {Genel}, {Torrey}, {Vogelsberger},
  {Kauffmann}, {Marinacci}, \& {Naiman}}]{NelsonD_17a}
{Nelson}, D. {et~al.} 2018, \mnras, 475, 624, [arXiv:1707.03395]

\bibitem[{{Palanque-Delabrouille} {et~al.}(2015){Palanque-Delabrouille},
  {Y{\`e}che}, {Baur}, {Magneville}, {Rossi}, {Lesgourgues}, {Borde}, {Burtin},
  {LeGoff}, {Rich}, {Viel}, \& {Weinberg}}]{Palanque_2015}
{Palanque-Delabrouille}, N. {et~al.} 2015, \jcap, 11, 011, [arXiv:1506.05976]

\bibitem[{{Pillepich} {et~al.}(2018{\natexlab{a}}){Pillepich}, {Nelson},
  {Hernquist}, {Springel}, {Pakmor}, {Torrey}, {Weinberger}, {Genel}, {Naiman},
  {Marinacci}, \& {Vogelsberger}}]{PillepichA_17a}
{Pillepich}, A. {et~al.} 2018{\natexlab{a}}, \mnras, 475, 648,
  [arXiv:1707.03406]

\bibitem[{{Pillepich} {et~al.}(2018{\natexlab{b}}){Pillepich}, {Springel},
  {Nelson}, {Genel}, {Naiman}, {Pakmor}, {Hernquist}, {Torrey}, {Vogelsberger},
  {Weinberger}, \& {Marinacci}}]{PillepichA_16a}
------. 2018{\natexlab{b}}, \mnras, 473, 4077, [arXiv:1703.02970]

\bibitem[{{Planck Collaboration} {et~al.}(2016{\natexlab{a}}){Planck
  Collaboration}, {Ade}, {Aghanim}, {Akrami}, {Aluri}, {Arnaud}, {Ashdown},
  {Aumont}, {Baccigalupi}, {Banday}, \& et~al.}]{Planck_isotropy}
{Planck Collaboration} {et~al.} 2016{\natexlab{a}}, \aap, 594, A16,
  [arXiv:1506.07135]

\bibitem[{{Planck Collaboration} {et~al.}(2016{\natexlab{b}}){Planck
  Collaboration}, {Ade}, {Aghanim}, {Arnaud}, {Arroja}, {Ashdown}, {Aumont},
  {Baccigalupi}, {Ballardini}, {Banday}, \& et~al.}]{Planck_non_Gaussianity}
------. 2016{\natexlab{b}}, \aap, 594, A17, [arXiv:1502.01592]

\bibitem[{{Planck Collaboration} {et~al.}(2016{\natexlab{c}}){Planck
  Collaboration}, {Ade}, {Aghanim}, {Arnaud}, {Ashdown}, {Aumont},
  {Baccigalupi}, {Banday}, {Barreiro}, {Bartlett}, \& et~al.}]{Planck_2015}
------. 2016{\natexlab{c}}, \aap, 594, A13, [arXiv:1502.01589]

\bibitem[{{Pontzen} {et~al.}(2016){Pontzen}, {Slosar}, {Roth}, \&
  {Peiris}}]{Pontzen_2016}
{Pontzen}, A., {Slosar}, A., {Roth}, N., \& {Peiris}, H.~V. 2016, \prd, 93,
  103519, [arXiv:1511.04090]

\bibitem[{{Rizzo} {et~al.}(2017){Rizzo}, {Villaescusa-Navarro}, {Monaco},
  {Munari}, {Borgani}, {Castorina}, \& {Sefusatti}}]{Rizzo_2017}
{Rizzo}, L.~A., {Villaescusa-Navarro}, F., {Monaco}, P., {Munari}, E.,
  {Borgani}, S., {Castorina}, E., \& {Sefusatti}, E. 2017, \jcap, 1, 008,
  [arXiv:1610.07624]

\bibitem[{{Schaan} {et~al.}(2014){Schaan}, {Takada}, \&
  {Spergel}}]{Schaan_2014}
{Schaan}, E., {Takada}, M., \& {Spergel}, D.~N. 2014, \prd, 90, 123523,
  [arXiv:1406.3330]

\bibitem[{{Schaye} {et~al.}(2015){Schaye}, {Crain}, {Bower}, {Furlong},
  {Schaller}, {Theuns}, {Dalla Vecchia}, {Frenk}, {McCarthy}, {Helly},
  {Jenkins}, {Rosas-Guevara}, {White}, {Baes}, {Booth}, {Camps}, {Navarro},
  {Qu}, {Rahmati}, {Sawala}, {Thomas}, \& {Trayford}}]{EAGLE}
{Schaye}, J. {et~al.} 2015, \mnras, 446, 521, [arXiv:1407.7040]

\bibitem[{{Scoccimarro} \& {Sheth}(2002)}]{PTHALOS}
{Scoccimarro}, R., \& {Sheth}, R.~K. 2002, \mnras, 329, 629,
  [arXiv:astro-ph/0106120]

\bibitem[{{Smith}(2009)}]{Smith_2009}
{Smith}, R.~E. 2009, \mnras, 400, 851, [arXiv:0810.1960]

\bibitem[{{Springel}(2005)}]{Gadget}
{Springel}, V. 2005, \mnras, 364, 1105, [arXiv:arXiv:astro-ph/0505010]

\bibitem[{{Springel}(2010)}]{Arepo}
------. 2010, \mnras, 401, 791, [arXiv:0901.4107]

\bibitem[{{Springel} {et~al.}(2018){Springel}, {Pakmor}, {Pillepich},
  {Weinberger}, {Nelson}, {Hernquist}, {Vogelsberger}, {Genel}, {Torrey},
  {Marinacci}, \& {Naiman}}]{SpringelV_17a}
{Springel}, V. {et~al.} 2018, \mnras, 475, 676, [arXiv:1707.03397]

\bibitem[{{Springel} {et~al.}(2001){Springel}, {White}, {Tormen}, \&
  {Kauffmann}}]{subfind}
{Springel}, V., {White}, S.~D.~M., {Tormen}, G., \& {Kauffmann}, G. 2001,
  \mnras, 328, 726, [arXiv:arXiv:astro-ph/0012055]

\bibitem[{{Taffoni} {et~al.}(2002){Taffoni}, {Monaco}, \&
  {Theuns}}]{Taffoni_2002}
{Taffoni}, G., {Monaco}, P., \& {Theuns}, T. 2002, \mnras, 333, 623,
  [arXiv:astro-ph/0109324]

\bibitem[{{Tassev} {et~al.}(2015){Tassev}, {Eisenstein}, {Wandelt}, \&
  {Zaldarriaga}}]{Tassev_2015}
{Tassev}, S., {Eisenstein}, D.~J., {Wandelt}, B.~D., \& {Zaldarriaga}, M. 2015,
  ArXiv e-prints, [arXiv:1502.07751]

\bibitem[{{Tassev} {et~al.}(2013){Tassev}, {Zaldarriaga}, \&
  {Eisenstein}}]{Tassev_2013}
{Tassev}, S., {Zaldarriaga}, M., \& {Eisenstein}, D.~J. 2013, \jcap, 6, 036,
  [arXiv:1301.0322]

\bibitem[{{Torrey} {et~al.}(2014){Torrey}, {Vogelsberger}, {Genel}, {Sijacki},
  {Springel}, \& {Hernquist}}]{Torrey_2014}
{Torrey}, P., {Vogelsberger}, M., {Genel}, S., {Sijacki}, D., {Springel}, V.,
  \& {Hernquist}, L. 2014, \mnras, 438, 1985, [arXiv:1305.4931]

\bibitem[{{Viel} {et~al.}(2010){Viel}, {Haehnelt}, \& {Springel}}]{Viel_2010}
{Viel}, M., {Haehnelt}, M.~G., \& {Springel}, V. 2010, \jcap, 6, 015,
  [arXiv:1003.2422]

\bibitem[{{Vogelsberger} {et~al.}(2013){Vogelsberger}, {Genel}, {Sijacki},
  {Torrey}, {Springel}, \& {Hernquist}}]{Vogelsberger_2013}
{Vogelsberger}, M., {Genel}, S., {Sijacki}, D., {Torrey}, P., {Springel}, V.,
  \& {Hernquist}, L. 2013, \mnras, 436, 3031, [arXiv:1305.2913]

\bibitem[{{Vogelsberger} {et~al.}(2014{\natexlab{a}}){Vogelsberger}, {Genel},
  {Springel}, {Torrey}, {Sijacki}, {Xu}, {Snyder}, {Bird}, {Nelson}, \&
  {Hernquist}}]{Vogelsberger_2014a}
{Vogelsberger}, M. {et~al.} 2014{\natexlab{a}}, \nat, 509, 177,
  [arXiv:1405.1418]

\bibitem[{{Vogelsberger} {et~al.}(2014{\natexlab{b}}){Vogelsberger}, {Genel},
  {Springel}, {Torrey}, {Sijacki}, {Xu}, {Snyder}, {Nelson}, \&
  {Hernquist}}]{Vogelsberger_2014b}
------. 2014{\natexlab{b}}, \mnras, 444, 1518, [arXiv:1405.2921]

\bibitem[{{Weinberger} {et~al.}(2017){Weinberger}, {Springel}, {Hernquist},
  {Pillepich}, {Marinacci}, {Pakmor}, {Nelson}, {Genel}, {Vogelsberger},
  {Naiman}, \& {Torrey}}]{WeinbergerR_16a}
{Weinberger}, R. {et~al.} 2017, \mnras, 465, 3291, [arXiv:1607.03486]

\bibitem[{{Zennaro} {et~al.}(2017){Zennaro}, {Bel}, {Villaescusa-Navarro},
  {Carbone}, {Sefusatti}, \& {Guzzo}}]{Zennaro_2017}
{Zennaro}, M., {Bel}, J., {Villaescusa-Navarro}, F., {Carbone}, C.,
  {Sefusatti}, E., \& {Guzzo}, L. 2017, \mnras, 466, 3244, [arXiv:1605.05283]

\end{thebibliography}
\bibliographystyle{hapj}

\end{document}